\documentclass[10pt,letterpaper]{article}
\usepackage{geometry}

\usepackage{amsmath,amssymb}

\usepackage{booktabs}

\usepackage{changepage}

\usepackage{textcomp,marvosym}

\usepackage{cite}

\usepackage{nameref}
\usepackage[colorlinks=true,urlcolor=blue]{hyperref} 
\hypersetup{
  colorlinks,
  citecolor=blue,
  linkcolor=blue,
  urlcolor=blue}

\usepackage{microtype}
\DisableLigatures[f]{encoding = *, family = * }

\usepackage[table]{xcolor}

\usepackage{array}

\newcolumntype{+}{!{\vrule width 2pt}}

\newlength\savedwidth

\usepackage[aboveskip=1pt,labelfont=bf,labelsep=period,justification=raggedright,singlelinecheck=off]{caption}

\makeatletter
\renewcommand{\@biblabel}[1]{\quad#1.}
\makeatother

\usepackage{lastpage,fancyhdr,graphicx}
\usepackage{epstopdf}
\usepackage{comment}

\usepackage[listings,breakable,skins,theorems]{tcolorbox}
\usepackage[c]{esvect}
\usepackage{calrsfs} 
\usepackage{stmaryrd}

\newcommand\ten[1]{\overline{\overline{#1}}}
\newcommand{\Id}{\mathbb{I}}

\begin{document}
\vspace*{0.2in}

\begin{flushleft}
{\Large
\textbf\newline{Sensitivity analysis enlightens effects of connectivity in a Neural Mass Model under Control-Target mode}
}
\newline
\\
Vallet Anaïs\textsuperscript{1},
Blanco Stéphane\textsuperscript{2},
Chevallier Coline\textsuperscript{2,3},
Eustache Francis\textsuperscript{1},
Gautrais Jacques\textsuperscript{2,3,*},
Grandpeix Jean-Yves\textsuperscript{4},
Joly Jean-Louis\textsuperscript{2},
Segobin Shailendra\textsuperscript{1},
Gagnepain Pierre\textsuperscript{1}
\\
\bigskip
\textbf{1} Normandie Univ, UNICAEN, PSL Research University, EPHE, INSERM, U1077, CHU de Caen, GIP Cyceron, Neuropsychologie et Imagerie de la Mémoire Humaine, 14000 Caen, France
\\
\textbf{2} LAPLACE, Université de Toulouse, CNRS, INPT, UPS, Toulouse, France
\\
\textbf{3} Centre de Recherches sur la Cognition Animale (CRCA), Centre de Biologie Intégrative (CBI), Université de Toulouse, CNRS, UPS, France
\\
\textbf{4} LMD/IPSL, Sorbonne Université, CNRS, École Polytechnique, ENS, Paris, France
\\

\bigskip
* jacques.gautrais@cnrs.fr
\end{flushleft}

\section*{Abstract}
\emph{Biophysical models} of human brain represent the latter as a graph of inter-connected neural regions.
Building from the model by Naskar et al. \cite{Naskar2021}, our motivation was to understand how these brain regions can be connected at neural level to implement some inhibitory control, which calls for inhibitory connectivity rarely considered in such models.
In this model, regions are made of inter-connected excitatory and inhibitory pools of neurons, but are long-range connected only via excitatory pools (mutual excitation).
We thus extend this model by generalizing connectivity, and we analyse how  connectivity affects the behaviour of this model.

Focusing on the simplest paradigm made of a Control area and a Target area,
we explore four typical kinds of connectivity: mutual excitation, Target inhibition by Control, Control inhibition by Target, and mutual inhibition.
For this, we build an analytical sensitivity framework, nesting up sensitivities of isolated pools, of isolated regions, and of the full system.
We show that  inhibitory control can emerge only in Target inhibition by Control and mutual inhibition connectivities.

We next offer an analysis of how the model sensitivities depends on connectivity structure, depending on a parameter controling the strength of the self-inhibition within Target region.
Finally, we illustrate the effect of connectivity structure upon control effectivity in response to an external forcing in the Control area. 

Beyond the case explored here, our methodology to build analytical sensitivities by nesting up levels (pool, region, system) lays the groundwork for expressing nested sensitivities for more complex network configurations, either for this model or any other one.

\clearpage

\tableofcontents

\clearpage

\section{Introduction}

\subsection{Motivation}

In recent years, a large effort has been made to design and implement large-scale simulations of human brain, in relation to functional neuroimaging data (EEG, MEG, functional MRI, ...) in order to draw inferences regarding neurophysiological mechanisms \cite{Sanzleon2013,Sanz-Leon2015,Amunts2022,Schirner2022}.
In this domain, the class of so-called biophysical models represent the brain as a graph of inter-connected neural masses, and the neural dynamics are used to link structural connectivity  (aka anatomical connectivity, or structural connectome) to functional imaging data (e.g. BOLD signals from fMRI) \cite{Honey2009,Deco2012,Deco2014a,Hansen2015,Castro2020,Kobeleva2022}.
A widespread practice is to use recordings of fMRI to infer the so-called functional connectivity from correlations between time courses of areas' activity level.
This functional connectivity can then be instrumental to tune biophysical models' parameters \cite{Kobeleva2022}.
In such approaches, structural connectivity among cortical areas plays then a crucial role into shaping dynamics and a huge effort has been made to obtain and characterize a precise mapping of brain connectivity by \emph{in vivo} tractography \cite{Sporns2005,Bullmore2009,Rubinov2010,Betzel2016,Elam2021}.

Structural connectivity among cortical areas is known to be supported by long-range connections that can be only excitatory, and in most biophysical models where neural masses activity level are represented by one state variable, they are considered as mutually excitatory.
There are however some contexts in which the influence from one area upon another one is clearly considered as inhibitory by nature, the most prominent paradigm being the tasks involving inhibitory control \cite{Munakata2011,Anderson2021,Apsvalka2022,Wessel2024}.
Inhibitory control is a mechanism depending on prefrontal executive functions, that enables the brain to override or cancel reflexive actions, memories, or emotions by deactivating their associated representations or processes \cite{Apsvalka2022}.

Functional MRI (fMRI) reveals that inhibitory control is characterized by a decrease in BOLD activity in some Target brain regions when some Control brain regions are subject to an increase in BOLD activity.
Analyses of effective connectivity further reveals that the downregulation of the Target regions is mediated by a negative top-down coupling originating from Control regions (e.g. \cite{Gagnepain2017}).

In such context, it must be considered that the activity coming from a control region can have an inhibitory effect upon target populations, namely inhibitory control should translate increasing activities in some control areas into decreased activities in their target areas, an effect that is currently lacking in recent dynamic mean field (DMF) model 
(e.g. \cite{Deco2018,Naskar2021}).

These models should therefore be extended to include long-range connectivity reflecting the feedforward activation of inhibitory interneurons via polysynaptic pathways, and drive suppression of activity in Target region.

As a consequence, there is a clear need to analyse and understand the effect of generalizing signs of connectivity upon the behaviour of such systems.
In this paper, we offer an analytical account of how changing the signs of connectivity can affect activity levels in a widely used biophysical model \cite{Deco2012}. 
We consider the simplest configuration with two areas connected with four scenarios of connectivity, within a control area - target area paradigm.

For this purpose, we develop a methodological framework for building a hierarchy of nested sensitivities and
we analyse the structure of the obtained sensitivities of target area responses, depending on connectivity.
We present our full analysis, in the hope that our results may enlighten modelling choices for connectivity, in the community dealing with biophysical models of human brain.

\subsection{The model}

\subsubsection{Modeling background}

We start from the dynamic mean field (DMF) model of excitatory/inhibitory neural populations first introduced by Deco et al. \cite{Deco2012} (after \cite{Wong2006}), in an effort to account for resting-state networks (RSNs). 
For each area, mean-field description summarises the corresponding underlying spiking neural network, in which excitatory pyramidal neurones and GABAergic inhibitory neurones are mutually interconnected.
At the mean field scale of description, areas are interconnected by long range excitatory signals (NMDA-mediated connections), weighted by anatomical connectivity and a global coupling strength. 

These mean field equations are then integrated numerically, spanning initial conditions, in order to study the stationary state landscape. 
They tuned the global inter-areal coupling strength in the model to fMRI-based functional connectivity in humans and showed that resting brain operates at the edge of multi-stability \cite{Deco2012}.
This DMF model was then used to explore more exhaustively the dynamics of the system \cite{Deco2013a}, especially how noise propagation reflects the double effect of anatomical connectivity and slow dynamics around the spontaneous low-firing stationary state, resulting in the functional connectivity (i.e. pairwise correlations between cortical areas). 
Here, they show that global inter-areal coupling strength fitted to fMRI data rather indicates that brain operates at full multi-stability, yet with resting state being a stable stationary state.
Furthermore, they show that pairwise correlations depend on the sensitivities of each area to the inputs coming from other areas. 
Since these sensitivities depend in turn upon the activity levels at the considered stationary state, affecting these activity levels (e.g. by over-activating a subset of areas) results in a different stationary state, and hence to a different functional connectivity, while anatomical connectivity remains the same  \cite{Deco2013a,Deco2013}. 
The model has then been used to infer anatomical connectivity from functional connectivity, relating structure and function \cite{Deco2014}, to explore the possibility of dynamical transitions between multiple RSNs depending on noise level \cite{Hansen2015}, to suggest that human brain anatomical connectivity may be evolutionary tuned to display the largest diversity in activities of area clusters in response to an increased inter-regional coupling \cite{Castro2020}.

In a refinement of this model, Deco et al.  \cite{Deco2014a} propose to better describe neural mass areas by explicitly considering that they are composed of two pools of neurones: an excitatory one and an inhibitory one, so that the local feedback inhibition can be locally constrained. Under these constraints, the empirical finding that intra-areal correlations should remain poor is recovered, whereas it is violated when using inter-areal coupling in single-pool modelling.
With this choice, the intra-areal activity can be made decorrelated while the inter-areal correlations keep reflecting functional connectivity.
Moreover, and as a consequence of this modelling choice, long range excitatory signals can now target either excitatory pools in other areas or their inhibitory pools, or both.
We then turned to this version of the model since it allows to represent feedforward inhibition, when one area can stimulate the inhibitory pool of another area.
Also, Deco et al.  \cite{Deco2014a} consider the effect of stimulating some random subset (10\%) of areas by exogenous stimuli, representing task-related signals.
Interestingly, in this version, the dynamics yield only one stable stationary state at resting (for reasonable coupling strength avoiding epileptical divergence), and must be stimulated by these exogenous stimuli to switch state.
They advocate that this refinement yield more robust prediction of functional connectivity, and more realistic responses to external stimuli  \cite{Deco2014a}.
This two-pool area model at resting state has been used by Glomb et al. to recover the spatio-temporal structure of RSNs from fluctuations around the single stationary state~\cite{Glomb2017}.
As a final extension, Naskar et al.~ justify the kinetic parameters driving the average synaptic gating variable based on GABA and glutamate concentrations~\cite{Naskar2021}.
We start from this version.

\subsubsection{Formal expression of the model}

In this model, the state variables at neural mass scale, $sn(t)$ and $sg(t)$, are called \emph{ average  synaptic gating variables} and represent the open fraction of NMDA (resp. GABA) synaptic channels for the excitatory (resp. inhibitory) pools.
For each area $i$, their dynamics are given by the following ODEs~:

\begin{equation}
  \left\{
      \begin{aligned}
        \dot{sn_i}(t) &= -\beta^E sn_i(t) &+\alpha^E T_{glu}rn_i(t)(1-sn_i(t)) &+ \sigma \nu_i(t) &\\
        \dot{sg_i}(t) &=-\beta^I sg_i(t) &+\alpha^I T_{gaba} rg_i(t)(1-sg_i(t)) &+\sigma \nu_i(t) &
      \end{aligned}
    \right.
    \label{NaskarSnSg}
\end{equation}

where the intermediate variables $rn(t)$ and $rg(t)$ represent the mean-field firing rates of each pool, which are given by input-output functions  (inspired from~\cite{Abbott2005}):

\begin{equation}
    \left\{
        \begin{aligned}
          rn_i(t) &= \frac{a_E xn_i(t)-b_E}{1-e^{-d_E (a_E xn_i(t)-b_E)}}\\
          rg_i(t) &= \frac{a_I xg_i(t)-b_I}{1-e^{-d_I (a_I xg_i(t)-b_I)}}
        \end{aligned}
      \right.
     \label{NaskarRnRg}
  \end{equation}
 
and depend, in turn, upon the input currents $xn(t)$ and $xg(t)$, that sum up all incoming contributions:

\begin{equation}
  \left\{
      \begin{aligned}
        xn_i(t) &= W_E I_0 &&+W_{+} J_{nmda} sn_i(t) &&-  J_{gaba_i} \ sg_i(t) &+ G J_{nmda}\sum_{j\neq i} C_{ij} sn_j(t)\\
        xg_i(t) &= W_I I_0  &&+ J_{nmda} sn_i(t)  &&-  sg_i(t) &
      \end{aligned}
    \right.
    \label{NaskarXnXg}
\end{equation}

In these representations, coupling are made explicit in Eq~\ref{NaskarXnXg}, namely coupling are modelled as flows of information about fractions of open channels in the excitatory pools of other areas.

\begin{figure}[ht]
\includegraphics[scale=1.]{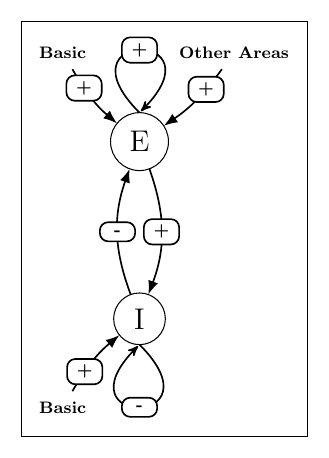}
\centering
\caption{Graphical representation of Eqs. \ref{NaskarSnSg} to \ref{NaskarXnXg} for a given area.}
\label{fig:oneareaNaskar}
\end{figure}

At the area scale, there are four couplings: self-excitatory coupling of the excitatory pool, excitatory coupling from the excitatory pool upon the inhibitory pool, self-inhibitory coupling of the inhibitory pool and inhibitory coupling from the inhibitory pool upon the excitatory pool  (see Fig~\ref{fig:oneareaNaskar}).\\

Coupling among areas are represented  through the sum term where $C_{ij}$ represent normalised anatomical connectivity, $J_{nmda}$ represents excitatory synaptic coupling strength, and $G$ is a positive free parameter which scales the global coupling strength between areas. 
The parameter $J_{gaba_i}$ represents inhibitory synaptic coupling strength, and can be locally adjusted to regulate the balance between long-range excitation and local feedback inhibition so as to ensure homeostasis \cite{Deco2014a}.

Note that without the so-called \emph{basic input current}, $I_0$, the system activity would tend to zero and only noise $\sigma$ would remain, and it is set to a negligible level throughout the papers considered above.

\subsubsection{Generalizing connectivity}

In the model above, coupling between areas is only between excitatory pools.
To allow flexibility, we introduce a further parameter $k_{E_{ij}} \in [0;1]$ which modulates the fraction of long range excitatory input coming from area $j$ that will projects onto the excitatory pool of area $i$, the remaining part being considered to project onto its inhibitory pool.
We also denote $B_{E_i}$ and $B_{I_i}$ the effective external input current so that they can vary among areas, and we introduce a parameter $J_{-}$ to get an homogenous expression.

With these extensions, our system only modifies Eq \ref{NaskarXnXg}, which now reads:

\begin{equation}
  \left\{
      \begin{aligned}
        xn_i(t)&= B_{E_i}&&+ W_{+} J_{nmda} sn_i(t) &&-  J_{gaba_i} sg_i(t) &&+ \sum_{j\neq i} k_{E_{ij}} \kappa_{ij} sn_j(t) &&\\
        xg_i(t)&= B_{I_i} &&+J_{nmda} sn_i(t) &&- J_{-}  sg_i(t) &&+ \sum_{j\neq i} (1-k_{E_{ij}}) \kappa_{ij} sn_j(t) &&
      \end{aligned}
    \right.
\end{equation}

where $\kappa_{ij} \equiv G \ J_{nmda} C_{ij}$ and $J_{-}=1$ nA in \cite{Naskar2021}.

\subsection{Analysis of connectivity effect based on sensitivities}

\subsubsection{Control-Target mode}
Since we are interested in inhibitory control process, we can distinguish two kinds of areas, splitting the system into a subset of control areas, and a subset of target areas.

In the aim to analyse how connectivity among areas will shape the behavioral response of the system, the main picture is then to understand how target areas respond to upregulation applied to control areas, once we take into account that target areas can themselves project back to control areas and have a feedback effect upon their upregulation.

For a single area, a typical time course of the response of the excitatory pool to inputs is shown in Fig~\ref{fig:isole4portes} (purple line) together with the corresponding BOLD signal (green line), as modelled in \cite{Wang2019}.
In response to alternate switching of $B_E$ value (between 0.382 and 0.482 nA), one can see that the activity level $sn(t)$ adjusts very quickly, whereas BOLD signal follows with an observable delay.
This time scale separation holds in the control-target system (see Fig~\ref{fig:BoldSignals} (C) and section \ref{fig:BoldSignalsComments}).
At the neural mass level, we can then study the behavior of the system considering only the value $sn(t)$ at steady states, which are fixed points of the dynamics.
Hence, we can address the question of the effect of connectivity looking only how the fixed points are affected by parameters, namely to predict how the target activity level will be affected.

For this, we will develop a \emph{sensitivity} analysis, in which we build an analytical expression for how the system responds to an infinitesimal perturbation upon some control areas, and how it affects activity levels in all areas (including those upon wich the perturbation is applied).
The point of interest is then how the activity levels of target areas are modified, whether they are up-regulated or down-regulated compared to the baseline, which is given by the sign of sensitivities (i.e. the sign of the derivative of activity level with regards to the perturbation).
Once such sensitivities expressions are established, they are used to analyze the behavior of the system, depending on the kind of connectivity.

\begin{figure}[ht]
\centering
\includegraphics[scale=0.61]{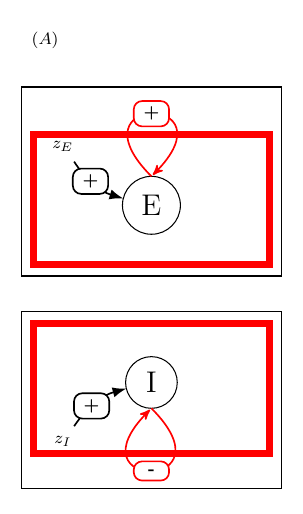}
\includegraphics[scale=0.61]{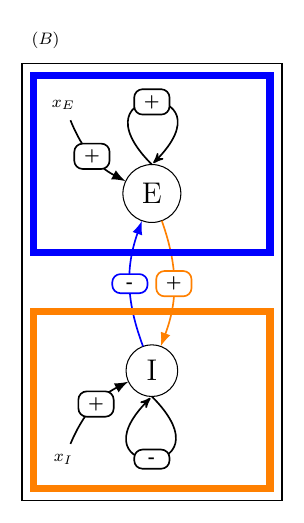}
\includegraphics[scale=0.58]{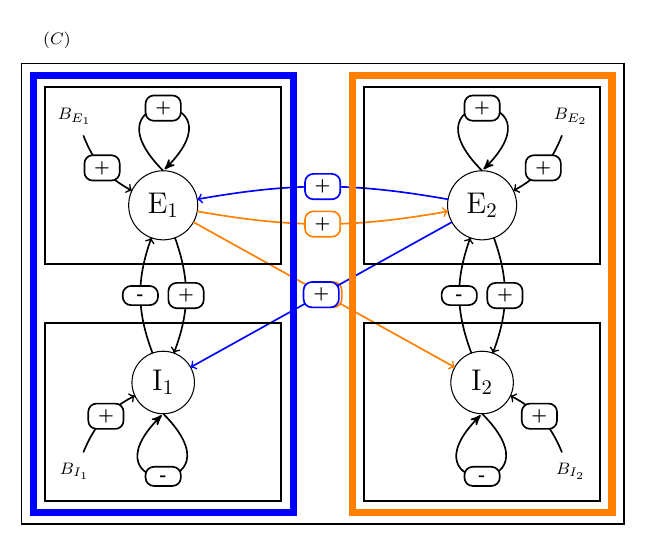}
\caption{\emph{Principle of the method.} (A) One excitatory pool, either excitatory or inhibitory. (B) One area with two coupled pools. (C) Two coupled areas.  \\
In case (A), the sensitivity of the activity level of the pool to perturbation upon its external forcing ($z_E$ or $z_I$) is expressed as function of its sensitivity when uncoupled with itself.\\
In case (B), the sensitivities of the activity level of each pool to either perturbation upon an external forcing ($x_E$ or $x_I$) are expressed as a function of the uncoupled pools'sensitivities, namely, the expressions found in case (A).\\
In case (C), the sensitivities of activity level of the excitatory pool of each area upon an external forcing ($B_{E_1}$, $B_{I_1}$, $B_{E_2}$ or $B_{I_2}$) are expressed as a function of the uncoupled area's sensitivities, namely, the expressions found in case (B). \\
In the three cases, sensitivities at system level are expressed as functions of sensitivities at component level (colored boxes), by opening the feedback loop (cutting colored connections).}
\label{fig:3systems}
\end{figure}

\subsubsection{Formalizing the sensitivity of interest as a hierarchy of nested sensitivities}

The sensitivities are to be estimated at the fixed point corresponding to the resting state. 
Following Naskar et al. \cite{Naskar2021}, the parameters $J_{gaba_i}$ will be adjusted on a per-area basis such that resting state corresponds to a firing rate of the excitatory pools prescribed at $rn_i^*=3$ Hz.

Once this fixed point is established, we apply an infinitesimal perturbation $\delta B_{E_i}$ upon $B_{E_i}$ in control areas, and we will observe its effects upon state variables $sn_j^*$ in target areas.

Hence, basically sensitivities are expressed as $\frac{\delta sn_j^* }{\delta B_{E_i}}$

\bigskip

To derive the analytical expression, we will proceed with a hierarchy of sensitivities, illustrated in Fig~\ref{fig:3systems}. 
Starting from the sensitivities of excitatory and inhibitory pools when considered isolated, we can express the sensitivity for one area where the two pools are coupled.
At the next level, the sensitivity of coupled areas can be expressed as a function of sensitivities of each area when considered uncoupled.
At the next level, the sensitivity for areas belonging to two coupled subsystems can be expressed as functions of sensitivities of areas when considered in their subsystem alone.
In the present paper, we will focus on a system consisting of two areas, one control (denoted as $C$) and one target (denoted as $T$), and  leave  generalization to larger systems to a future companion paper.

Hence, we look for a formal expression for $\frac{\delta sn_T^* }{\delta B_{E_C}}$.


\section{Building a hierarchy of nested sensitivities}

The complete system can be regarded as made of three nested levels : the system itself, which contains areas, which contain pools.
Here, we will expose how we build the analytical expressions of the sensitivities at one level as functions of sensitivities at lower level.
Since, the principle to build sensitivities at one level from sensitivities at the lower level is generic across levels, we will fully expose this principle in the simplest case (from pool level to area level).
For further levels integration, the formal details are given in \nameref{S1_Appendix} and we only report the corresponding results. 

\subsection{At the pool level}
Focusing on one isolated pool (here, one excitatory pool, see Fig~\ref{fig:3systems} (A)) at stationary state, its state variables must obey:

\begin{equation}
  \left\{
      \begin{aligned}
        &0= -\beta^E sn^*+\alpha^E T_{glu}(1-sn^*)rn^* \equiv fn(sn^*,rn^*)\\
        &rn^*= \frac{a_E xn^*-b_E}{1-e^{-d_E (a_E xn^*-b_E)}} \equiv hn(xn^*)\\
        &xn^*= {W_{+}}J_{nmda} sn^* + z_E \equiv wn(sn^*, z_E)
      \end{aligned}
    \right.
\label{eq:PoolEFixedPoints}
\end{equation}

where the \emph{star} notation is to designate fixed point values and $z_E$ represents the external forcing. Below we omit the star notation for the sake of clarity.

If we now consider a perturbation upon $z_E$, we get a new model, which reads: 

\begin{equation}
  \left\{
      \begin{aligned}
        &fn(sn,rn)= 0\\
        &rn= hn(xn)\\
        &xn= wn(sn, z_E + \delta z_E)
       \end{aligned}
    \right.
\end{equation}

and we want the formal expression for the sensitivity $\frac{\delta sn}{\delta z_E}$.

\bigskip

The linearization for the perturbation yields:

\begin{equation}
  \left\{
      \begin{aligned}
      	\frac{\partial fn}{\partial sn} \delta sn + \frac{\partial fn}{\partial rn} \delta rn &= 0\\
      	\delta rn &= \frac{\mathrm{d} hn}{\mathrm{d} xn} \delta xn \equiv hn'\ \delta xn\\
      	\delta xn &= \frac{\partial wn}{\partial sn} \delta sn + \frac{\partial wn}{\partial z_E} \delta z_E
      \end{aligned}
    \right.
\end{equation}

where the derivatives are to be evaluated at fixed point values ( the expression for $hn'$ is given in Sec \ref{sumupIsoPools}).

Plugging the last two equations into the first, we get:

\begin{equation}
 	\frac{\partial fn}{\partial sn}\delta sn 
	   + \frac{\partial fn}{\partial rn} \frac{\mathrm{d} hn}{\mathrm{d} xn}
		\left[
		 \frac{\partial wn}{\partial sn} \delta sn 
		+ \frac{\partial wn}{\partial z_E} \delta z_E
		\right ]= 0
\end{equation}

Rearranging to read how $\delta sn$ depends upon $\delta z_E$:

\begin{equation}
 	\big( 1 + \frac{\partial fn}{\partial sn}^{-1}\frac{\partial fn}{\partial rn} \frac{\mathrm{d} hn}{\mathrm{d} xn} \frac{\partial wn}{\partial sn} \big) \delta sn = -\frac{\partial fn}{\partial sn}^{-1}\frac{\partial fn}{\partial rn} \frac{\mathrm{d} hn}{\mathrm{d} xn}\frac{\partial wn}{\partial z_E} \delta z_E
\end{equation}

This expression is of a primary interest in the methodology to build hierarchical sensitivities, as it represents the \emph{canonical} form for identifying open loop and close loop responses to perturbation upon $z_E$,  that are related through the feedback gain $g_{sn,z_E}$, that would be identified as:

\begin{equation}
(1-g_{sn,z_E}) \delta sn = \delta sn^O 
\end{equation}

Indeed, in the present case of the isolated excitatory pool, what appears as the support of non linearity  is the self-amplification of the excitatory pool by the recurrent connectivity, which is expressed in the function $wn$ through the parameter $W_+$.

In the absence of this recurrent connectivity (setting $W_+=0$, hence, \emph{opening the loop} between $sn^*$ and itself), we would have $\frac{\partial wn}{\partial sn}=0$, so that the only effect of perturbating $z_E$ would translate directly into a perturbation of $sn^*$ in the absence of the feedback loop.
This would amount to consider \emph{opening} the feedback loop of $sn$ upon itself and we can denote the open loop response to perturbation as $\delta sn^O$.

Then, the open loop sensitivity is defined as:

\begin{equation}
 	\mathcal{P}^O_{sn,z_E} 
	\equiv \frac{\delta sn^O}{ \delta z_E}
	= -\frac{\partial fn}{\partial sn}^{-1}\frac{\partial fn}{\partial rn} \frac{\mathrm{d} hn}{\mathrm{d} xn}\frac{\partial wn}{\partial z_E}
\label{eq:PoolEOL}
\end{equation}

Then, we can write the \emph{close loop} sensitivity $\mathcal{P}_{sn,z_E}$ as a function of the \emph{open loop} sensitivity, following:

\begin{equation}
 	\mathcal{P}_{sn,z_E} \equiv \frac{\delta sn}{ \delta z_E} =   \frac{1}{ 1 - g_{sn,z_E} } \mathcal{P}^O_{sn,z_E}
\label{eq:PoolECL}
\end{equation}

where 

\begin{equation}
g_{sn,z_E} \equiv -\frac{\partial fn}{\partial sn}^{-1}\frac{\partial fn}{\partial rn} \frac{\mathrm{d} hn}{\mathrm{d} xn} \frac{\partial wn}{\partial sn}
\label{eq:PoolEgain}
\end{equation}

defines the feedback gain for $sn^*$ upon perturbation of $z_E$ through the feedback loop between  $sn^*$ and itself. 
From here, we can analyze the role of closing the loop, depending on the value of $g$, hence predicting the behaviour of the system in response to a perturbation on external input, namely:
\begin{itemize}
	\item If $g<0$, then $\frac{1}{1-g} \in ]0;1[$, so that $\mathcal{P}_{sn,z_E} < \mathcal{P}^O_{sn,z_E}$. 
    Coupling with a negative feedback gain dampens the sensitivity to inputs.

    \item If $0<g<1$, then $\frac{1}{1-g} \in ]0;+\infty[$, so that $\mathcal{P}_{sn,z_E} > \mathcal{P}^O_{sn,z_E}$. 
    Coupling with a positive feedback gain lower than 1 enhances the sensitivity to inputs.
    
    \item $g \ge 1$ leads the system in uncharted territories that have to be studied on a per-case basis.

\end{itemize}

In the case of the isolated excitatory pool, the feedback gain is positive by construction (see Fig~\ref{fig:PoolE}).

\bigskip

Finally, we can conclude that (see \ref{IsoExcPool} for details):
\begin{equation}
 	\mathcal{P}_{sn,z_E} 
	= 
	\left( 
		\frac{\beta_E + \alpha_E T_{glu} hn(wn(sn^*,z_E))}
		       {\alpha_E T_{glu} (1-sn^*) hn'(wn(sn^*,z_E))} 
		-  {W_{+}}J_{nmda} 
	\right)^{-1} 
	\equiv \varphi n_{_E} (sn^*, z_E)
\label{eq:phiE}
\end{equation}

where we define the function $\varphi n_{_E}$ so as to explicitly express the dependencies of the sensitivity for the isolated excitatory pool. 
 
 \bigskip
 The main point here is to consider that this analytical expression for the sensitivity depends both upon the external forcing, and upon the fixed point under consideration (there can be more than one).
 The expression will remain the same when we will connect the excitatory pool to the inhibitory one, even if its value will change, since the coupling will modify the values of $sn^*$ and $z_E$ at which to compute it.
In the case of the isolated pool, $z_E$ is prescribed as the effective external input, which in turn fixes the fixed point value $sn^*$, whereas in coupled situations, $z_E$ will also depend upon the inputs coming from other parts of the system, which will depend, in turn, upon the stationary values they take in the coupled situation. 

\bigskip
Regarding the inhibitory pool, we proceed the same way (see \ref{IsoInhPool}), and we get:
 
\begin{equation}
	\mathcal{P}_{sg,z_I} 
	= 
 	\left( \frac{\beta_I + \alpha_I T_{gaba} hg(wg(sg^*,z_I))}{\alpha_I T_{gaba} (1-sg^*) hg'(wg(sg^*,z_I))} +  J_{-}\right)^{-1} 	\equiv \varphi g_{_I} (sg^*, z_I)
\label{eq:phiI}
\end{equation}

where function $\varphi g_{_I}$ explicitly express the dependencies of the sensitivity for the isolated inhibitory pool.

In the case of the isolated inhibitory pool, the feedback gain is negative by construction (see Fig~\ref{fig:PoolI}).

\subsection{At the area level}

Each area contains coupled excitatory and inhibitory pools (see Fig~\ref{fig:3systems} (B)).
Here, we are interested in the sensitivity of excitatory and inhibitory pool activity to positive external forcings.

In order to build an analytical expression of these sensitivities, we will consider here again open loop condition versus close loop, now considering that the loop to be opened is the coupling between the two pools.

Open loop sensitivities become the sensitivities of the pools to a perturbation of either $x_E$ or $x_I$ when the connectivity between the two pools is nullified (i.e. in Fig~\ref{fig:3systems} (B), when the colored connectivity is cut), i.e. the sensitivities to a perturbation upon external forcings (now denoted $x_E$ and $x_I$) when the feedback between the two pools is absent, namely the pool-level sensitivities that have just been built in the previous section.

\bigskip

We denote $\mathcal{A}_{sn,x_E}$ (respectively $\mathcal{A}_{sn,x_I}$) the sensitivity of the excitatory pool activity $sn^*$ to the forcing $x_E$ applied upon the excitatory pool (respectively $x_I$ applied upon the inhibitory pool), and $\mathcal{A}_{sg,x_E}$ (respectively $\mathcal{A}_{sg,x_I}$) the sensitivity of the inhibitory pool activity $sg^*$ to the same forcing. We denote $\mathcal{A}_{sn,x_E}^{O}$ (respectively $\mathcal{A}_{sn,x_I}^{O}$,$\mathcal{A}_{sg,x_E}^{O}$,$\mathcal{A}_{sg,x_I}^{O}$) the corresponding open loop sensitivities.\\

Following the same lines as for the isolated pools (see \ref{Sec:SensiUneRegion}), we express close loop sensitivities (sensitivities for the pools when they are coupled) as functions of open loop sensitivities (sensitivities for the pools with only their recurrent coupling).

\bigskip

Finally, we obtain:

\begin{equation}
\begin{aligned}
     \begin{pmatrix}
         \mathcal{A}_{sn,x_E} &  \mathcal{A}_{sn,x_I}\\
         \mathcal{A}_{sg,x_E} & \mathcal{A}_{sg,x_I}
    \end{pmatrix} 
    &=
    \frac{1}{1 + J_{nmda} J_{gaba}\mathcal{A}_{sn,x_E}^{O} \mathcal{A}_{sg,x_I}^{O} }\\
    &\times 
    \begin{bmatrix}
        \mathcal{A}_{sn,x_E}^{O} & -J_{gaba}\mathcal{A}_{sn,x_E}^{O} \mathcal{A}_{sg,x_I}^{O} \\[3mm]
         J_{nmda} \mathcal{A}_{sn,x_E}^{O} \mathcal{A}_{sg,x_I}^{O} & \mathcal{A}_{sg,x_I}^{O} 
    \end{bmatrix}
\end{aligned}
\end{equation}

\bigskip

As mentioned above, the main point here is to consider that open loop sensitivities have the same functional form as the isolated pool sensitivities, only they have to be evaluated at the fixed points $sn^*$ and $sg^*$ yielded by the coupled system, and considering the total amount of external forcings (i.e. basic forcing $x_E$ or $x_I$ together with the forcing coming from the alternate pool: $z_E = -J_{gaba} sg^* + x_E $ et $z_I = J_{nmda} sn^* + x_I$). 

Hence, we can then write the functional forms $\Phi n_{_E}$, $\Phi g_{_E}$, $\Phi n_{_I}$ and $\Phi g_{_I}$ of the sensitivities for an isolated region to explicitly express their dependencies :

\begin{equation}
  \left\{
  \begin{aligned}
       &\mathcal{A}_{sn,x_E}^{O}  &=&\  \varphi n_{_E} (sn^*, -J_{gaba} sg^* + x_E)\\
	&\mathcal{A}_{sg,x_I}^{O} &=& \ \varphi g_{_I} (sg^*, J_{nmda} sn^* + x_I)\\  
     &\Phi n_{_E}(\vv{s^{*}},x_E,x_I) &=& \frac{\mathcal{A}_{sn,x_E}^{O}}{1 + J_{nmda} J_{gaba}\mathcal{A}_{sn,x_E}^{O} \mathcal{A}_{sg,x_I}^{O} }\\[2mm]
     &\Phi g_{_E}(\vv{s^{*}},x_E,x_I) &=& \frac{J_{nmda} \mathcal{A}_{sn,x_E}^{O} \mathcal{A}_{sg,x_I}^{O}}{1 + J_{nmda} J_{gaba}\mathcal{A}_{sn,x_E}^{O} \mathcal{A}_{sg,x_I}^{O} }\\[2mm]
     &\Phi n_{_I}(\vv{s^{*}},x_E,x_I) &=& \frac{-J_{gaba}\mathcal{A}_{sn,x_E}^{O} \mathcal{A}_{sg,x_I}^{O}}{1 + J_{nmda} J_{gaba}\mathcal{A}_{sn,x_E}^{O} \mathcal{A}_{sg,x_I}^{O} }\\[2mm]
     &\Phi g_{_I}(\vv{s^{*}},x_E,x_I) &=& \frac{\mathcal{A}_{sg,x_I}^{O} }{1 + J_{nmda} J_{gaba}\mathcal{A}_{sn,x_E}^{O} \mathcal{A}_{sg,x_I}^{O} }\\
  \end{aligned}
  \right.
\label{eq:sensisUneRegion}
\end{equation}
 
where $\varphi n_{_E}$ and $\varphi g_{_I}$ are given by Eq \ref{eq:phiE} and \ref{eq:phiI}.

\subsection{At two-area system level}

We now turn to the coupling between two areas (see Fig~\ref{fig:3systems} (C)).

In the same way that we have built isolated area sensitivities using functions $\varphi n_{_E}$ and $\varphi g_{_I}$ of isolated pool sensitivities (see \eqref{eq:sensisUneRegion}), we will use the functions $\Phi n_{_E}$, $\Phi g_{_E}$, $\Phi n_{_I}$ and $\Phi g_{_I}$ defined in \eqref{eq:sensisUneRegion} to express open loop sensitivities of areas in the two-area system. Hence the loop to be opened is now the connections between the two areas.

In a first step, we have built, in full generality for any two-area system, the expression of the matrix of sensitivities to a perturbation upon the input current in one of the four pools, and they are expressed as functions of the sensitivities in the single-area system (see \ref{twocoupledareas}).

\subsection{Control-Target system}
In a second step, we use this general expression to address the question of how sensitivities would drive the response of the excitatory pool of one area to the activation of the excitatory pool of the other one, depending on the connectivity between the two areas.
Hence, we attribute a role to each area: the area receiving a perturbated input current will be called "Control" area (denoted by C), and the other area will be called "Target area" (denoted by T).
Since we are interested in understanding how BOLD signals can be affected by connectivity, we will focus on the expression of the perturbations of their excitatory pools $\delta sn_C$ and $\delta sn_T$ (which drive changes in BOLD signals, see \cite{Wang2019}) in response to a perturbation $\delta B_{E_C}$ applied to the input current of Control Area.

We obtain (see \ref{twocoupledareas}):

\begin{align}
 \begin{pmatrix}
        \delta sn_C \\
        \delta sn_T \\
    \end{pmatrix}
    &=\frac{1}{1 -G_{CT} G_{TC} }
    \begin{pmatrix}
        \mathcal{A}_{sn_C,x_{E_C}} \\
         G_{TC} \mathcal{A}_{sn_C,x_{E_C}} 
    \end{pmatrix} 
    \delta B_{E_C}
\end{align}

with

\begin{equation}
    \left\{
    \begin{aligned}
        G_{CT} &= \mathcal{A}_{sn_C,x_{E_C}}  k_{E_{CT}}\kappa_{CT} + \mathcal{A}_{sn_C,x_{I_C}} (1-k_{E_{CT}})\kappa_{CT} \\
        G_{TC} &= \mathcal{A}_{sn_T,x_{E_T}}  k_{E_{TC}}\kappa_{TC} +  \mathcal{A}_{sn_T,x_{I_T}}  (1- k_{E_{TC}})\kappa_{TC} 
    \end{aligned}
    \right.
\end{equation}

\bigskip

Written in the canonical form, we then have for Control area:

    \begin{equation}
            (1 - G_{CT} G_{TC} ) \delta sn_C = \mathcal{A}_{sn_C,x_{E_C}} \delta B_{E_C}
    \end{equation}

so we can define $g_{sn_C,B_{E_C}} = G_{CT} G_{TC}$ the feedback gain for $\delta B_{E_C} $ acting on $\delta sn_C$ through the feedback loop, from Control to Control area via Target area.

\bigskip

For Target area, we have:

    \begin{equation}
            (1 -G_{CT} G_{TC} ) \delta sn_T = G_{TC} \mathcal{A}_{sn_C,x_{E_C}} \delta B_{E_C}
    \end{equation}
    
 so that the feedback gain $g_{sn_T,B_{E_C}}$ acting on $\delta sn_T$  is the same, namely:
 \begin{equation}
 \begin{aligned}
    g_{sn_{C|T},B_{E_C}} 
    & = \big(\mathcal{A}_{sn_C,x_{E_C}}  k_{E_{CT}}\kappa_{CT} + \mathcal{A}_{sn_C,x_{I_C}} (1-k_{E_{CT}})\kappa_{CT} \big)\\ 
    &\times \big(\mathcal{A}_{sn_T,x_{E_T}}  k_{E_{TC}}\kappa_{TC} +  \mathcal{A}_{sn_T,x_{I_T}}  (1- k_{E_{TC}})\kappa_{TC} \big)
\end{aligned}
\label{eq:fbgain}
\end{equation}

\subsection{Effect of connectivity upon Target sensitivity}
\label{subsec:ConnectivityTargetSensi}

So far, we have derived the formal expressions for sensitivities and gains for any connectivity, i.e. for any values of $k_{E_{ij}}$ that governs how excitatory and inhibitory pools are connected between areas.
We will now consider archetypal possibilities by assigning binary values to $k_{E_{CT}}$ and $k_{E_{TC}}$.

\begin{figure}[ht]
\begin{tabular}{p{0.1cm} l p{0.1cm} l}
(A)
&\raisebox{-\height}{\includegraphics[scale=0.5]{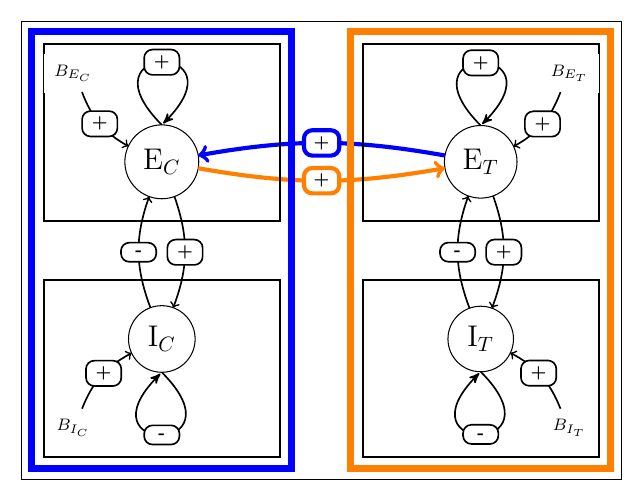}}
&(B)
&\raisebox{-\height}{\includegraphics[scale=0.5]{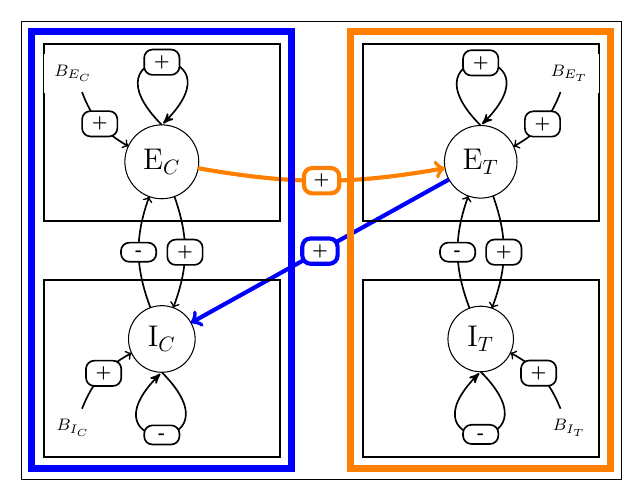}}\\
(C)
&\raisebox{-\height}{\includegraphics[scale=0.5]{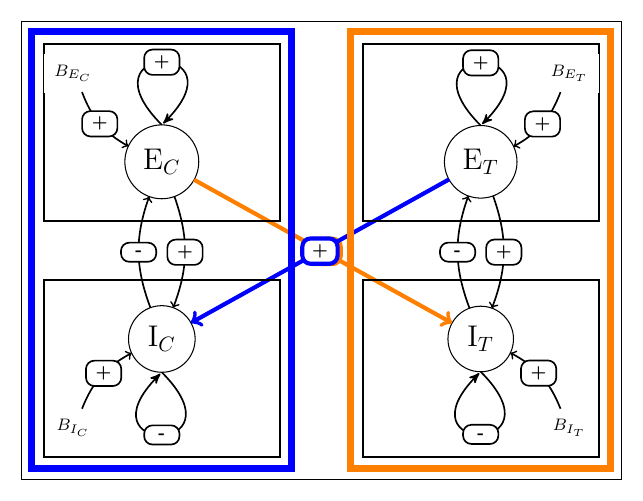}}
&(D)
&\raisebox{-\height}{\includegraphics[scale=0.5]{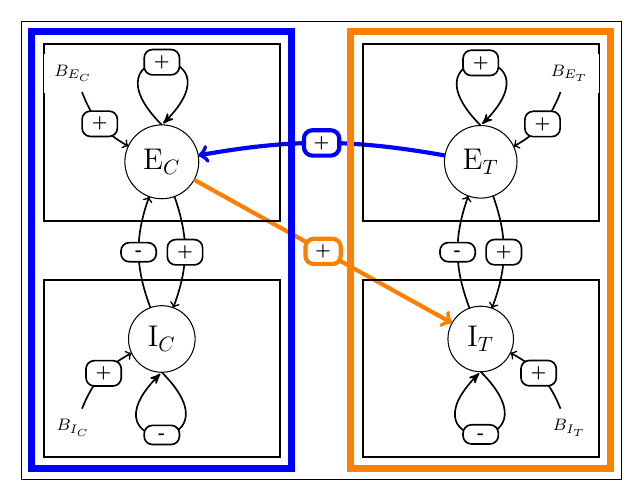}}\\
\end{tabular}
\centering
\caption{\emph{Connectivity that are considered}.
(A)~Mutual excitation (E-E), (B)~Target-to-Control inhibition (I-E), (C)~Mutual inhibition (I-I), (D)~Control-to-Target inhibition (E-I).
We denote the connectivity by $\square - \square$ where first $\square$ denotes the reception pools ($E$ or $I$) for the Control (in blue) and second one for the Target (in orange) areas respectively.}
\label{fig:connectivities}
\end{figure}

Considering that (long-range) connexions always originate from excitatory pools, we then have four possibilities, depending for each area on whether it receives excitatory signal upon its excitatory pool or its inhibitory one (see Fig~\ref{fig:connectivities}).
We can then produce predictions on system behavior for the schemes of connectivity representing inhibitory control models: Target-to-Control inhibition (I-E), mutual inhibition (I-I) or Control-to-Target inhibition (E-I),  w.r.t. the "model of reference" which is mutual excitation (E-E).

We convene to denote the connectivity by $\square - \square$ where $\square$ denotes the reception pools ($E$ or $I$) for the Control and Target areas respectively.

Accordingly, we will denote the corresponding feedback gain $g_{sn_T,B_{E_C}}$ as: $g_{_{\square - \square}}$.

\bigskip

We can then write a generic expression of feedback gains and sensitivities for any connectivity:

\begin{equation}
\boxed{
    \left\{
    \begin{aligned}
        g_{_{\square - \square}} & = \mathcal{A}_{sn_C,x_{\square_C}}\kappa_{CT} \mathcal{A}_{sn_T,x_{\square_T}}\kappa_{TC} \\
        \delta sn_C & =\frac{\mathcal{A}_{sn_C,x_{E_C}} }{1-g_{_{\square - \square}}} \delta B_{E_C} \\
        \delta sn_T & = \frac{ \mathcal{A}_{sn_T,x_{\square_T}} \kappa_{TC} \mathcal{A}_{sn_C,x_{E_C}}     }{1- g_{_{\square - \square}}} \delta B_{E_C} = \mathcal{A}_{sn_T,x_{\square_T}} \kappa_{TC}  \delta sn_C
    \end{aligned}
    \right.
    }
\end{equation}

yielding the sensitivities given in table \ref{tab:sensisCT} for each connectivity.
 
From this, we can now predict the signs of sensitivities:
\begin{itemize}
  
\item For $g_{_{\square - \square}}<1$, since $\mathcal{A}_{sn_C,x_{E_C}}>0$ (see section \ref{subsec:coupling2pools}), we always have $\delta sn_C>0$.

\item Sensitivities for Target area are given by:

\begin{center}
\begin{tabular}{|c|c|c|c|} 
    \hline $\square - \square$ &  $\mathcal{A}_{sn_T,x_{\square_T}}$   & $\delta sn_T$ \\[4mm] 

    \hline $E-E$ & $\mathcal{A}_{sn_T,x_{E_T}} > 0$ & $> 0$  \\[4mm] 

    \hline $I-E$ & $\mathcal{A}_{sn_T,x_{E_T}} > 0$ & $>0$ \\[4mm] 

    \hline $I-I$ & $\mathcal{A}_{sn_T,x_{I_T}} < 0$ & $< 0 $  \\[4mm] 

    \hline $E-I$ & $\mathcal{A}_{sn_T,x_{I_T}} < 0$ & $<0$\\[4mm] 

    \hline 
\end{tabular}
 \end{center}

\end{itemize}

We conclude that only I-I or E-I connectivity could translate upregulation in Control area into downregulation in Target area.

\clearpage

\bigskip

\begin{table}[h!]
\begin{tabular}{|c|c|c|c|} 
    \hline $\square - \square$ & $g_{_{\square - \square}}$ & $\delta sn_C$ & $\delta sn_T$ \\[4mm] 

    \hline $E-E$ & $\mathcal{A}_{sn_C,x_{E_C}} \kappa_{CT} \mathcal{A}_{sn_T,x_{E_T}}  \kappa_{TC}$ &  $\frac{\mathcal{A}_{sn_C,x_{E_C}}}{1 -g_{E-E} } \delta B_{E_C}$  & $\mathcal{A}_{sn_T,x_{E_T}}  \kappa_{TC}  \delta sn_C$  \\[4mm] 

    \hline $I-E$ & $\mathcal{A}_{sn_C,x_{I_C}}  \kappa_{CT} \mathcal{A}_{sn_T,x_{E_T}} \kappa_{TC}$ & $\frac{\mathcal{A}_{sn_C,x_{E_C}}}{1 - g_{I - E}} \delta B_{E_C}$ & $\mathcal{A}_{sn_T,x_{E_T}} \kappa_{TC}  \delta sn_C$ \\[4mm] 

    \hline $I-I$ & $\mathcal{A}_{sn_C,x_{I_C}} \kappa_{CT} \mathcal{A}_{sn_T,x_{I_T}}  \kappa_{TC}$ & $\frac{\mathcal{A}_{sn_C,x_{E_C}}}{1 -g_{I - I}} \delta B_{E_C}$ & $\mathcal{A}_{sn_T,x_{I_T}}  \kappa_{TC}  \delta sn_C$  \\[4mm] 

    \hline $E-I$ & $\mathcal{A}_{sn_C,x_{E_C}}  \kappa_{CT}\mathcal{A}_{sn_T,x_{I_T}} \kappa_{TC}$ & $\frac{\mathcal{A}_{sn_C,x_{E_C}}}{1 -g_{E - I} } \delta B_{E_C}$ & $\mathcal{A}_{sn_T,x_{I_T}} \kappa_{TC}  \delta sn_C$\\[4mm]

    \hline 
    \end{tabular}

    \caption{Sensitivities for $sn^*$ in Control and Target area to perturbation upon $B_{E_C}$, the input current into Control Area.  }
    \label{tab:sensisCT}
\end{table}

\clearpage

\section{Interpretation and discussion}

\subsection{Effect of self-coupling in pools}
\label{subsec:selfcoupling}

\begin{figure}[ht]
\includegraphics[scale=0.8]{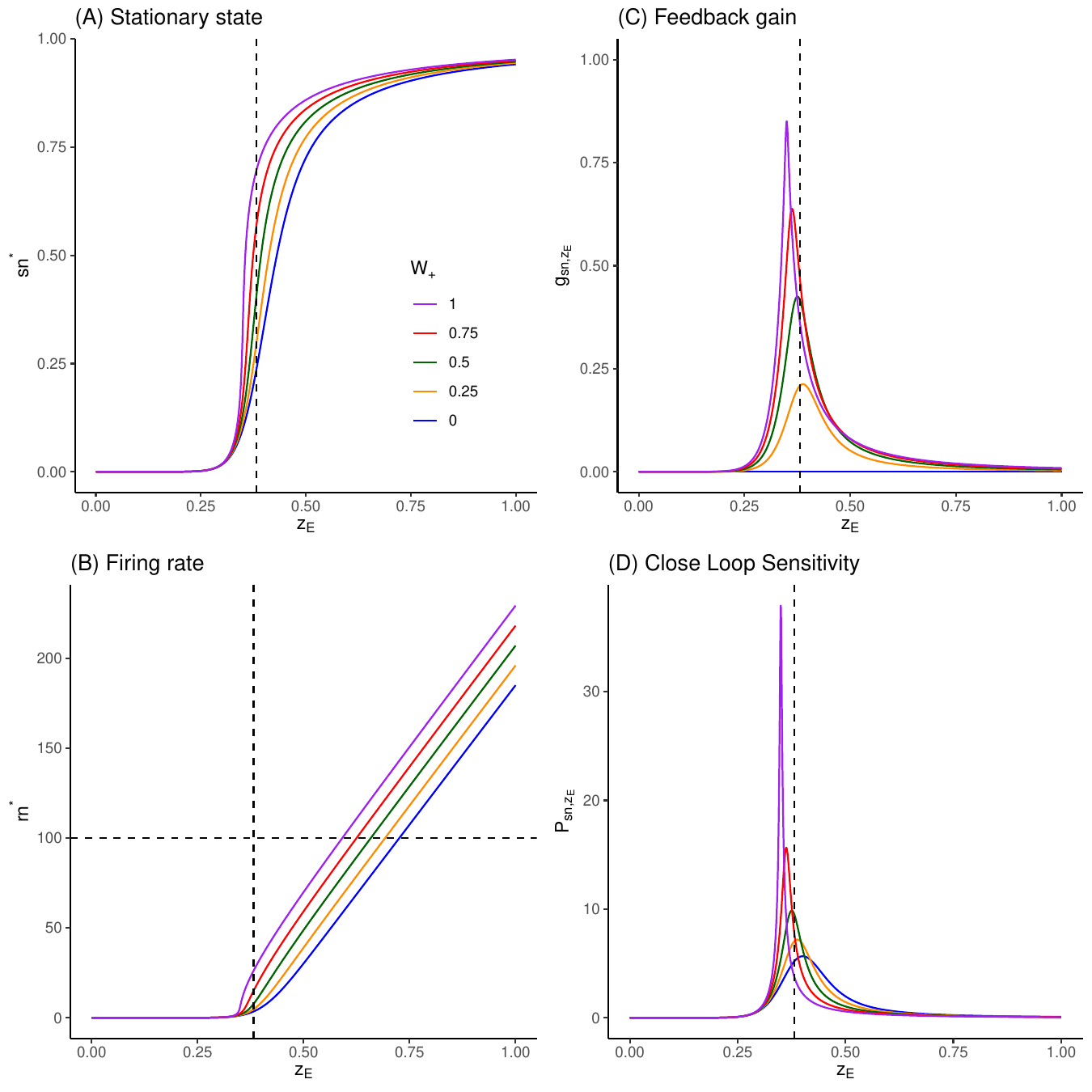}
\centering
\caption{\emph{Response of excitatory pool to inputs, feedback gain and sensitivities.} (A) Fixed point values of $sn^*$ are numerically extracted by root-finding of Eq \ref{eq:PoolEFixedPoints}, for values of $z_E \in [0;1.0]$ and $W_+ \in (0,0.25,0.5,0.75,1)$. Vertical dashed line correspond to $z_E=x_{E,\mathrm{ref}}=0.382$ nA. (B) Corresponding firing rates, following Eq \ref{NaskarRnRg}. (C) Corresponding feedback gain values, following Eq \ref{eq:PoolEgain}. (D) Corresponding close loop sensitivities,  following Eq \ref{eq:phiI}. See comments in Sec~\ref{subsec:selfcoupling}.
}
\label{fig:PoolE}
\end{figure}

To illustrate the behavior of the basic unit of the system, the effect of incoming input $z_E$ upon the activity level $sn^*$ in an isolated excitatory pool is illustrated in Fig~\ref{fig:PoolE} for increasing values of its self-excitatory coupling parameter $W_+$.

As $z_E$ increases, fixed points $sn^*$ increase monotonically up to the saturating value 1 (Fig~\ref{fig:PoolE} (A)).
With the set of parameters given in \cite{Naskar2021} (listed in Sec \ref{S2_Appendix}), we observe a threshold effect, with an absence of reaction to input values lower than about $z_E \simeq 0.3$ nA.
This threshold effect is due to the filtering effect in the translation from input to $sn$ through firing rate $rn$ (Eq \ref{NaskarRnRg}, Fig~\ref{fig:PoolE} (B)), so that if $rn^*\rightarrow 0$ in Eq \ref{eq:PoolEFixedPoints}, so must do $sn^*$.

Beyond this threshold, fixed points values increase strongly with small increments of $z_E$, such that the range of relevant input values is pretty narrow.
If we consider a firing rate at 100 Hz as the reasonable maximal value for a highly activated pool, this range would span $[0.3;0.75]$ nA in absence of self-excitatory loop ($W_+=0$) down to $[0.3;0.6]$ nA for $W_+=1$ (Fig~\ref{fig:PoolE} (B)).

This high reactivity to input is explained by the feedback gain (Fig~\ref{fig:PoolE} (C)), which is positive by construction and can get strong values as the self-excitatory coupling parameter $W_+$ is increased.
When $W_+=0$, the gain is analytically null (blue curves), and the corresponding sensitivity  is the open loop one (the lowest one in Fig~\ref{fig:PoolE} (D)).
As $W_+$ increases, the gain remains null for $sn^* \rightarrow 0$ corresponding to $z_E < 0.3$ (since there is no input for self-amplification), and also saturates towards 0 for $sn^*$ saturating to 1 (since there is no more possible gain). 
In between, the gain reaches a maximal value in the range of $z_E$ which has a maximal impact upon $sn^*$. 
The close loop sensitivity is then scaled accordingly.

\begin{figure}[ht]
\includegraphics[scale=0.8]{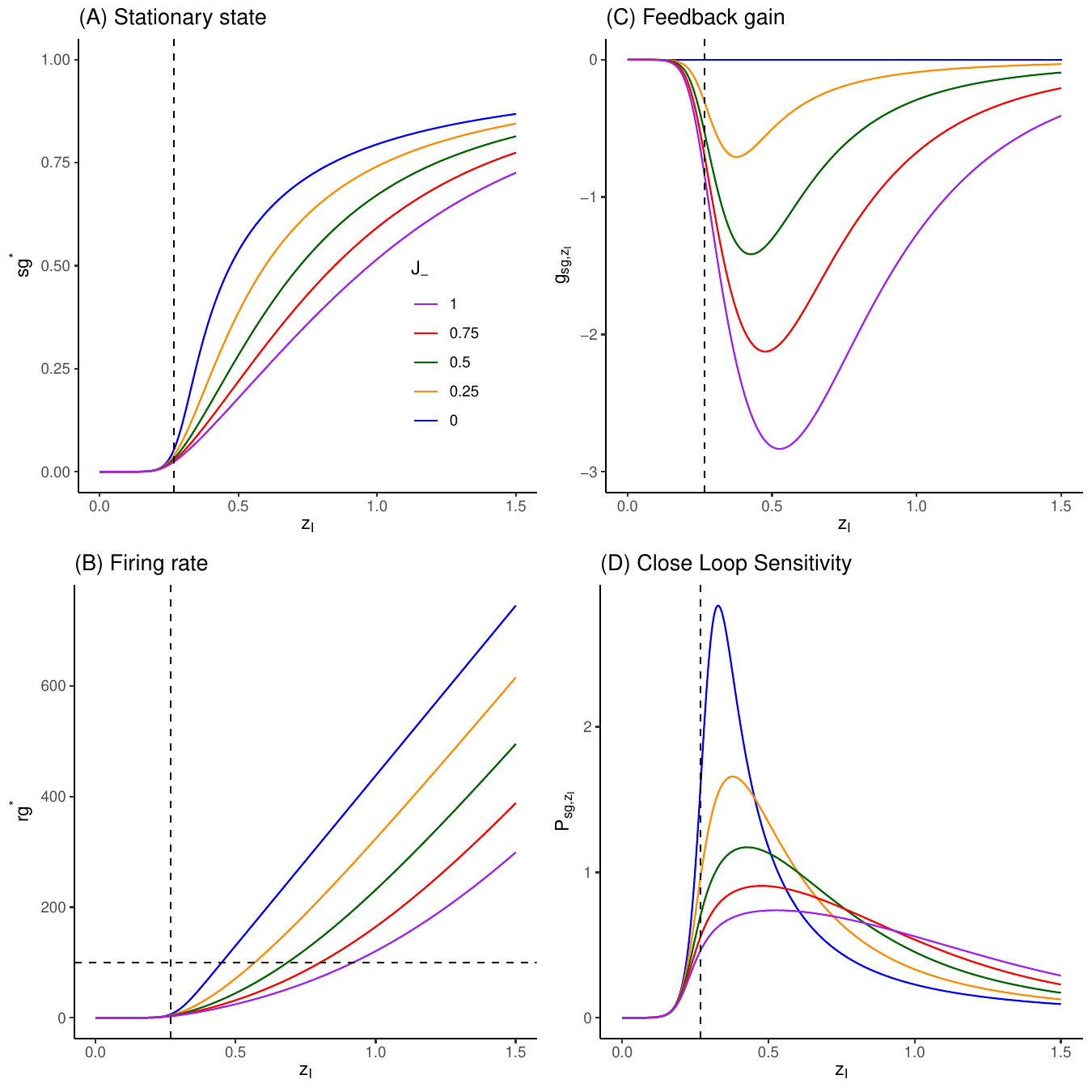}
\centering
\caption{\emph{Response of inhibitory pool to inputs, feedback gain and sensitivities.} (A) Fixed point values of $sg^*$ are numerically extracted by root-finding, for values of $z_I \in [0;1.5]$ and $J_{-} \in (0,0.25,0.5,0.75,1)$. Vertical dashed line correspond to $z_I=x_{I,\mathrm{ref}}=0.382*0.7$ nA. (B) Corresponding firing rates, following Eq \ref{NaskarRnRg}. (C)  Corresponding feedback gain values. (D) Corresponding (close loop) sensitivities, following \ref{eq:phiI}. See comments in Sec~\ref{subsec:selfcoupling}.
}
\label{fig:PoolI}
\end{figure}

\bigskip
The effect of incoming input $z_I$ upon the activity level $sg^*$ in an isolated inhibitory pool is illustrated in Fig~\ref{fig:PoolI} for increasing values of its self-inhibitory coupling parameter $J_{-}$.
As expected, increasing $z_I$ increases fixed points $sg^*$ values monotonically, but here, at a far lower rate than in excitatory pool (Fig~\ref{fig:PoolI} (A)).
Since the self-coupling is inhibitory, the feedback gain is negative (Fig~\ref{fig:PoolI} (C)), hence the sensitivity is the highest in absence of it (blue curves, Fig~\ref{fig:PoolI} (D)).
As a consequence, the self-inhibitory coupling will extend the  range of relevant input values $z_I$ for firing rates below 100 Hz (Fig~\ref{fig:PoolI} (B)), from about $[0.2;0.45]$ nA  to about $[0.2;0.9]$ nA.

\subsection{Effect of coupling two pools}
\label{subsec:coupling2pools}

On Figs.~\ref{fig:PoolE} (A) and \ref{fig:PoolI} (A), vertical dashed lines report the values representing \emph{basic external inputs} $x_E$ and $x_I$ when the pools are connected to form an area, and that are given in \cite{Naskar2021}.
From their location in the range of relevant values, we can anticipate the role of the inhibitory pool upon the basic level of activity in one isolated area: since $x_I$ is close to the threshold, it will be up-regulated by the excitatory input from the excitatory pool, so that its induced inhibitory input backward to the excitatory pool will actually lower $z_E$ towards values near the threshold.
Namely, with the value of $x_E$ prescribed by \cite{Naskar2021}, the excitatory pool \emph{needs} to be down-regulated to have reasonable excitation level at the resting state.

\begin{figure}[ht]
\begin{tabular}{p{0.1cm} l p{0.1cm} l}
(A)
&\raisebox{-\height}{\includegraphics[scale=0.5]{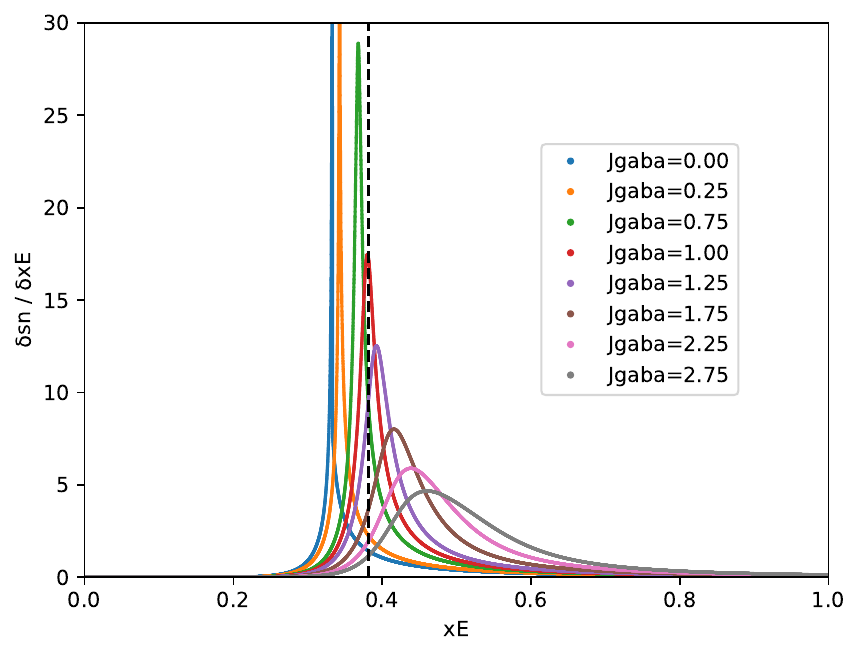}}
&(D)
&\raisebox{-\height}{\includegraphics[scale=0.5]{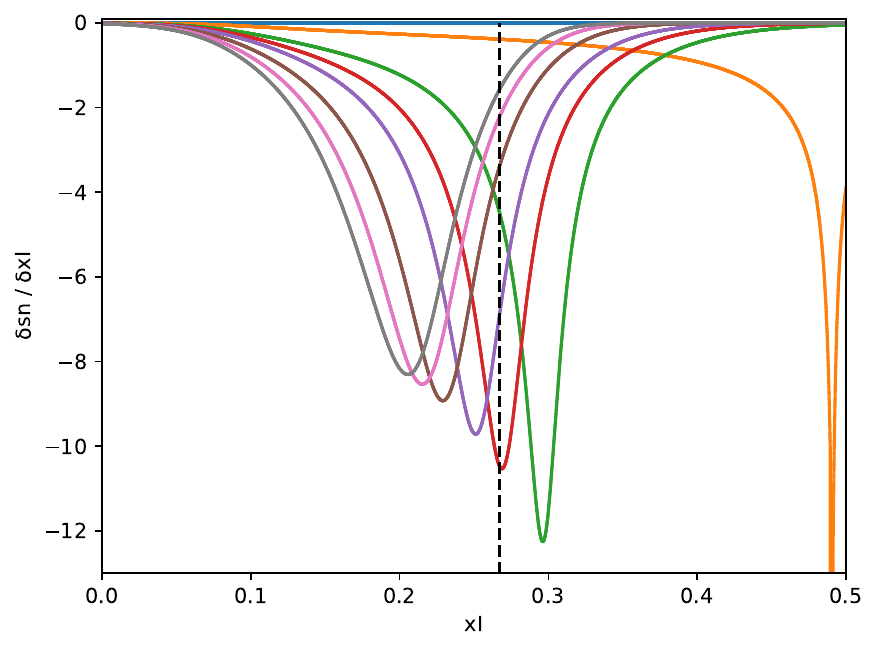}}\\
(B)
&\raisebox{-\height}{\includegraphics[scale=0.5]{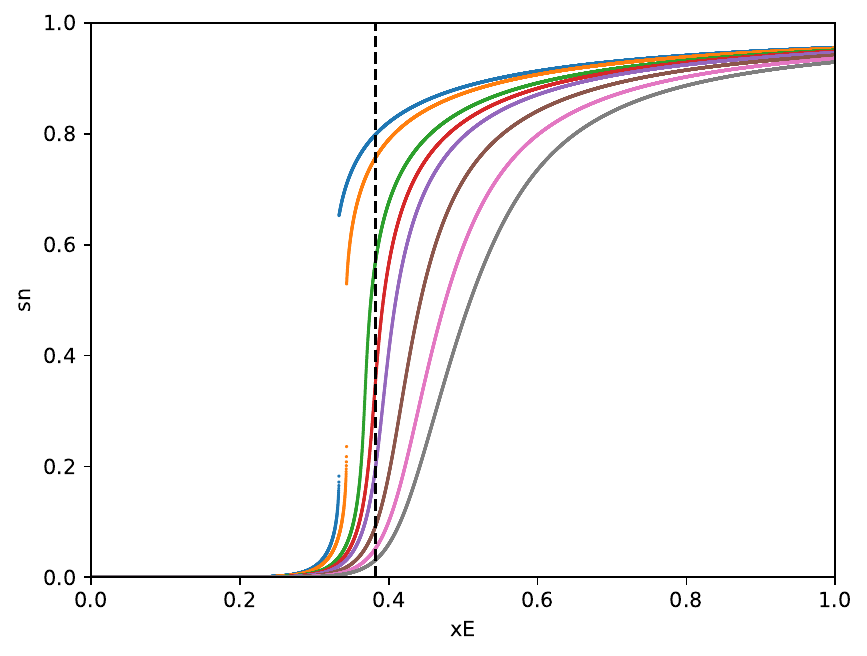}}
&(E)
&\raisebox{-\height}{\includegraphics[scale=0.5]{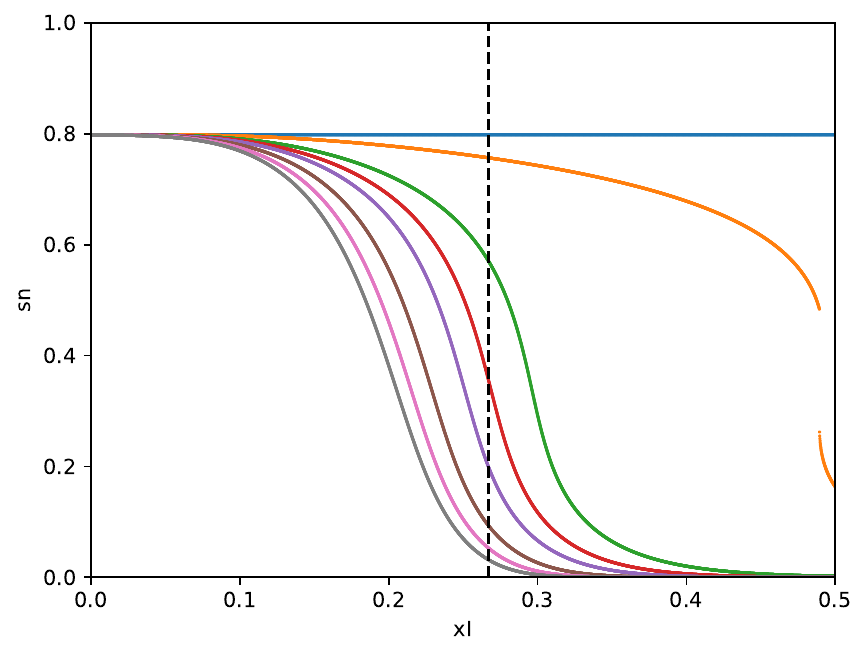}}\\
(C)
&\raisebox{-\height}{\includegraphics[scale=0.5]{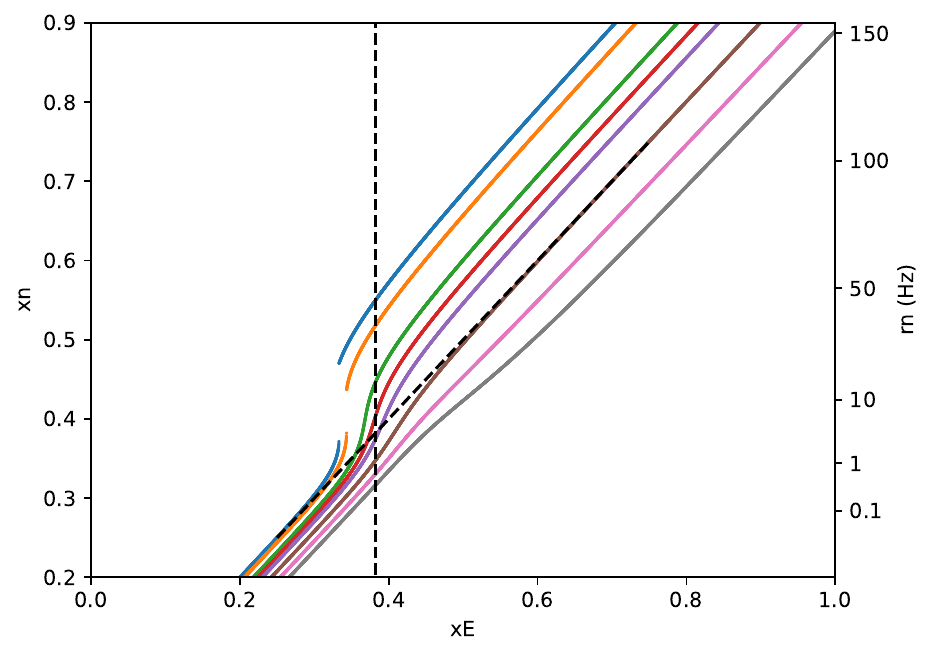}}
&(F)
&\raisebox{-\height}{\includegraphics[scale=0.5]{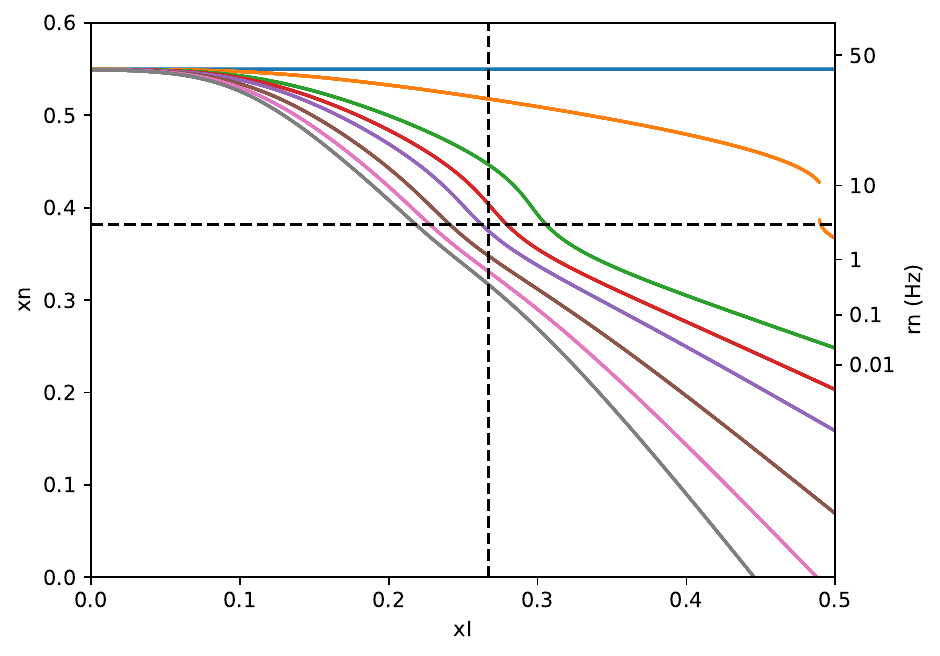}}
\end{tabular}
\centering
\caption{\emph{Sensitivities and responses of excitatory pool in an isolated area.} (A---C)  Response to external input upon the excitatory pool, (D---F) Response to external input upon the inhibitory pool. (A, D) Sensitivities, (B, E) $sn^*$, (C, F) $xn^*$. See comments in Sec~\ref{subsec:coupling2pools}.
}
\label{fig:RegionIsolee}
\end{figure}

This effect of coupling the two pools is illustrated in Fig~\ref{fig:RegionIsolee}, where we submit the area to varied values of $x_E$ or $x_I$ around the values of references (vertical dashed lines) and for a range of $J_{gaba}$ which controls the strength of feedback inhibition from inhibitory pool to excitatory pool upon activation of the latter. This feedback inhibition then acts as a self-inhibition at the area level.

We report on the left column how varying $x_E$, while keeping $x_I$ at the reference value, affects the activity level of the excitatory pool.
In this case, increasing $x_E$ will, on the one hand, directly enhance activity level in the excitatory pool, on the other hand, this enhanced activity level will in turn enhance activity level in the inhibitory pool (by feedforward activation), which will in turn exert a feedback inhibition effect upon the excitatory pool. 
All feedbacks taken into account (self-excitation and feedback inhibition), the sensitivity $\mathcal{A}_{sn,x_E}$ remains positive (Fig~\ref{fig:RegionIsolee} (A)), albeit with a responsiveness that can be far smoother than for the isolated pool.

We first note that sensitivities have here again bell-shape curves.
Indeed, for lower values of external input ($x_E < 0.25$), $sn^*$ values would slowly tend to 0, whereas for too high values of external input ($x_E > 0.6$) it would slowly saturate to 1 (Fig~\ref{fig:RegionIsolee} (B)).
In between, sensitivities must have a peak value. 

The responsiveness to external inputs depends upon the strength of the self-inhibition at the area level, which is governed by  $J_{gaba}$.
For lower values of $J_{gaba}$, the responsiveness is steep, and the range of operational values of total input remains narrow.
As a matter of fact, in the case $J_{gaba}=0$, we recover the sensitivity of the isolated excitatory pool (Eq \ref{eq:phiE}), yet for the high value $W_{+}=1.4$ that is prescribed when coupling pools in \cite{Naskar2021}, and we observe a divergence of the sensitivity around $x_E\simeq 0.33$, which translates into a jump for $sn^*$ in Fig~\ref{fig:RegionIsolee} (B).

This means that, in absence of a sufficient down-regulation of the excitatory pool by the feedback inhibition, the slightest variation in total input could translate into an all-or-nothing response, which is not a property we would consider as reasonable for a neural pool.
This divergence appears no more for $J_{gaba}$ greater than 0.75 nA.

As $J_{gaba}$ value becomes larger and larger, the responsiveness is smoother and smoother as the peak of sensitivity to inputs tends to decrease, and the operational range of input values increases.
Since sensitivities can be regarded as the derivatives of $sn^*$ with respect to the input, their bell-shaped curves translate into \emph{sigmoidal} shapes when considering $sn^*$  (Fig~\ref{fig:RegionIsolee} (B)) and the enlargement of the operational range translates into a response of the pool that becomes gradual.

Depending on $J_{gaba}$, the $sn^*$ level at the value of reference for $x_E$ spreads over quite large a range (from about 0.6 for $J_{gaba}=0.75$ down to about 0.05 for $J_{gaba}=2.25$).

 This spread in the excitatory pool $sn^*$ is however largely reduced when considering the total effective input $xn^*$, as it is tempered by the negative contribution from the inhibitory input.
 For instance, in the case $J_{gaba}=0.75$ where $sn^*\simeq0.6$, the total input current is about $xn^*\simeq0.45$ (Fig~\ref{fig:RegionIsolee} (C)) and would translate into a limited firing rate ($rn^*\simeq 16$ Hz whereas it would be $rn^*=45$ Hz for $J_{gaba}=0$).
 
 Actually, for large enough a value for $J_{gaba}$ (around 1.75nA), inhibitory feedback control at the area level would compensate exactly for self-excitatory feedback in the excitatory pool (Fig~\ref{fig:RegionIsolee} (C), where oblique dotted line corresponds to $xn^*=x_E$).
With such a high value of $J_{gaba}$, an external input $x_E=0.75$ nA would be fully counter-balanced and yield a firing rate at 100 Hz, as if in an isolated excitatory pool with no self-excitation ($W_{+}=0$, blue curve in Fig~\ref{fig:PoolE} (B)).

\bigskip

On the right column of Fig~\ref{fig:RegionIsolee}, we report the symmetrical effect of varying $x_I$ while keeping $x_E$ at the value of reference.
The sensitivity $\mathcal{A}_{sn,x_I}$ of the activity ($sn^*$) in the excitatory pool with regards to positive perturbation upon input current into the inhibitory pool is, as expected, negative (Fig~\ref{fig:PoolE} (D)) since increasing input current into the inhibitory pool will lower the input current into the excitatory pool (Fig~\ref{fig:PoolE} (F)).
We note that in absence of external input current into the inhibitory pool ($x_I=0$), we have an almost null activity in the inhibitory pool (i.e. $sg^*\simeq0$, not shown), meaning that the forward excitation from the excitatory pool is not enough \emph{per se} to activate it (with the current set of parameters) so that we recover the high stationary state of the isolated excitatory pool ($sn^*=0.8$).

On the other hand, with too strong an activation of the inhibitory pool ($x_I>0.4$), $sn^*$ would be flattened to 0 and loose any responsiveness.

In between, the shape of sensitivities curves are poorly affected by $J_{gaba}$ (for $J_{gaba} \ge 0.75$) so that the range of operational values for $x_I$ remains about the same, yet with a median value which is lower and lower as $J_{gaba}$ is increased.

In Fig~\ref{fig:RegionIsolee} (F), vertical dashed line represents $x_{I,\mathrm{ref}}=0.382 \times 0.7$ nA and horizontal dashed line represents the cases $x_n =x_{E,\mathrm{ref}}=0.382$ nA, corresponding to situations when inhibitory feedback compensates self-excitation.
When $xn < x_{E,\mathrm{ref}}$ (below horizontal dashed line), inhibitory feedback over compensates self-excitation. In the opposite case, inhibitory feedback only acts as a brake on self-excitation.
A crossing value between the two behaviours appears to be around $J_{gaba}=1.25$ nA.
An important point is then that, with $J_{gaba} \ge 1.25$
, any additional input current to the inhibitory pool 
will translate into a decrease of the total input current into the excitatory pool (i.e. $xn < x_{E,\mathrm{ref}}$).

To summarize,  we will point out at four main points:

\begin{enumerate}

\item The sensitivities to positive perturbation of input current to either excitatory ($\mathcal{A}_{sn,x_E}$) or inhibitory pool ($\mathcal{A}_{sn,x_I}$) are respectively positive and negative. This point will be of importance for the interpretation of the Control-Target system behaviour.

\item At state of reference (i.e. setting $x_E$ and $x_I$ at their values of reference representing \emph{basic inputs} with no additional input from other areas), the stationary values of the excitatory pool ($sn^*$ and $rn^*$) can be directly adjusted by tuning $J_{gaba}$. 

\item With the set of parameters prescribed by \cite{Naskar2021}, the responsiveness of the area can be steep when the inhibitory feedback, driven by $J_{gaba}$ from the inhibitory pool to the excitatory pool upon positive perturbation of the latter, is too low.
Moreover, even with a high strength of the self-inhibitory feedback loop within the area, the operational range for further inputs is not so large: with the given values of references, higher values of $J_{gaba}$ would at most allow a doubling of input current with regards to basic input current. 

\item Given the structure of the model, activation of inhibitory pool by external input can either translate into a lower excitation of the excitatory pool (a brake on self-excitation), or into a inhibition of the excitatory pool (inhibition becomes greater than self-excitation), depending upon how the excitation of the inhibitory pool can have an effect upon total current input into the excitatory pool through $J_{gaba}$. A crossing value between the two behaviors appears to be around $J_{gaba}=1.25$ nA.

\end{enumerate}

\subsection{Effect of connectivity upon sensitivities in Control-Target system}
\label{subsec:CTsys}

We now turn to illustrations of the Control-Target system's behaviour, depending upon the connectivity, as described in Fig~\ref{fig:connectivities}.

In all cases, $J_{gaba_C}$ in the Control area and $J_{gaba_T}$ in the Target area are set such that 
the firing rate $rn_C = rn_T = 3$ Hz for the input currents of reference, i.e. at rest.
For an area connected by pool E $J_{gaba_{C|T}} = 1.758$ nA and for an area connected by pool I $J_{gaba_{C|T}} = 0.929$ nA.
Then we study the effect of varying the excitatory input current $B_{E_C}$ into the Control area, for a range of $J_{gaba_T}$ in the Target area.
Note that changing $J_{gaba_T}$ while keeping $J_{gaba_C}$ affects the firing rates at rest.

\begin{figure}[ht]
\begin{tabular}{p{0.1cm} l p{0.1cm} l}
(A)
&\raisebox{-\height}{\includegraphics[scale=0.7]{EE}}
&(B)
&\raisebox{-\height}{\includegraphics[scale=0.55]{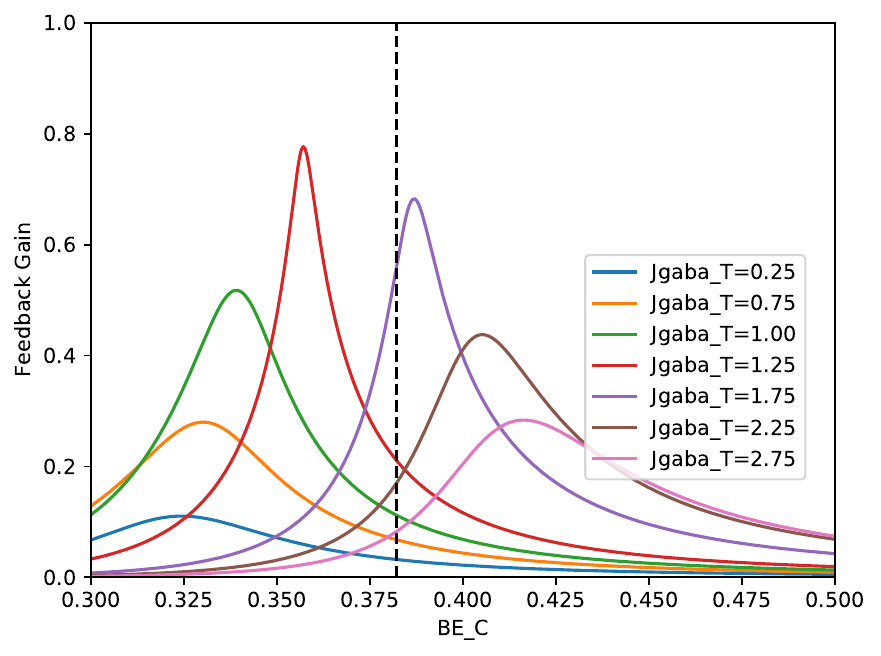}}\\
(C)
&\raisebox{-\height}{\includegraphics[scale=0.55]{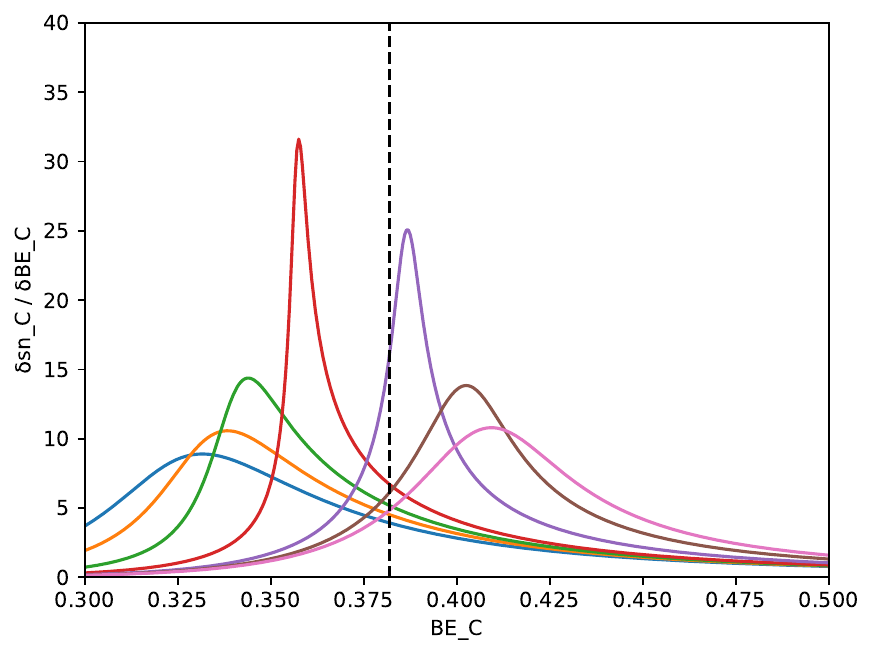}}
&(D)
&\raisebox{-\height}{\includegraphics[scale=0.55]{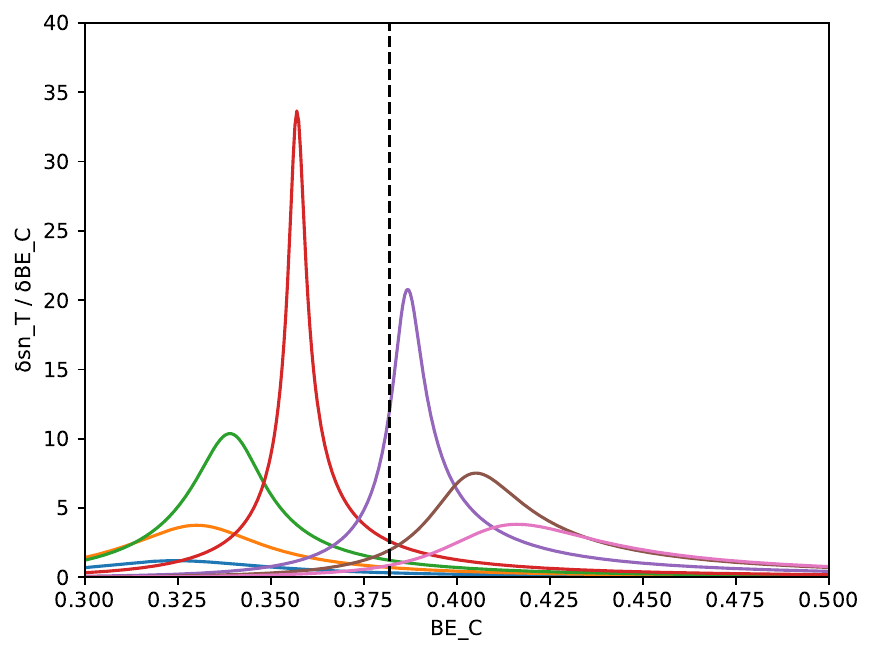}}\\
(E)
&\raisebox{-\height}{\includegraphics[scale=0.55]{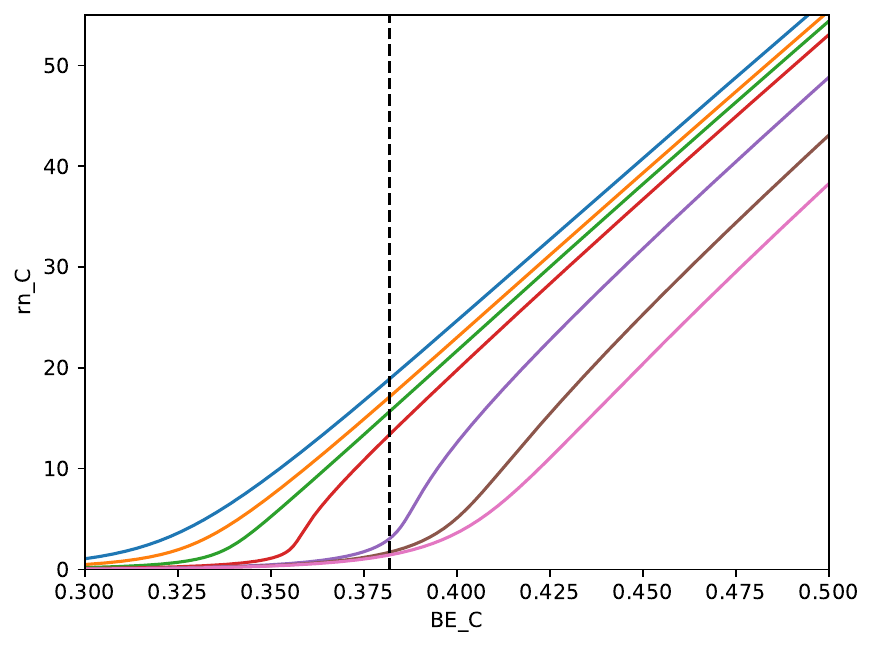}}
&(F)
&\raisebox{-\height}{\includegraphics[scale=0.55]{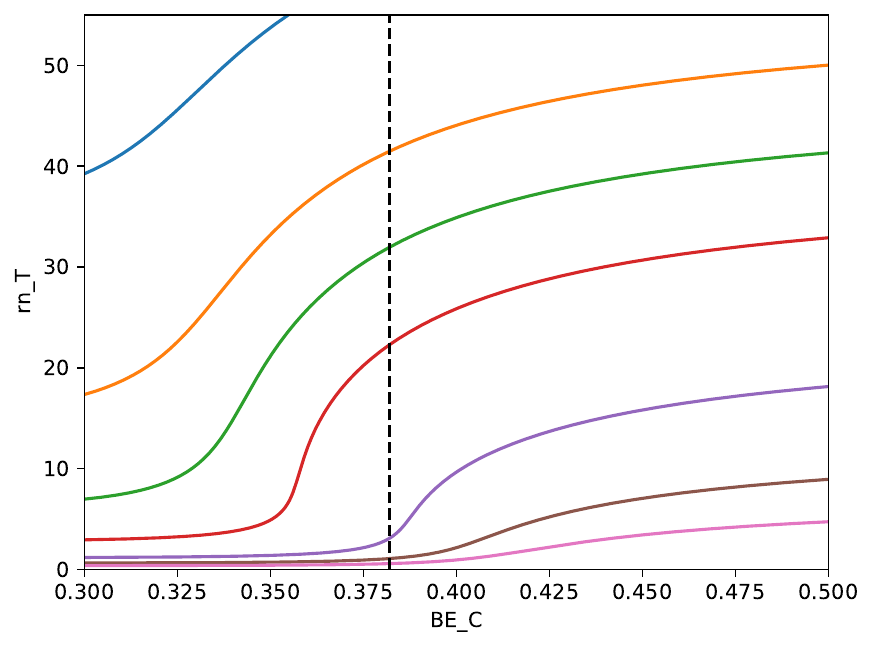}}\\
\end{tabular}
\centering
\caption{\emph{Responses and sensitivities of excitatory pools in two-area system with connectivity E-E.} (A) Illustration of the connectivity. (B)  Feedback gain $g_{sn_{C|T},B_{E_C}}$ (Eq \ref{eq:fbgain}). (C, D) Sensitivities of $sn$ to perturbations upon excitatory pool of Control Area, in Control and Target areas respectively. (E,F) Corresponding firing rates. See comments in Sec~\ref{subsec:CTsys}.
}
\label{fig:EE}
\end{figure}

\begin{figure}[ht]
\begin{tabular}{p{0.1cm} l p{0.1cm} l}
(A)
&\raisebox{-\height}{\includegraphics[scale=0.7]{IE}}
&(B)
&\raisebox{-\height}{\includegraphics[scale=0.55]{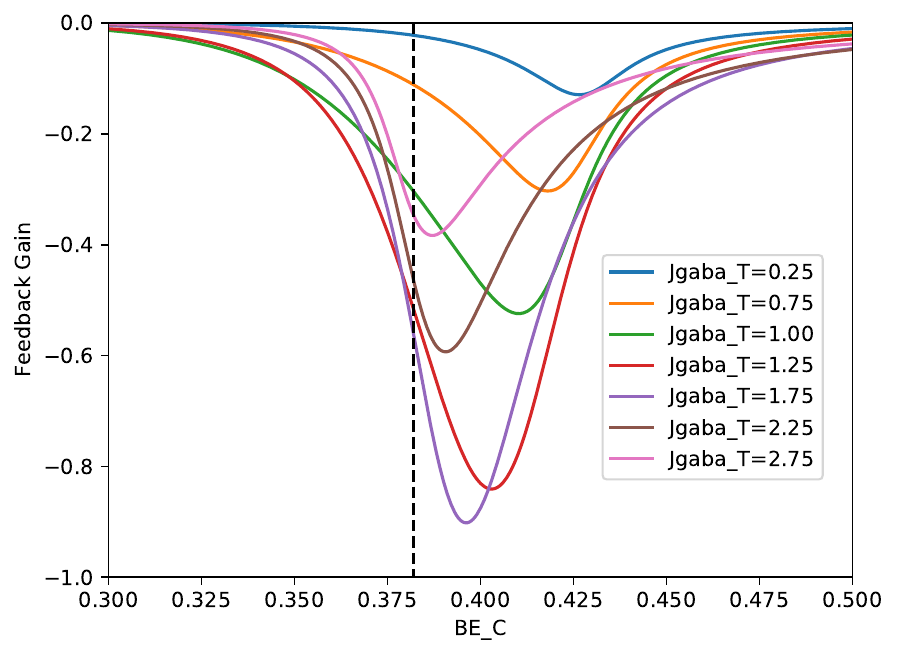}}\\
(C)
&\raisebox{-\height}{\includegraphics[scale=0.55]{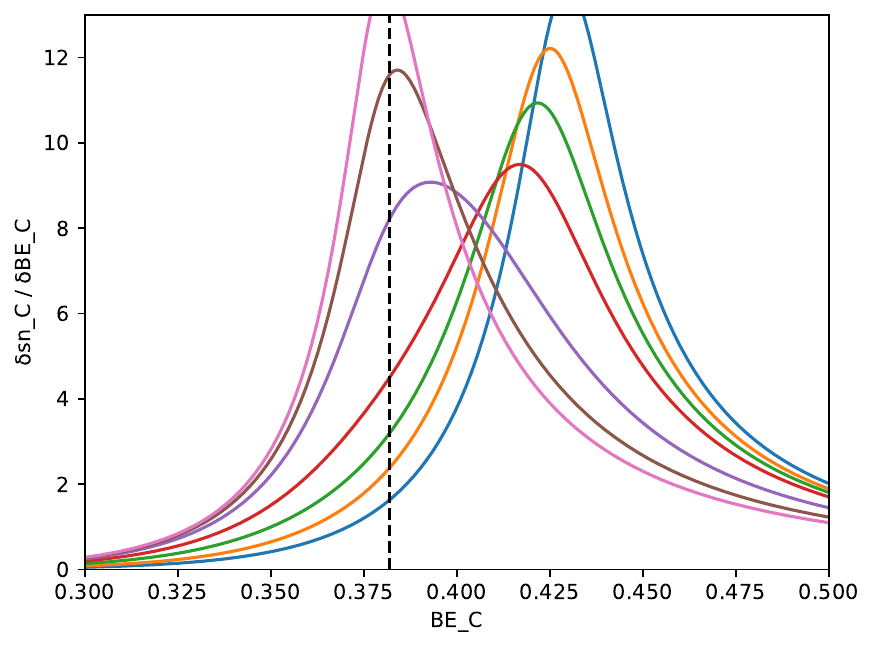}}
&(D)
&\raisebox{-\height}{\includegraphics[scale=0.55]{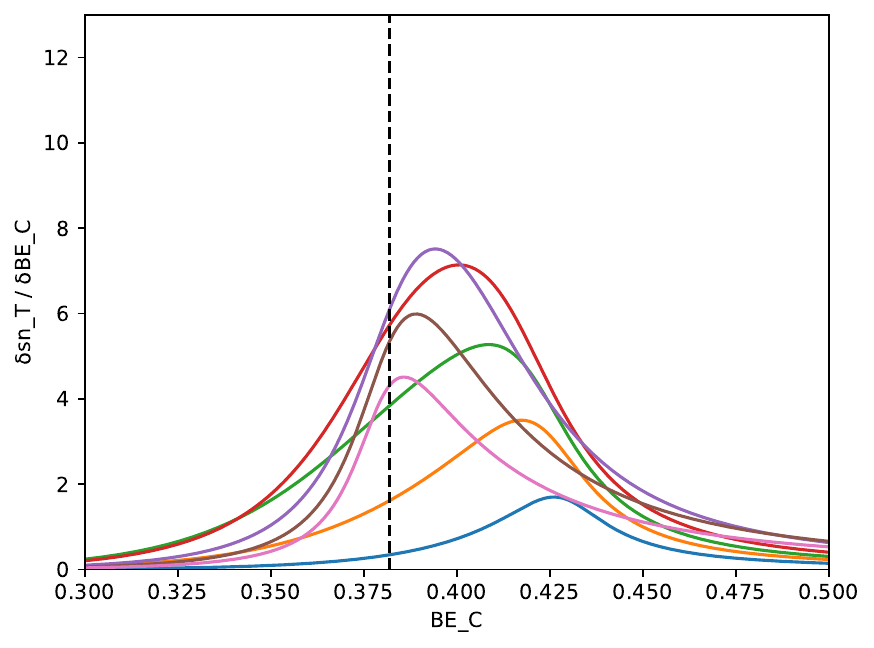}}\\
(E)
&\raisebox{-\height}{\includegraphics[scale=0.55]{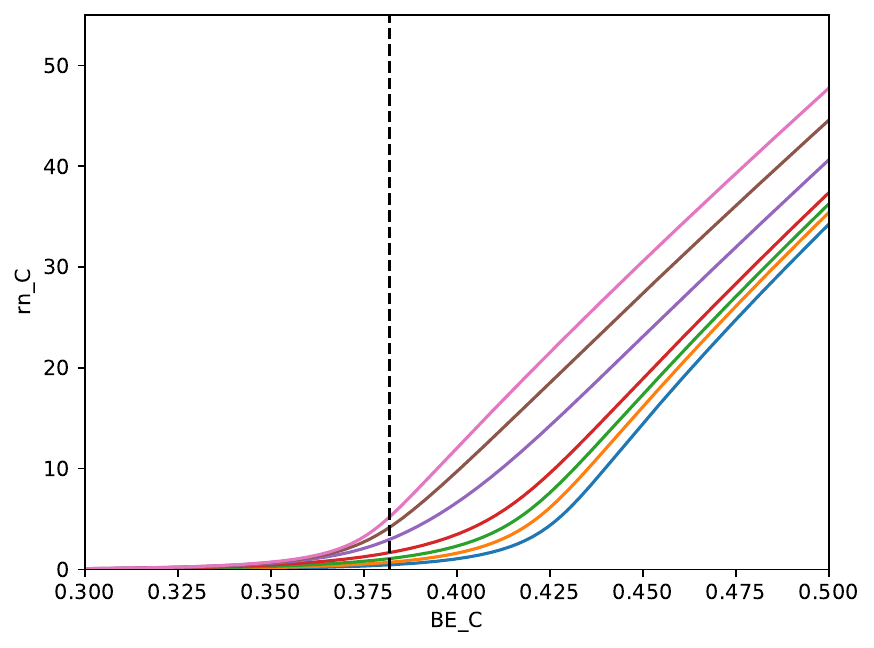}}
&(F)
&\raisebox{-\height}{\includegraphics[scale=0.55]{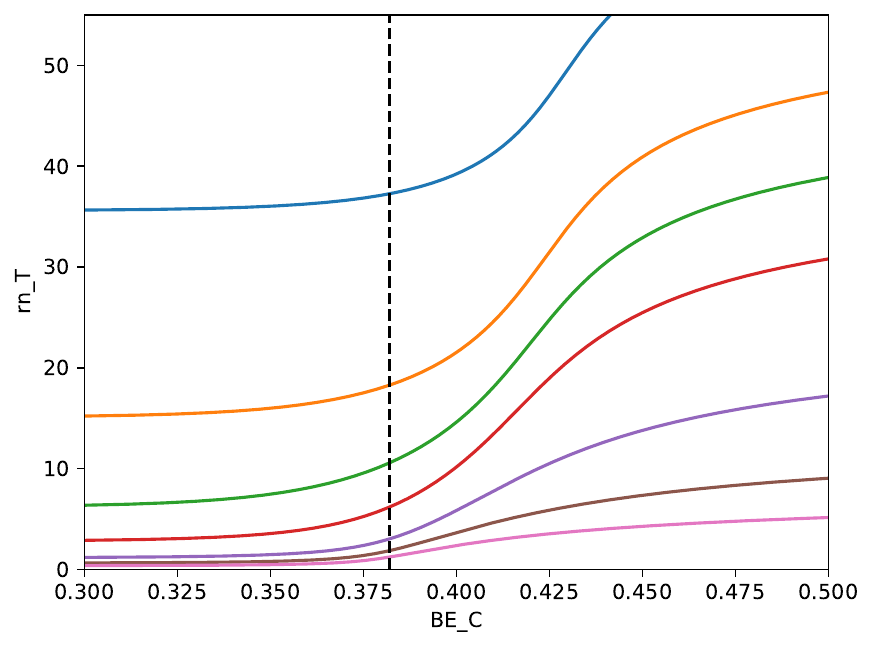}}\\
\end{tabular}
\centering
\caption{\emph{Responses and sensitivities of excitatory pools in two-area system with connectivity I-E.} Same legend as in Fig~\ref{fig:EE}. See comments in Sec~\ref{subsec:CTsys}.
}
\label{fig:IE}
\end{figure}

\begin{figure}[ht]
\begin{tabular}{p{0.1cm} l p{0.1cm} l}
(A)
&\raisebox{-\height}{\includegraphics[scale=0.7]{II}}
&(B)
&\raisebox{-\height}{\includegraphics[scale=0.55]{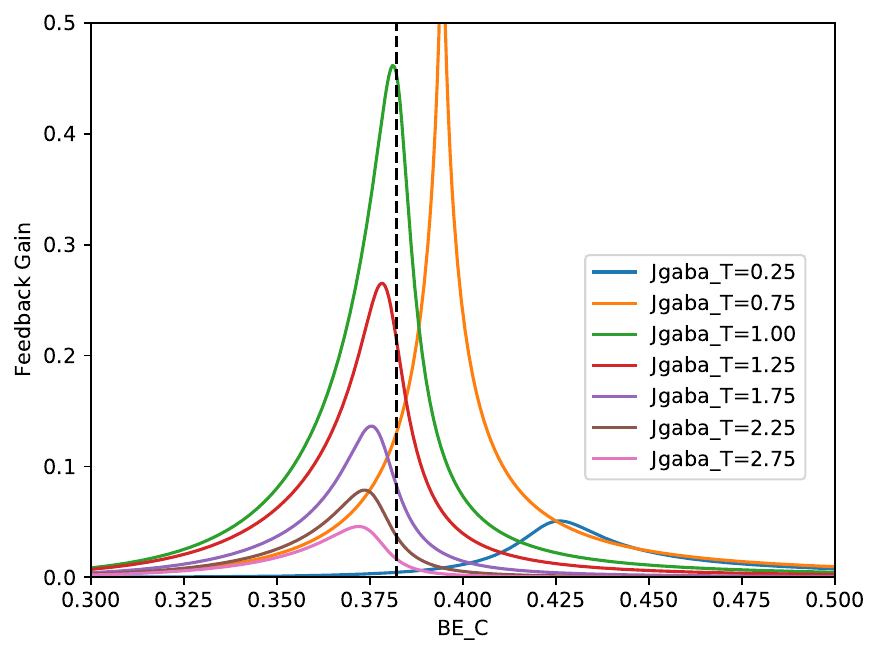}}\\
(C)
&\raisebox{-\height}{\includegraphics[scale=0.55]{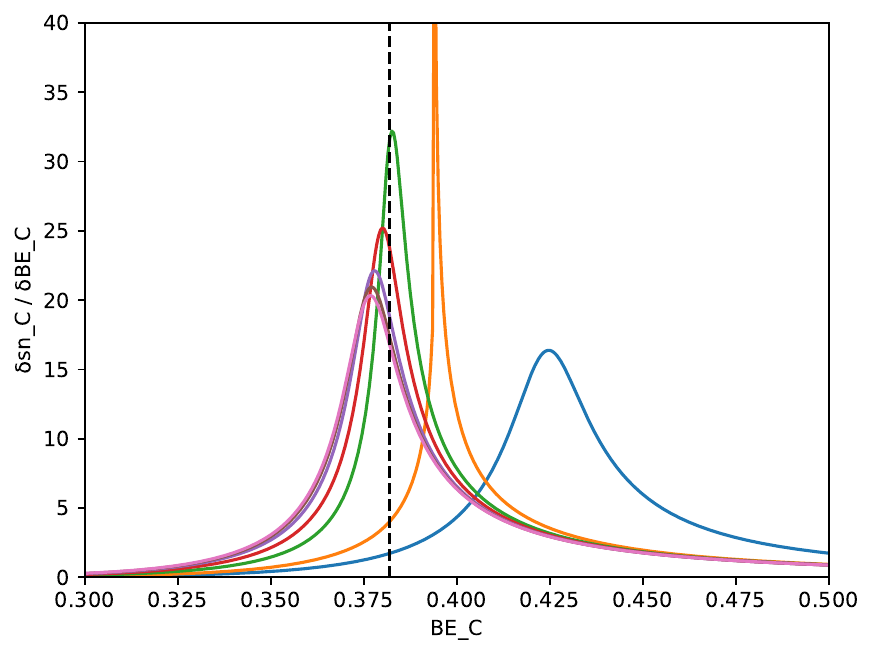}}
&(D)
&\raisebox{-\height}{\includegraphics[scale=0.55]{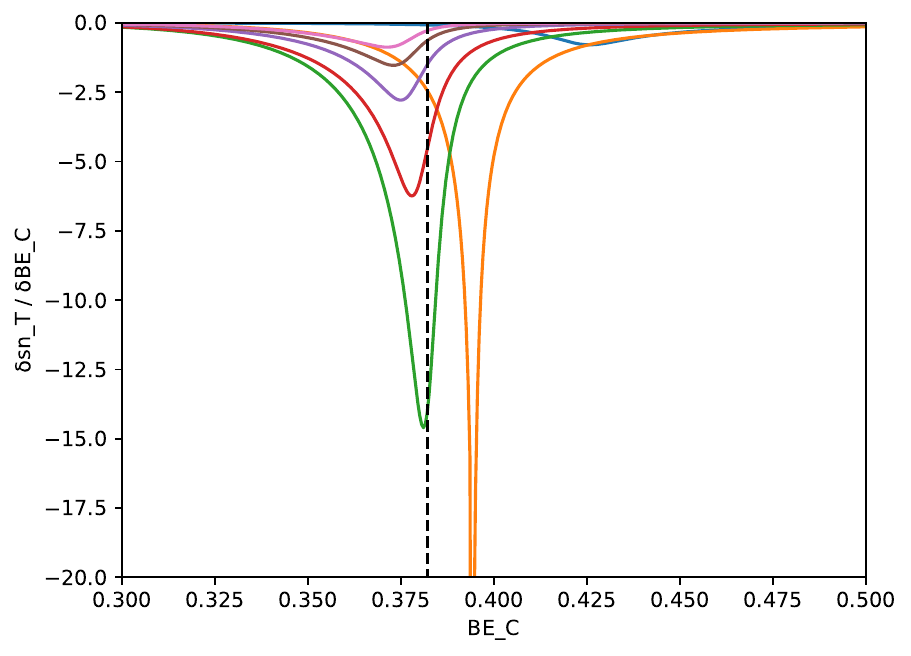}}\\
(E)
&\raisebox{-\height}{\includegraphics[scale=0.55]{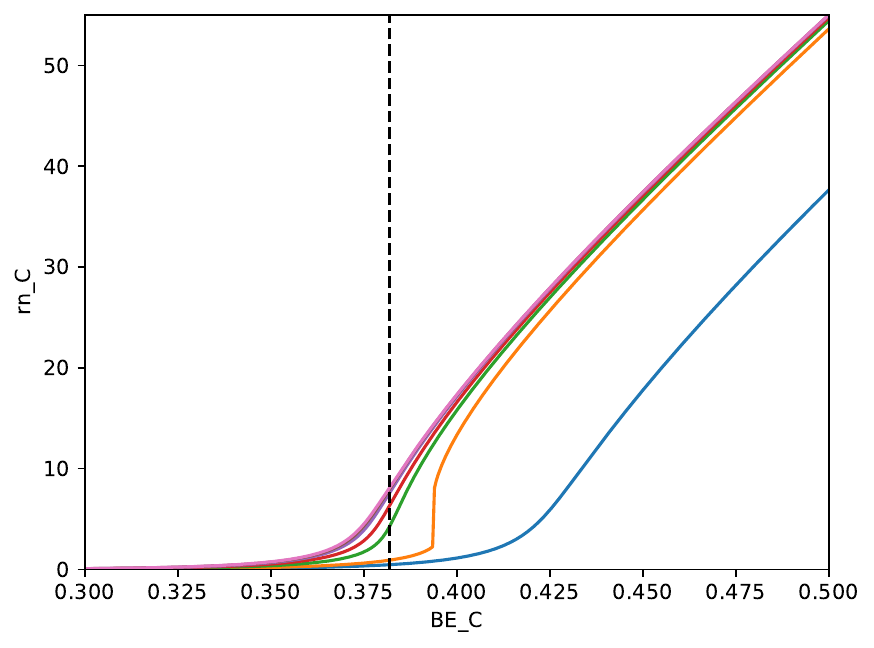}}
&(F)
&\raisebox{-\height}{\includegraphics[scale=0.55]{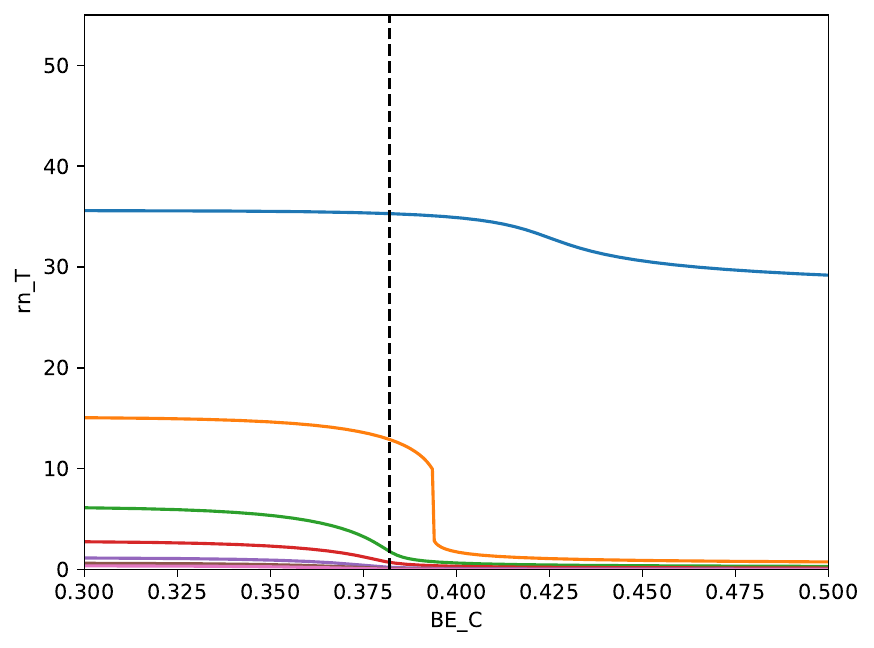}}\\
\end{tabular}
\centering
\caption{\emph{Responses and sensitivities of excitatory pools in two-area system with connectivity I-I.} Same legend as in Fig~\ref{fig:EE}. See comments in Sec~\ref{subsec:CTsys}.
}\label{fig:II}
\end{figure}

\begin{figure}[ht]
\begin{tabular}{p{0.1cm} l p{0.1cm} l}
(A)
&\raisebox{-\height}{\includegraphics[scale=0.7]{EI}}
&(B)
&\raisebox{-\height}{\includegraphics[scale=0.55]{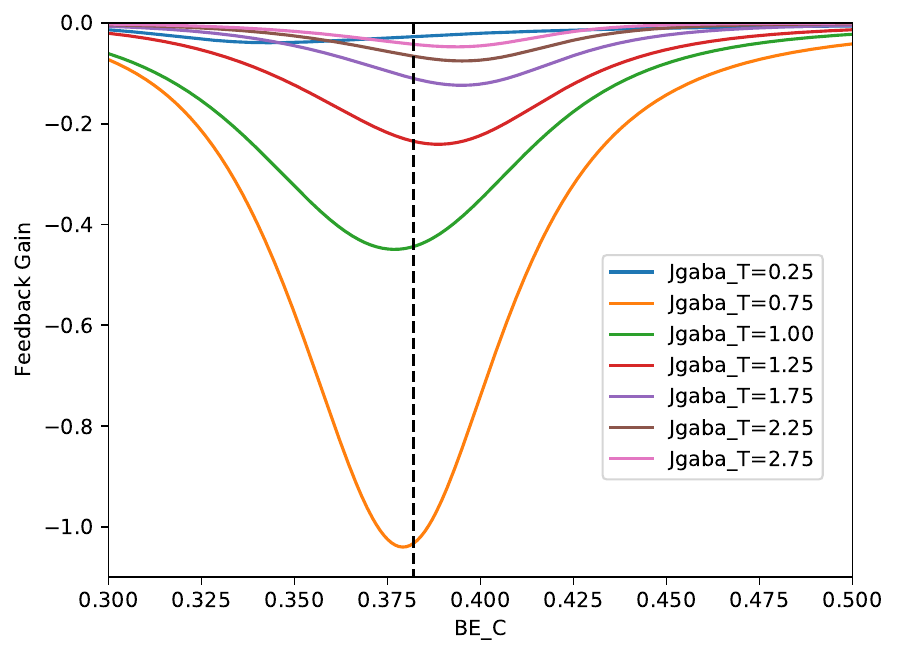}}\\
(C)
&\raisebox{-\height}{\includegraphics[scale=0.55]{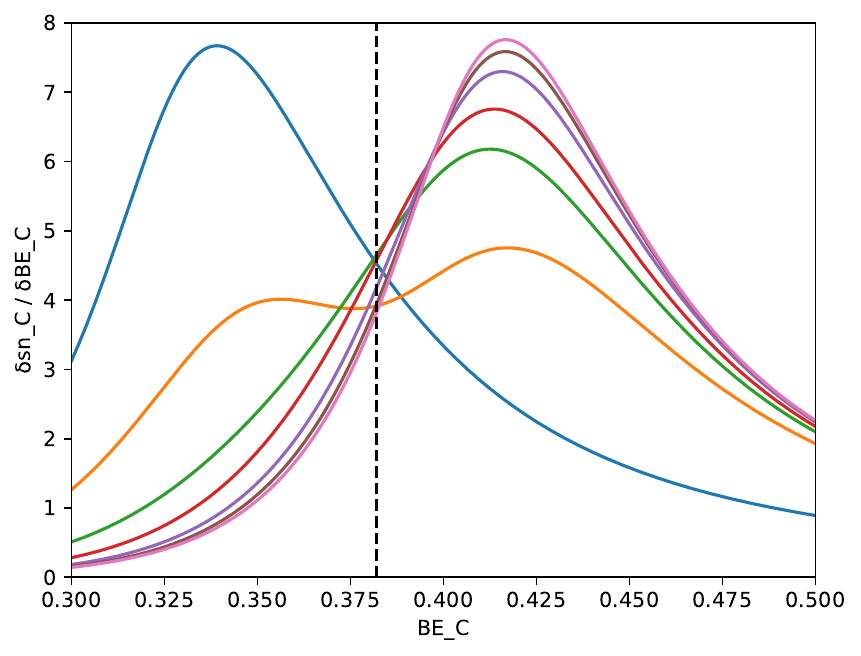}}
&(D)
&\raisebox{-\height}{\includegraphics[scale=0.55]{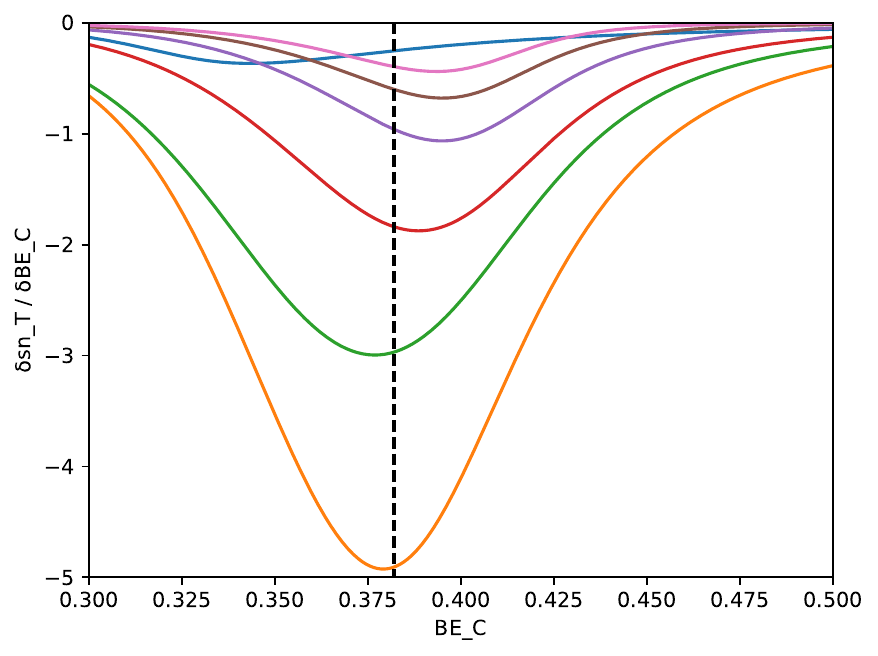}}\\
(E)
&\raisebox{-\height}{\includegraphics[scale=0.55]{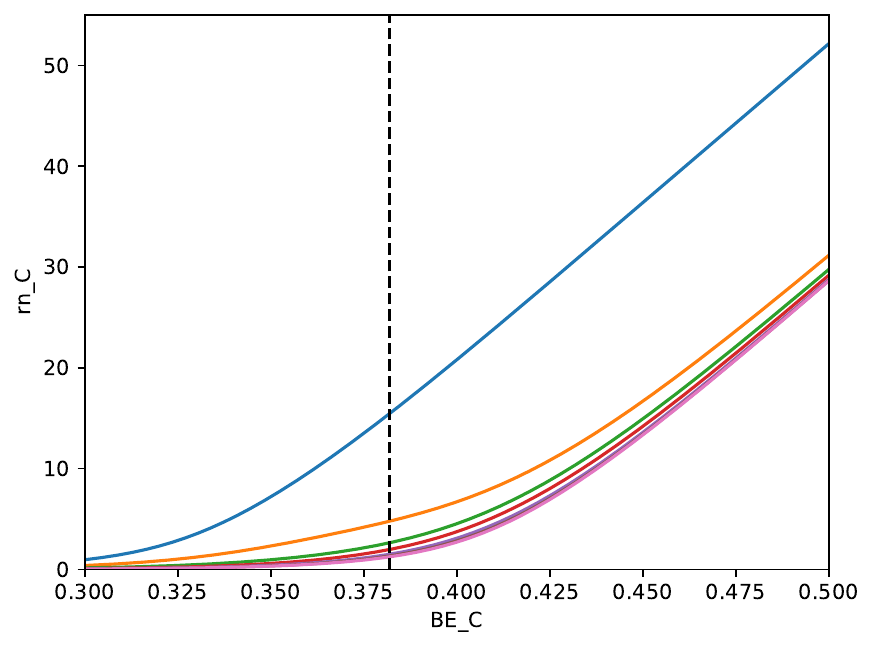}}
&(F)
&\raisebox{-\height}{\includegraphics[scale=0.55]{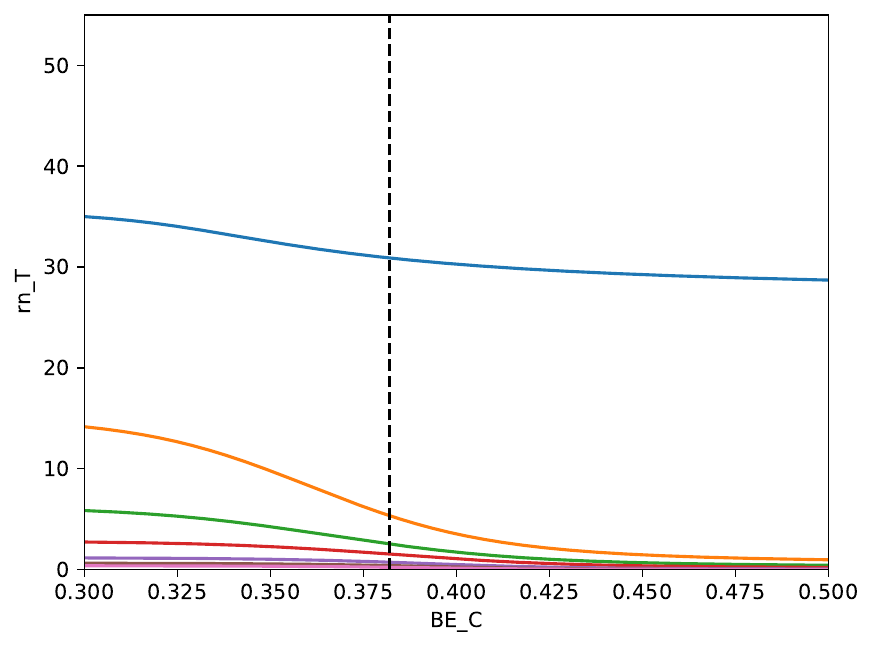}}\\
\end{tabular}
\centering
\caption{\emph{Responses and sensitivities of excitatory pools in two-area system with connectivity E-I.} Same legend as in Fig~\ref{fig:EE}. See comments in Sec~\ref{subsec:CTsys}.
}
\label{fig:EI}
\end{figure}

In all cases, we verify that the signs of sensitivities (see Figs.~\ref{fig:EE}, \ref{fig:IE}, \ref{fig:II}, \ref{fig:EI} (C,D)) are the same as those predicted in section \ref{subsec:ConnectivityTargetSensi}.

In E-E connectivity (mutual excitation, Fig~\ref{fig:EE}), the excitatory pools are in a mutual excitation regime (Fig~\ref{fig:EE} (A)) so that both sensitivities to perturbation upon $B_{E_C}$ are positive (Fig~\ref{fig:EE} (C,D)), and Target activity can only increase upon Control excitation.
Moreover, the feedback gain associated with the inter-area connection is positive (Fig~\ref{fig:EE} (B)). 
Upon stimulation of Control, this positive feedback will then amplify the effect of the stimulation, and stabilize the activity levels of both areas at higher levels than in absence of mutual excitation. 
The capacity of Control area to increase Target activity is modulated by the value set for $J_{gaba_T}$ (Fig~\ref{fig:EE} (E,F)).
For a low value of $J_{gaba_T}$ (blue curve in Fig~\ref{fig:EE} (F)), the self-inhibition at Target area level is less operative so that excitation of Control can translate in higher activity in Target (here illustrated by firing rates).
We note however that the capacity to increase Target activity can be limited to narrow ranges of $B_{E_C}$.
For example, for $J_{gaba_T} = 1.75$ nA, the effective range of control is about $[0.35,0.425]$, i.e. a 20\% increase, to take firing rate in Target from about 2 Hz to about 15 Hz.
Beyond this interval, further increase of Control will have poor effect (e.g. doubling the firing rate in Control, from 25 Hz to 50 Hz, will translate into an additional 5 Hz in Target).
Strikingly, sensitivities peak values, as core values of range for effective control, are quite disparate, so that control efficacy would highly depend upon the value set for basic input current into Control area, which should become a function of $J_{gaba_T}$.  

In I-E connectivity (Target-to-Control inhibition, Fig~\ref{fig:IE}), the Control area has an excitatory effect upon Target area, which in turn has an inhibitory effect upon Control area.
Here again, both sensitivities to perturbation upon $B_{E_C}$ are positive (Fig~\ref{fig:EI} (C,D)), and Target activity can only increase upon Control excitation.
However, the feedback gain associated with the inter-area connection becomes negative (Fig~\ref{fig:EI} (B)) so that the effect of the Control stimulation will be dampened in both areas, resulting in activity levels lower than in absence of the feedback loop.  
The capacity of Control area to increase Target activity is still modulated by the value set for $J_{gaba_T}$ (Fig~ \ref{fig:EI} (E,F)), but, by contrast with E-E connectivity, the range of effective control is enlarged and become pretty similar across $J_{gaba_T}$ values (Fig~\ref{fig:EI} (C,D)), so that changing $J_{gaba_T}$ mainly regulates the range for firing rates in presence versus in absence of Control activation.

\bigskip

Direct projection of Control area to inhibitory pool of Target area (I-I and E-I connectivity) completely changes the picture: activating Control can now \emph{decrease} activity in Target area, because sensitivities of $sn_T$ to perturbation upon $B_{E_C}$ become negative in both cases.

In I-I connectivity (mutual inhibition, Fig~\ref{fig:II}), the feedback gain is positive and sensitivities can reach high values and span narrow ranges of $B_{E_C}$, as in E-E connectivity (note that the case $J_{gaba_T}=0.75$ nA yields a jump for $B_{E_C}$ close to 0.4 nA and can be disregarded), here again yielding narrow ranges for control.
For instance, for the condition ensuring resting state at 3 Hz in both areas for basic input current set to their value of reference ($J_{gaba_T}=J_{gaba_C}=0.929 \simeq 1$ nA, hence near the green curve in Fig~\ref{fig:II} (F)), a slight activation of the Control area by $\Delta B_{E_C} = 0.425-x_{E,\mathrm{ref}} = 0.043$ nA, i.e. a small 10\% increase,  would be sufficient to shut the activity ot Target down to 0.5 Hz.  

In E-I connectivity (Control-to-Target inhibition, Fig~\ref{fig:EI}), by contrast, sensitivities of Target to activation of Control area span the full range considered for $B_{E_C}$  and their peak values are quite lower (in absolute value), so that effect of Control is smoother and can be nicely gradual (as in I-E connectivity).
For instance, for the 3 Hz condition ($J_{gaba_C}=1.758$ and $J_{gaba_T}=0.929$, green curves), there would need now about $\Delta B_{E_C} = 0.5-x_{E,\mathrm{ref}} = 0.118$ nA in Control to get 0.5 Hz in the Target, i.e. a 30\% increase.

Note that in both cases however, the 10\% increase in I-I and the 30\% increase in E-I, the activity level in Control area would shift from 3 Hz to  about the same 30 Hz.

\bigskip

To summarize,  there are here two main points:
\begin{enumerate}
\item In mutual excitation E-E, the upregulation of Control is amplified by the feedback loop in both areas; in Target-to-Control inhibition I-E, the upregulation of Control is dampened in both areas; in mutual inhibition I-I, the feedback loop amplifies the downregulation of Target triggered by upregulation of Control, and in Target-to-Control inhibition E-I, it is dampened,
\item In I-E and E-I connectivity,  control effectivity is more gradual with regards to $J_{gaba_T}$ than in E-E and I-I.
\end{enumerate}

\subsection{Consequences for system responses upon step activation of Control area}

So far, we have analyzed the response of the system in terms of sensitivities, namely how $sn_C$ and $sn_T$ would be affected by an infinitesimal (positive) perturbation applied upon the excitatory forcing of the excitatory pool of Control area.
Formal expression of sensitivities are the only path to derive analytical results that allow to understand the dynamical behavior of the system in full generality.

We now turn to illustrations that extend this analysis by direct comparison between states of the system at rest versus when submitted to a finite step of activation, as it is a typical case in macroscopic imagery such as fMRI, where the BOLD signal is recorded during tasks like Go-NoGo or Think-NoThink in inhibitory control studies (\cite{Mary2020, Apsvalka2022}).

\subsubsection{Illustrations with BOLD signals}
\label{fig:BoldSignalsComments}

\begin{figure}[ht]
\begin{tabular}{p{0.1cm} l p{0.1cm} l}
(A)
&\raisebox{-\height}{\includegraphics[scale=0.35]{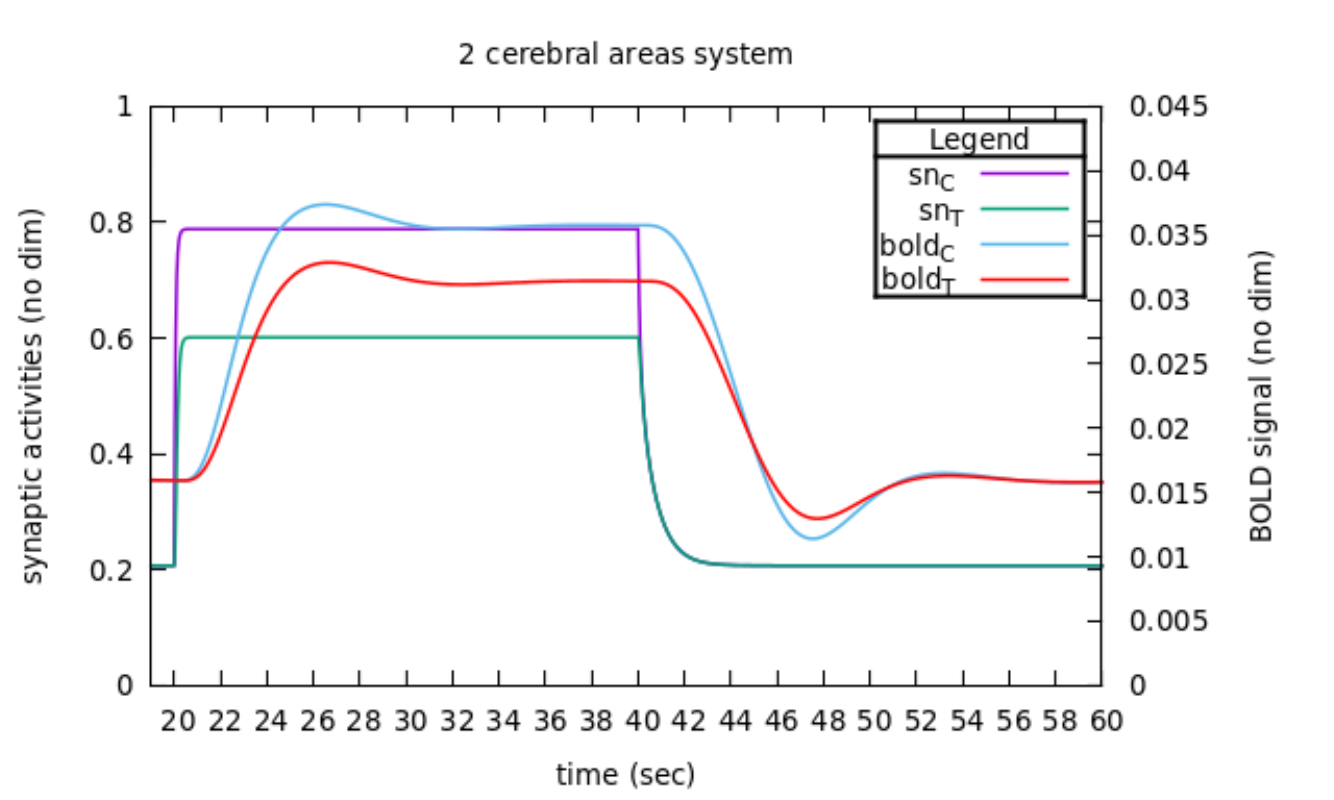}}
&(B)
&\raisebox{-\height}{\includegraphics[scale=0.35]{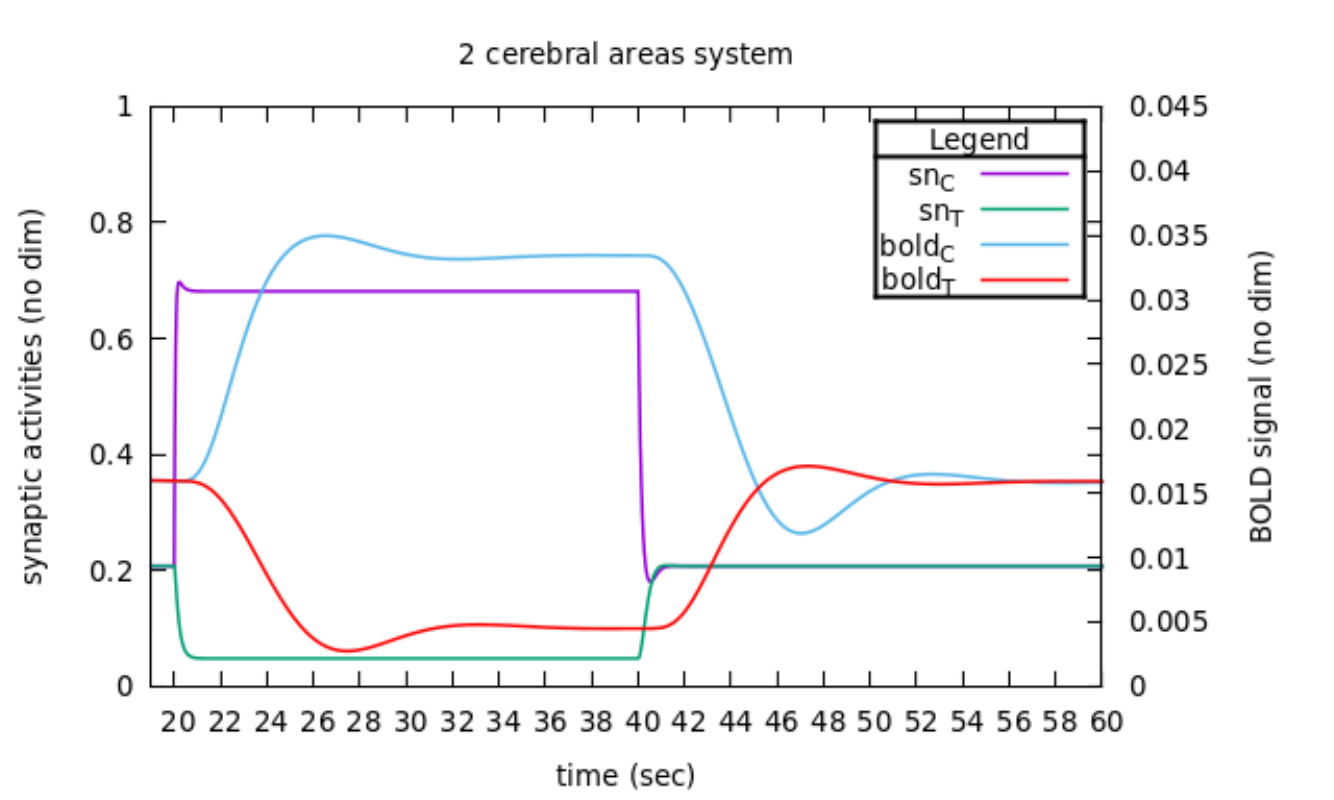}}\\
(C)
&\raisebox{-\height}{\includegraphics[scale=0.35]{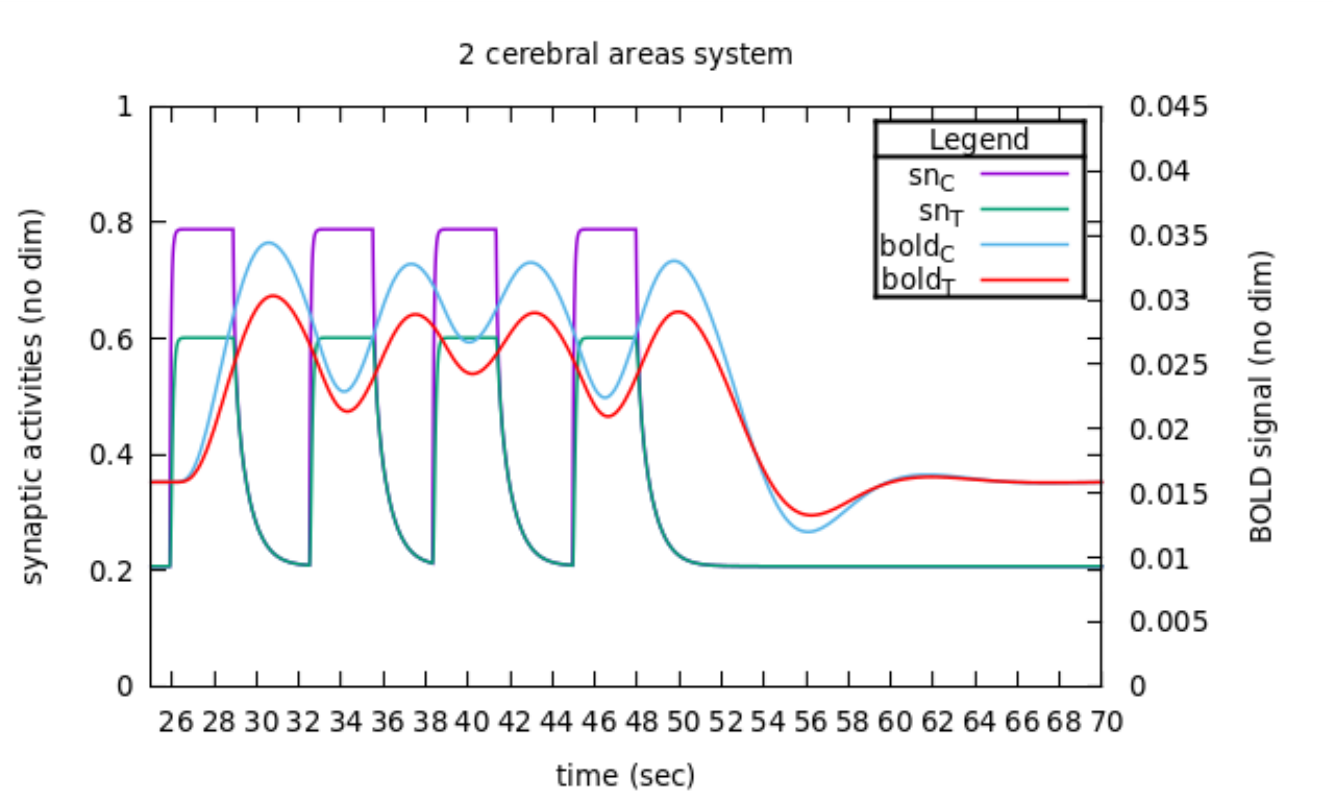}}
&(D)
&\raisebox{-\height}{\includegraphics[scale=0.35]{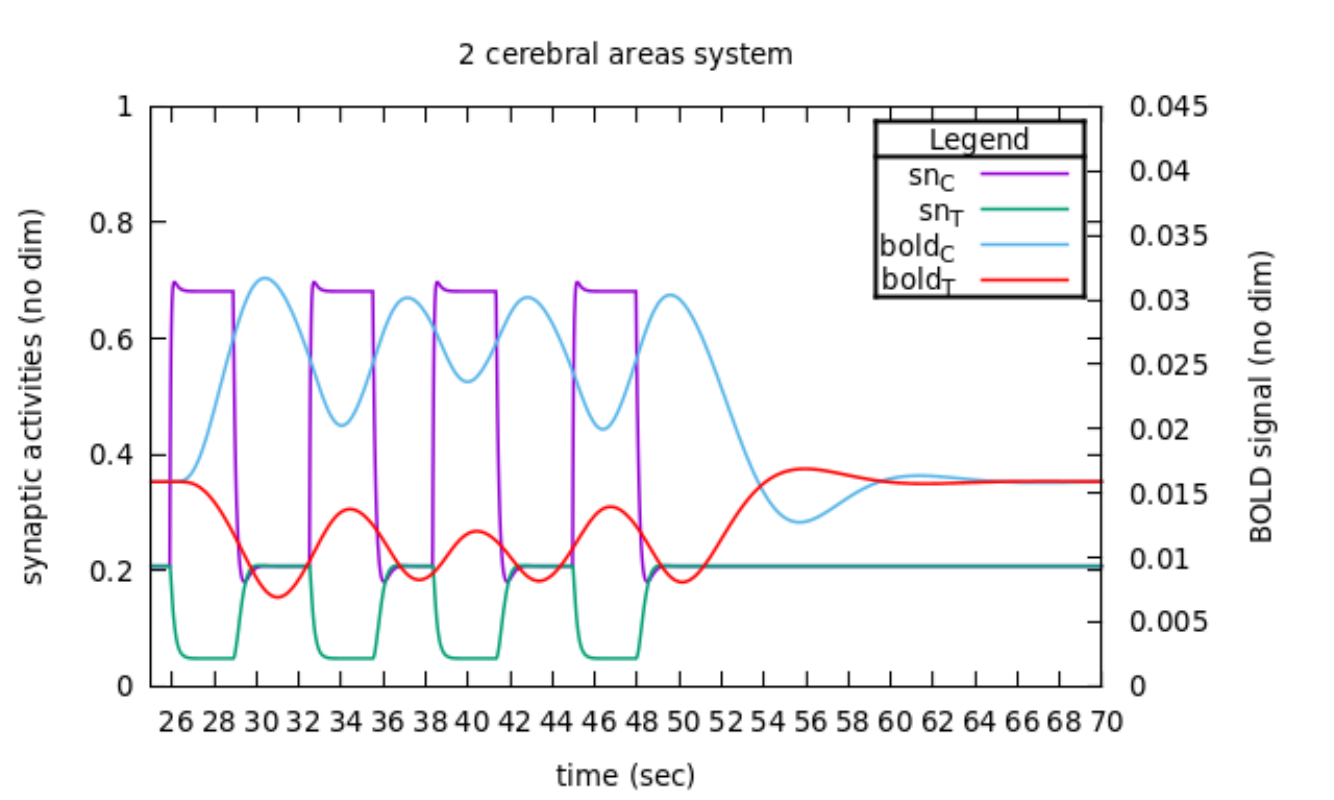}}\\
\end{tabular}
\centering
\caption{\emph{Response as BOLD signals in E-E and E-I connectivities}. Left : E-E connectivity, right E-I connectivity.\\ (A, B) A stimulation of 0.1 nA is applied upon the excitatory pool of the Control area during 20 s. (C, D) Time course of stimulations are made typical of a TNT task with stimulation steps during 3 s, separated by random intervals between 2.4 and 3.6 s. See comments in Sec~\ref{fig:BoldSignalsComments}.
}
\label{fig:BoldSignals}
\end{figure}

In Fig~\ref{fig:BoldSignals}, we illustrate the system dynamical response for E-E (left) and E-I (right) connectivities, when submitted to a step stimulation of $\Delta B_{E_C}=0.1$ nA upon the excitatory pool of Control area.
We report the behavior of synaptic activities $sn(t)$ and of the BOLD signal in both areas, when submitted to a fictitious experiment with a 20s long stimulation (upper row) or a more realistic experiment (as, e.g., in \cite{Mary2020} ) with four 3s long stimulations interspaced by random delay (lower row).

\begin{figure}[ht]
\includegraphics[scale=0.5]{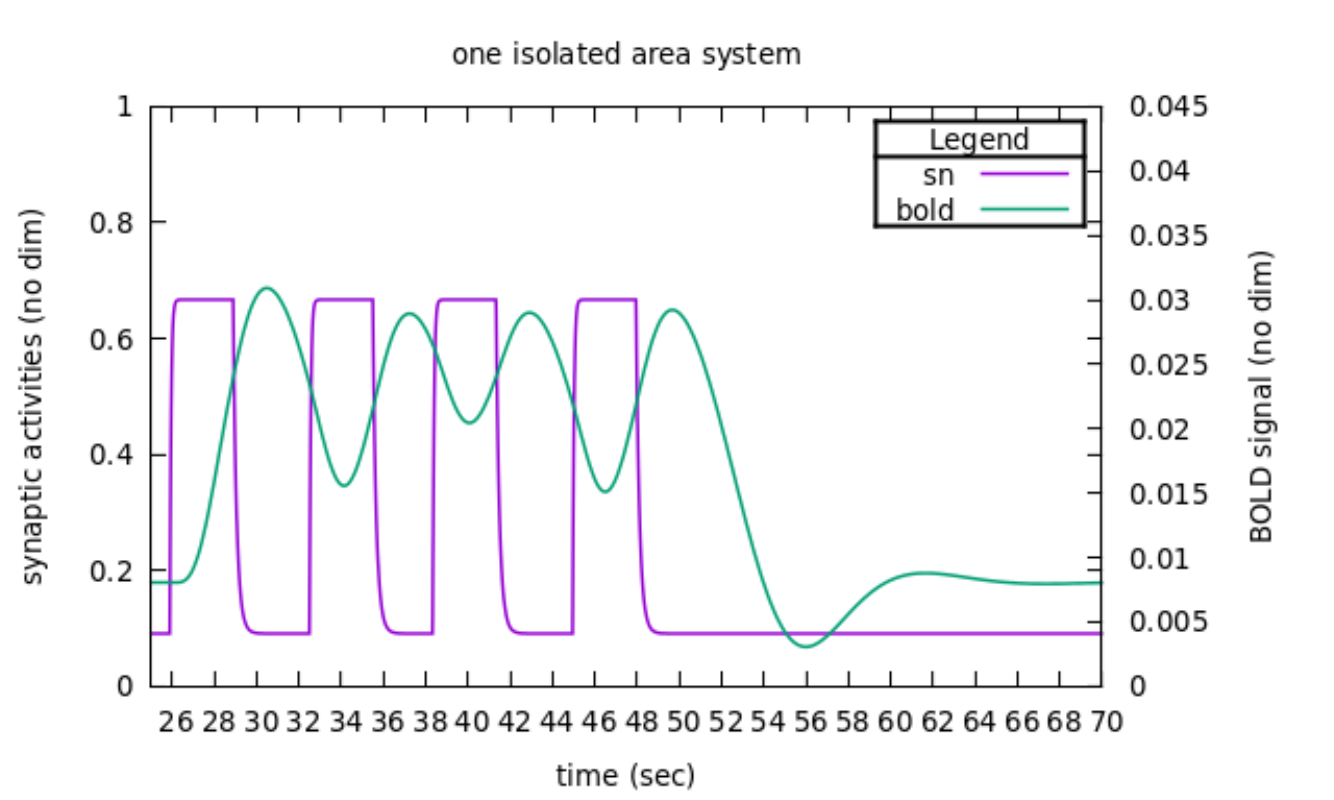}
\centering
\caption{Basic response of an isolated area to inputs. The fraction of open channels $sn$ and the corresponding BOLD signal are reported in reaction to successive stimulations. Time course of stimulations are made typical of a TNT task with stimulation steps during 3 s, separated by random intervals between 2.4 and 3.6 s (as in \cite{Mary2020}). The BOLD signal is driven by $sn$ as modelled in \cite{Wang2019}. }
\label{fig:isole4portes}
\end{figure}

As expected from Fig~\ref{fig:isole4portes}, the dynamical response of the synaptic activity to stimulus onset is step-like in all case, and, as expected from our sensitivities analysis, the Target responds by a decrease of activity (both in synaptic activity, and then in BOLD signal) in E-I connectivity.

The long stimulation gives an idea of the characteristic times of the dynamics.
Contrasting to synaptic activities, the BOLD signal needs about 15s to reach its stimulus-induced steady state value
and about the same delay to return to its rest value after the stimulus stop.

If, in both connectivities, synaptic activities adjust to stimulus onset within less than a second, their relaxation back to resting state after stimulus stops is longer in E-E connectivity (about 6s) than in E-I connectivity (less than a second).
This longer delay in E-E connectivity is due to the positive feedback gain between the two areas: if we focus for instance on Control area, even after the stimulus has stopped, there is still some overactivation (w.r.t. basic forcing) due to stimulation by Target area, which in turn is sustained by the remaining activity level of Control area (and vice versa). 
Here, the positive feedback gain acts as a brake on relaxation.

By contrast, in E-I connectivity, the high activity level of Control area tends to decrease Target activity (w.r.t. resting state) so that there is an under-activation of Control area (even if less and less as relaxation proceeds) so that nothing slows down the return to resting state (a negative feedback gain would not act as a brake).

As a consequence of its long characteristic time, the BOLD signal does not have  time enough to reach its steady state value in the case of a series of short stimulations: it oscillates around values slightly below the steady state value observed for the long stimulation.
In such stimulation series pattern, the time-shifted bijectivity between synaptic activity and BOLD signal is lost.

Still, considering the short time adjustment of synaptic activities levels to stimulus onset, it remains relevant to consider finite differences between stationnary synaptic activities at stimulated state vs. non stimulated state as a valuable observable of system response once all feedback loops are taken into account. For example, in Fig~\ref{fig:BoldSignals}, $\Delta sn_C^* =(0.8 -0.2)=0.6$ and $\Delta sn_T ^*=(0.6-0.2)=0.4$ are a good identification of the system's response to a stimulus of $\Delta B_{E_C}=0.1$ nA in Control area in an E-E connectivity.

\subsubsection{Effect of self-inhibitory strength within Target area depending on connectivity}
\label{fig:jgabComments}

To understand how parameters can drive system dynamics through such feedback loops, 
we chose to look at the cross-effect between the self-inhibitory parameter $J_{{gaba}_T}$ of the Target area and the ability of the Control area to have an effect upon the Target area, when submitted to a step activation $\Delta B_{E_C} \equiv I_{stim}$.

\begin{figure}[ht]
\includegraphics[scale=0.7]{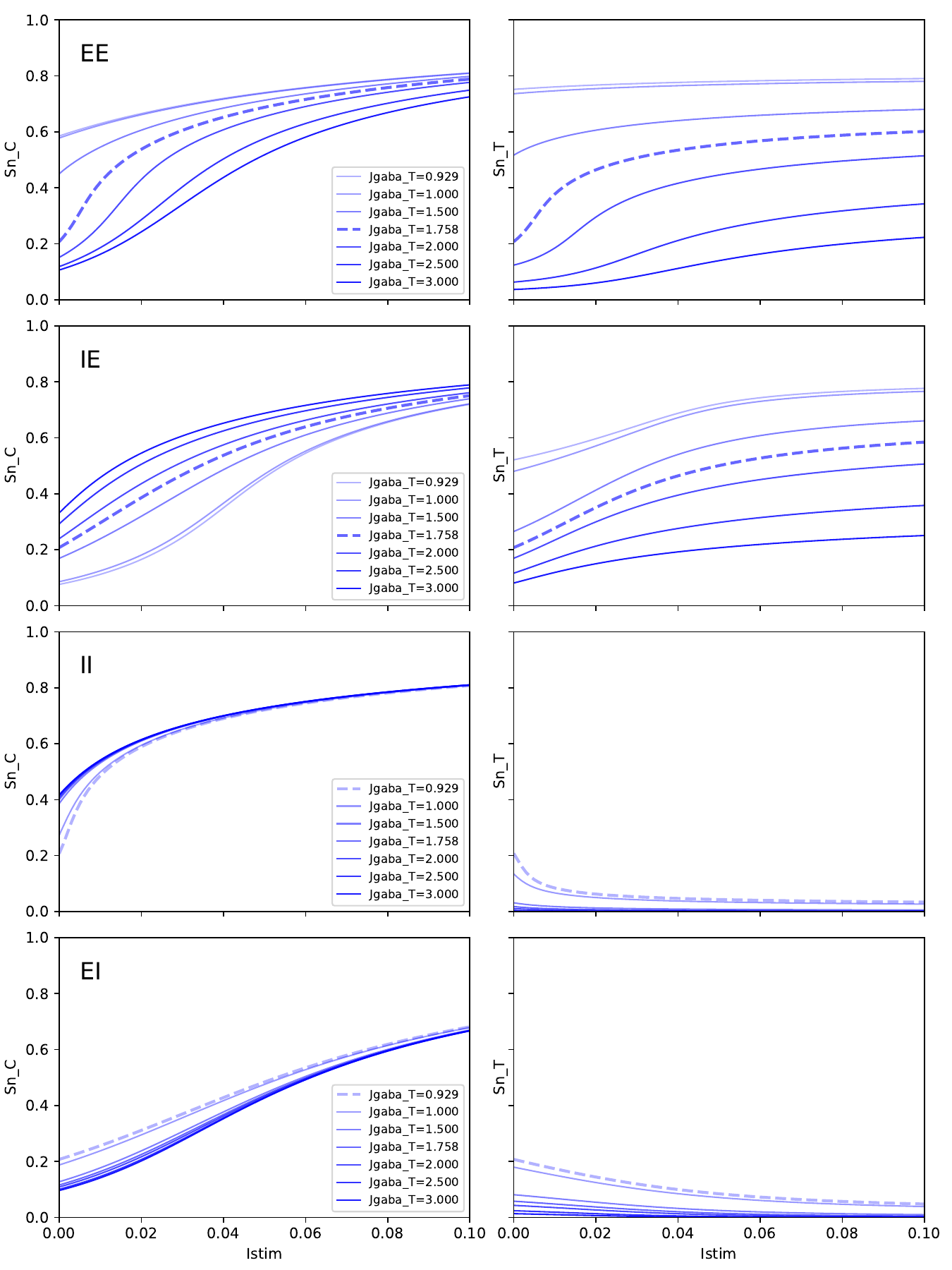}
\centering
\caption{\emph{Effect of varying $J_{{gaba}_T}$, the self-inhibitory strength within Target area, upon response to step activation of Control area}.
In order by lines: EE, IE, II, EI. Left : Control area, Right : Target area. Dotted lines correspond to the value of $J_{{gaba}_T}$ yielding firing rates at 3 Hz in both areas when the system is at rest ($I_{stim}=0$). \\
See comments in Sec~\ref{fig:jgabComments}. 
}
\label{fig:jgab}
\end{figure}

As for the range explored for $J_{{gaba}_T}$, we set it with regards to the values that would ensure resting states at 3 Hz in both areas, from the minimal value for EI and II connectivities (0.929 nA) up to about twice the value for EE and IE connectivities (1.758 nA). 
The response $sn_C$ and $sn_T$  are presented in Fig~\ref{fig:jgab} for the four connectivities (BOLD signal would display corresponding changes).

To make the link with previous sections, we outline, as a first point, that the response of $sn_C$ and $sn_T$ to a finite step of activation of Control area can be read as 
a continuous sum over $I_{stim}$ 
of the sensitivities that we have established analytically in previous section.
We can then understand that, for any value of $J_{{gaba}_T}$, integrating always positive or always negative sensitivities, the sign depending on the kind of connectivity, will translate into the same sign of response of Target area: E-E and E-I connectivities will trigger Target up-regulation, and only I-E and I-I connectivities can account for inhibitory control, where Target activity is down-regulated by activation of Control activity.

Focusing now on responses with $J_{{gaba}_T}$ set to values of reference (those ensuring 3 Hz in both areas at rest, dotted lines in Fig~\ref{fig:jgab}), we observe well the non linear effect of stimulation upon responses in both area, resulting from the integration of sensitivities.
For instance, in E-E connectivity, the sharp increase in $sn_C$ and $sn_T$ for a slight value of $I_{stim} \simeq 0.02$ nA results for the integration of the sharp curves of sensitivities in Fig~\ref{fig:EE} (C,D) just beyond $x_{E,\mathrm{ref}}$ and for $J_{{gaba}_T}=1.75$ nA.
As a consequence, further steps of activation can not but have a lower and lower effect on target area $sn_T$, and the range for control is limited. It is also the case for I-I connectivity.
As anticipated from sensitivities, responses in I-E and E-I connectivities are more gradual.

\bigskip

If we now consider affecting $J_{{gaba}_T}$ while keeping $J_{{gaba}_C}$ unaffected (namely set at the value of reference w.r.t. 3 Hz resting state), the effect of feedback loop from Target to Control can be significant upon the activity of Control area, even at resting state. 
It can even lead to unrealistic values for resting firing rates. For instance in EE connectivity, for $J_{{gaba}_T} = 1.0$ nA (instead of $1.758$), we get $sn_C=0.58$ and $sn_T=0.74$, corresponding respectively to $rn_C=15.7$ Hz and $rn_T=32.0$ Hz. 

In the same way, upon stimulation, applying the same step activation (e.g. $I_{stim}=0.02$ nA) not only will affect differently $sn_T$ depending on $J_{{gaba}_T}$, but will also affect the corresponding $sn_C$ in Control Area, as a result of feedback loops. 

As a consequence, since the control effectivity is to be understood as the contrast between resting values and activated values, a major point is that changing $J_{{gaba}_T}$ not only affects stimulated states but also affects resting values.
Hence, $J_{{gaba}_T}$ affects control effectivity through two factors: activity levels before stimulation is applied, and the margin for increase or decrease.

Overall, all these results show the major role of the within-area self-inhibition in Target area, depending on the connectivity.

\section{Conclusion and perspectives}
In this study, our aim was to understand how brain regions can be connected at neural level so that, in a Target area, a decrease in BOLD activity would be observed when a Control area is subjected to an increase in BOLD activity. 
For the neural model, we took inspiration from the model by Naskar et al.~\cite{Naskar2021}, which describes the mean synaptic activities, both excitatory and inhibitory within each area. 
More specifically, we used the same differential equations and set of neurobiological parameters, including their proposed range of values. 
Their model includes an implicit type of connectivity which we termed E-E. 
For the coupling from neural state to BOLD signal, we used the Extended Balloon-Windkessel model as proposed by~\cite{Wang2019}. 
When this composite model (neural and BOLD) was studied as a dynamical model, it was observed to reach, quite rapidly, a stationary neural state (also called a fixed point, FP), as opposed to the BOLD signal which follows the fluctuations in neural signal with a certain delay and would hardly reach a FP. 
We have therefore focused on this neural model in its stationary state, as a first step. 
Using this as a foundation, we built a sensitivity analysis of the FP using two brain areas as a configuration model, more specifically, a control area and a target area. 
Using this two-areas model, we built a hierarchical system of nested sensitivities and we have shown that:

\begin{itemize}
    \item To have a decrease in synaptic activity in a Target area, we needed to consider other types of connectivity than E-E, as it is unable to cause an inhibitory response. Four different types of connectivities have been explored, and we show that only E-I and I-I connectivities can reduce the synaptic activity in the Target region. 

    \item The E-I and I-I connectivities, which both enable inhibitory control, contrast with each other in terms of the correlation between changes in activity in control and target areas, with possibly E-I allowing a smoother control.

    \item The intra-area self-inhibitory coefficient $J_{{gaba}_T}$ in the Target area modulates this decrease in synaptic activity when the Control region is stimulated. By analyzing the response to a finite perturbation of the external forcing in the Control area as a function of $J_{gaba_T}$ in the Target area, we observed that $J_{gaba_T}$ affects both resting and activated states. 

\end{itemize}

As for our perspectives, we currently work on the generalization of the method for building nested sensitivities for more complex network configurations, e.g. a system with several control and target areas grouped into two sub-systems: a sub-system of Control areas and a sub-system of Target areas. 
Such an approach would yield an \emph{analytical} access to sensitivities for some configurations proposed in the context of inhibitory control, such as the one by Mary et al. \cite{Mary2020} which points out a network of 14 regions, comprising of 9 Control and 5 Target areas as a drive for the inhibitory control of intrusive memory in Post-Traumatic Stress Disorder. 
Furthermore, this generalization to a system-level with more areas would allow for the integration of relays in the network, as observed in recent studies \cite{Anderson2016}.
Analytical sensitivities could also help identifying areas that mostly influence information flow in any brain network, with no need to fall back to numerical experiments \cite{Harush2017, Mohan2022}.


\clearpage

\section{Appendix 1~: Formal developments for sensitivities}
\label{S1_Appendix}

This section exposes the formal developments to build sensitivities in the coupled system as a hierarchy of nested sensitivities.

\subsection{Sensitivity analysis for the isolated pools}

We consider in this section one isolated excitatory pool and one isolated inhibitory pool. Each pool contains an excitatory or inhibitory recurrent coupling and an excitatory forcing. 
Here, we are interested in the sensitivity of each excitatory and inhibitory pool activity to excitatory forcing.

\subsubsection{Isolated excitatory pool}
\label{IsoExcPool}

At the fixed point, the system reads:

\begin{equation}
  \left\{
      \begin{aligned}
        &0= -\beta^E sn^*+\alpha^E T_{glu}(1-sn^*)rn^* \equiv fn(sn^*,rn^*)\\
        &rn^*= \frac{a_E xn^*-b_E}{1-e^{-d_E (a_E xn^*-b_E)}} \equiv hn(xn^*)\\
        &xn^*= {W_{+}}J_{nmda} sn^* + z_E \equiv wn(sn^*, z_E)
      \end{aligned}
    \right.
\end{equation}

With the perturbation of $z_E$, we get the new model: 

\begin{equation}
  \left\{
      \begin{aligned}
        &fn(sn^*,rn^*)= 0\\
        &rn^*= hn(xn^*)\\
        &xn^*= wn(sn^*, z_E + \delta z_E)
       \end{aligned}
    \right.
\end{equation}

We search the formal expression for the sensitivity $\frac{\delta sn}{\delta z_E}$.\\
The linearization for the perturbation around fixed point $sn^*$ yields:

\begin{equation}
  \left\{
      \begin{aligned}
      	\frac{\partial fn}{\partial sn} \delta sn + \frac{\partial fn}{\partial rn} \delta rn &= 0\\
      	\delta rn &= \frac{\mathrm{d} hn}{\mathrm{d} xn} \delta xn \equiv hn' \delta xn\\
      	\delta xn &= \frac{\partial wn}{\partial sn} \delta sn + \frac{\partial wn}{\partial z_E} \delta z_E
      \end{aligned}
    \right.
\end{equation}

Plugging the last two equations into the first, we get:

\begin{equation}
 	\big(\frac{\partial fn}{\partial sn} + \frac{\partial fn}{\partial rn} \frac{\mathrm{d} hn}{\mathrm{d} xn} \frac{\partial wn}{\partial sn} \big) \delta sn = -\frac{\partial fn}{\partial rn} \frac{\mathrm{d} hn}{\mathrm{d} xn}\frac{\partial wn}{\partial z_E} \delta z_E
\end{equation}

Rearranging to read how $\delta sn$ depends upon $\delta z_E$:

\begin{equation}
 	\big( 1 + \frac{\partial fn}{\partial sn}^{-1}\frac{\partial fn}{\partial rn} \frac{\mathrm{d} hn}{\mathrm{d} xn} \frac{\partial wn}{\partial sn} \big) \delta sn = -\frac{\partial fn}{\partial sn}^{-1}\frac{\partial fn}{\partial rn} \frac{\mathrm{d} hn}{\mathrm{d} xn}\frac{\partial wn}{\partial z_E} \delta z_E
\end{equation}

Then we get: 
\begin{align}
 	\frac{\delta sn}{\delta z_E} &= \frac{-\frac{\partial fn}{\partial sn}^{-1}\frac{\partial fn}{\partial rn} \frac{\mathrm{d} hn}{\mathrm{d} xn}\frac{\partial wn}{\partial z_E} } {1 + \frac{\partial fn}{\partial sn}^{-1}\frac{\partial fn}{\partial rn} \frac{\mathrm{d} hn}{\mathrm{d} xn} \frac{\partial wn}{\partial sn} } \\
 	&= - \frac{ 1 }{\frac{\partial fn}{\partial sn}\frac{\partial fn}{\partial rn}^{-1} \frac{\mathrm{d} hn}{\mathrm{d} xn}^{-1}\frac{\partial wn}{\partial z_E}^{-1} +  \frac{\partial wn}{\partial sn}\frac{\partial wn}{\partial z_E}^{-1} } 
\end{align}

Considering that

\begin{align}
	\frac{\partial fn}{\partial sn} &= -\beta_E - \alpha_E T_{glu} rn^*\\
	\frac{\partial fn}{\partial rn} &= \alpha_E T_{glu} (1-sn^*) \\
	\frac{\partial wn}{\partial sn} &= {W_{+}}J_{nmda}\\
	\frac{\partial wn}{\partial z_E} &= 1
\end{align}

We can calculate:
\begin{equation}
 	\frac{\delta sn}{\delta z_E} = \frac{ 1 }{\frac{\beta_E + \alpha_E T_{glu} rn^*}{\alpha_E T_{glu} (1-sn^*) \frac{\mathrm{d} hn}{\mathrm{d} xn}} -  {W_{+}}J_{nmda} }
\end{equation}

\subsubsection{Isolated inhibitory pool}
\label{IsoInhPool}

At the fixed point, the system reads:

\begin{equation}
  \left\{
      \begin{aligned}
        &0= -\beta^I sg^*+\alpha^I T_{gaba}(1-sg^*)rg^* \equiv fg(sg^*,rg^*)\\
        &rg^*= \frac{a_I xg^*-b_I}{1-e^{-d_I (a_I xg^*-b_I)}} \equiv hg(xg^*)\\
        &xg^*= -J_{-} sg^*  + z_I \equiv wg(sg^*, z_I)
      \end{aligned}
    \right.
\end{equation}

With the perturbation of $z_I$, we get the new model: 

\begin{equation}
  \left\{
      \begin{aligned}
        &fg(sg^*,rg^*)= 0\\
        &rg^*= hg(xg^*)\\
        &xg^*= wg(sg^*, z_I + \delta z_I)
       \end{aligned}
    \right.
\end{equation}

We search the formal expression for the sensitivity $\frac{\delta sg}{\delta z_I}$.\\
The linearization for the perturbation around fixed point $sg^*$ yields:

\begin{equation}
  \left\{
      \begin{aligned}
      	\frac{\partial fg}{\partial sg} \delta sg + \frac{\partial fg}{\partial rg} \delta rg &= 0\\
      	\delta rg &= \frac{\mathrm{d} hg}{\mathrm{d} xg} \delta xg \equiv hg' \delta xg\\
      	\delta xg &= \frac{\partial wg}{\partial sg} \delta sg + \frac{\partial wg}{\partial z_I} \delta z_I
      \end{aligned}
    \right.
\end{equation}

Plugging the last two equations into the first, we get:

\begin{equation}
 	\big(\frac{\partial fg}{\partial sg} + \frac{\partial fg}{\partial rg} \frac{\mathrm{d} hg}{\mathrm{d} xg} \frac{\partial wg}{\partial sg} \big) \delta sg = -\frac{\partial fg}{\partial rg} \frac{\mathrm{d} hg}{\mathrm{d} xg}\frac{\partial wg}{\partial z_I} \delta z_I
 	\end{equation}

Rearranging to read how $\delta sg$ depends upon $\delta z_I$:

\begin{equation}
 	\big( 1 + \frac{\partial fg}{\partial sg}^{-1}\frac{\partial fg}{\partial rg} \frac{\mathrm{d} hg}{\mathrm{d} xg} \frac{\partial wg}{\partial sg} \big) \delta sg = -\frac{\partial fg}{\partial sg}^{-1}\frac{\partial fg}{\partial rg} \frac{\mathrm{d} hg}{\mathrm{d} xg}\frac{\partial wg}{\partial z_I} \delta z_I
\end{equation}

Then we get: 
\begin{align}
 	\frac{\delta sg}{\delta z_I} &= \frac{-\frac{\partial fg}{\partial sg}^{-1}\frac{\partial fg}{\partial rg} \frac{\mathrm{d} hg}{\mathrm{d} xg}\frac{\partial wg}{\partial z_I} } {1 + \frac{\partial fg}{\partial sg}^{-1}\frac{\partial fg}{\partial rg} \frac{\mathrm{d} hg}{\mathrm{d} xg} \frac{\partial wg}{\partial sg} } \\
 	&= - \frac{ 1 }{\frac{\partial fg}{\partial sg}\frac{\partial fg}{\partial rg}^{-1} \frac{\mathrm{d} hg}{\mathrm{d} xg}^{-1}\frac{\partial wg}{\partial z_I}^{-1} +  \frac{\partial wg}{\partial sg}\frac{\partial wg}{\partial z_I}^{-1} } 
\end{align}

Considering that

\begin{align}
	\frac{\partial fg}{\partial sg} &= -\beta_I - \alpha_I T_{gaba} rg^*\\
	\frac{\partial fg}{\partial rg} &= \alpha_I T_{gaba} (1-sg^*) \\
	\frac{\partial wg}{\partial sg} &= -J_{-}\\
	\frac{\partial wg}{\partial z_I} &= 1
\end{align}

We can calculate:
\begin{equation}
 	\frac{\delta sg}{\delta z_I} = \frac{ 1 }{\frac{\beta_I + \alpha_I T_{gaba} rg^*}{\alpha_I T_{gaba} (1-sg^*) \frac{\mathrm{d} hg}{\mathrm{d} xg}} +  J_{-}} 
\end{equation}

\subsubsection{Functional expressions of the sensitivities for the isolated excitatory and inhibitory pools}
\label{sumupIsoPools}

We define the function $\varphi n_{_E}$ and $\varphi g_{_I}$ to explicitly express the dependencies of the sensitivities for an isolated excitatory pool and an isolated inhibitory pool: 

\begin{equation}
	\left\{
	\begin{aligned}
 	\varphi n_{_E} (sn^*, z_E) & \equiv
	\left( 
		\frac{\beta_E + \alpha_E T_{glu} hn(wn(sn^*,z_E))}
		       {\alpha_E T_{glu} (1-sn^*) hn'(wn(sn^*,z_E)} 
		-  {W_{+}}J_{nmda} 
	\right)^{-1} \\[2mm]
	\varphi g_{_I} (sg^*, z_I) & \equiv 
	\left( \frac{\beta_I + \alpha_I T_{gaba} hg(wg(sg^*,z_I))}{\alpha_I T_{gaba} (1-sg^*) hg'(wg(sg^*,z_I))} +  J_{-}\right)^{-1}  
		\end{aligned} 
	\right.
\label{eq:funSensiPools}
\end{equation}

where

\begin{equation}
h{\scriptstyle{\square}} '(x) = \frac{(a e^{d (a x - b)} (e^{d (a x - b)} - a d x + b d - 1))}{(e^{d (a d - b)} - 1)^2}
\end{equation}

with $a$, $b$ and $d$ to be taken accordingly.

\subsection{Sensitivity analysis for one isolated area}
\label{Sec:SensiUneRegion}

At the fixed point, the system with the two coupled pools reads:

\begin{equation}
  \left\{
      \begin{aligned}
        &0= -\beta^E sn^*+\alpha^E T_{glu}(1-sn^*)rn^* \equiv fn(sn^*,rn^*)\\
        &0=-\beta^I sg^*+\alpha^I T_{gaba}(1-sg^*)rg^*\equiv fg(sg^*,rg^*)
      \end{aligned}
    \right.
\end{equation}

with

\begin{equation}
  \left\{
      \begin{aligned}
        &rn^*= \frac{a_E xn^*-b_E}{1-e^{-d_E (a_E xn^*-b_E)}} \equiv hn(xn^*)\\
        &rg^*= \frac{a_I xg^*-b_I}{1-e^{-d_I (a_I xg^*-b_I)}} \equiv hg(xg^*)
      \end{aligned}
    \right.
\end{equation}

in which $xn^*$ and $xg^*$ represent the respective total input currents:

\begin{equation}
  \left\{
      \begin{aligned}
        &xn^*= {W_{+}}J_{nmda} sn^* - J_{gaba} sg^* + x_E\\
        &xg^*= J_{nmda} sn^* -J_{-} sg^*  + x_I 
      \end{aligned}
    \right.
\end{equation}

where $x_E$ and $x_I$ represent basal forcings (effective external inputs), that we will respectively perturbate.

\subsubsection{Splitting $x_{inter}$ from $x_{intra}$}

In order to express closed loop sensitivities of the pools as a function of their open loop sensitivities, we split explicitly the total input currents between an internal current, representing the effect of intra-pool recurrences: 

\begin{equation}
  \left\{
      \begin{aligned}
        &xn_{intra}^*= {W_{+}}J_{nmda} sn^* \equiv wn_{intra}(sn^*)\\
        &xg_{intra}^*= -J_{-} sg^* \equiv wg_{intra}(sg^*)
      \end{aligned}
    \right.
\end{equation}

and an external current, representing the total amount of forcing, due to the basic external forcing and the feedback current from the alternate pool:

\begin{equation}
  \left\{
      \begin{aligned}
        &xn_{inter}^*= - J_{gaba} sg^* + x_E  \equiv wn_{inter}(sg^*,x_E)\\
        &xg_{inter}^*= J_{nmda} sn^* + x_I \equiv wg_{inter}(sn^*,x_I)
      \end{aligned}
    \right.
    \label{eq:xinter}
\end{equation}

so we have:

\begin{equation}
  \left\{
      \begin{aligned}
        &xn^*= xn_{intra}^* + xn_{inter}^* \equiv wn(xn_{intra}^*, xn_{inter}^*)\\
        &xg^*= xg_{intra}^* + xg_{inter}^*  \equiv wg(xg_{intra}^*, xg_{inter}^* )
      \end{aligned}
    \right.
\end{equation}

\subsubsection{Intermediate variable elimination}

We define

\begin{equation}
    \vv{s}=
    \begin{pmatrix}
    sn^* \\
    sg^* \\
    \end{pmatrix}, \ \ \
    \vv{r}=
    \begin{pmatrix}
    rn^* \\
    rg^* \\
    \end{pmatrix}, \ \ \
    \vv{x}=
    \begin{pmatrix}
    xn^* \\
    xg^* \\
    \end{pmatrix}, \ \ \
    \vv{x_{intra}}=
    \begin{pmatrix}
    xn_{intra}^* \\
    xg_{intra}^* \\
    \end{pmatrix}, \ \ \
    \vv{x_{inter}}=
    \begin{pmatrix}
    xn_{inter}^* \\
    xg_{inter}^* \\
    \end{pmatrix}
\end{equation}

so that we can summarize the fixed point as:

\begin{equation}
  \left\{
      \begin{aligned}
        &\vv{f}(\vv{s},\vv{r})= \vv{0}\\
        &\vv{r}= \vv{h}(\vv{x})\\
        &\vv{x}= \vv{w}(\vv{x_{intra}},\vv{x_{inter}})\\
        &\vv{x_{intra}}= \vv{w_{intra}}(\vv{s})\\
        &\vv{x_{inter}}= \vv{w_{inter}}(\vv{s},x_E,x_I)
      \end{aligned}
    \right.
    \label{sys1region_1eForm}
\end{equation}

Plugging intermediate variables $\vv{r}$, $\vv{x}$, $\vv{x_{intra}}$  into the first equation, Eq \ref{sys1region_1eForm} can be rewritten as:

\begin{equation}
  \left\{
      \begin{aligned}
        &\vv{F}(\vv{s},\vv{x_{inter}})= \vv{0}\\
        &\vv{x_{inter}}= \vv{w_{inter}}(\vv{s},x_E ,x_I)
      \end{aligned}
    \right.
    \label{1reg_fullmodel}
\end{equation}

where 

\begin{equation}
  \left\{
      \begin{aligned}
        &Fn(sn^*,xn_{inter}^* ) \equiv -\beta^E sn^*+\alpha^E T_{glu}(1-sn^*) \frac{a_E (W_{+}J_{nmda} sn^* + xn_{inter}^*)-b_E}{1-e^{-d_E [a_E( W_{+}J_{nmda} sn^* + xn_{inter}^*)-b_E]}} \\
        &Fg(sg^*,xg_{inter}^* ) =-\beta^I sg^*+\alpha^I T_{gaba}(1-sg^*)\frac{a_I (-J_{-} sg^*+ xg_{inter}^* )-b_I}{1-e^{-d_I [a_I (-J_{-} sg^*+ xg_{inter}^* )-b_I]}} \\
      \end{aligned}
    \right.
\end{equation}

and where $\vv{x_{inter}}$, given by Eq \ref{eq:xinter}, will be the support of information transfer between the two pools (hence, denoted \emph{transfer variables}).

\subsubsection{Expressing sensitivities w.r.t. forcings}

We start from Eq \ref{1reg_fullmodel} where we consider the dependency to $x_E$.

With the perturbation, we get the new model:

\begin{equation}
  \left\{
      \begin{aligned}
        &\vv{F}(\vv{s},\vv{x_{inter}})= \vv{0}\\
        &\vv{x_{inter}}= \vv{w_{inter}}(\vv{s}, x_E + \delta x_E, x_I)
      \end{aligned}
    \right.
    \label{sys1regionxE}
\end{equation}

We search the formal expression for:

\begin{equation}
\begin{pmatrix}
        \mathcal{A}_{sn,x_E} \\[3mm]
        \mathcal{A}_{sg,x_E} \\
        \end{pmatrix} \equiv
        \begin{pmatrix}
        \frac{\delta sn }{\delta x_E} \\[3mm]
        \frac{\delta sg }{\delta x_E} \\
        \end{pmatrix} 
=\vv{\frac{\delta s}{\delta x_E}}   
\label{eq:sensiUneRegion}   
\end{equation}

The linearization for the perturbation around fixed points $\vec{s^*} = (sn^*,sg^*)$  and $\vv{x_{inter}} = (xn_{inter}^*,xg_{inter}^*)$ yields:

\begin{equation}
  \left\{
      \begin{aligned}
        &\ten{\frac{\partial F}{\partial s}} \vv{\delta s} + \ten{\frac{\partial F}{\partial x_{inter}}} \vv{\delta x_{inter}} = \vv{0}\\
        &\vv{\delta x_{inter}}= \ten{\frac{\partial w_{inter}}{\partial s}} \vv{\delta s}+\vv{\frac{\partial w_{inter}}{\partial x_E}} \delta x_E\\
      \end{aligned}
    \right.
\end{equation}

Plugging the last expression into the first, we get:

\begin{equation}
    \ten{\frac{\partial F}{\partial s}} \vv{\delta s} 
    + 
    \ten{\frac{\partial F}{\partial x_{inter}}} \ten{\frac{\partial w_{inter}}{\partial s}} \vv{\delta s} 
    +
    \ten{\frac{\partial F}{\partial x_{inter}}} \vv{\frac{\partial w_{inter}}{\partial x_E}} \delta x_E
    = \vv{0}
\end{equation}

Rearranging to read how $\vv{\delta s}$ depends upon $\delta x_E$:

\begin{equation}
    \big( \ten{\Id} 
    + 
    \ten{\frac{\partial F}{\partial s}} ^{-1} \ten{\frac{\partial F}{\partial x_{inter}}} \ten{\frac{\partial w_{inter}}{\partial s}} \big) \vv{\delta s} 
    =
    -\ten{\frac{\partial F}{\partial s}} ^{-1} \ten{\frac{\partial F}{\partial x_{inter}}} \vv{\frac{\partial w_{inter}}{\partial x_E}} \delta x_E
\end{equation}

Defining

\begin{equation}
    \ten{S} =  -\ten{\frac{\partial F}{\partial s}} ^{-1} \ten{\frac{\partial F}{\partial x_{inter}}}
\label{defMatS}
\end{equation}

we get:

\begin{equation}
    \big( \ten{\Id} - \ten{S} \ten{\frac{\partial w_{inter}}{\partial s}} \big) \vv{\delta s} = \ten{S} \vv{\frac{\partial w_{inter}}{\partial x_E}} \delta x_E
\label{Eq:deltaXe}
\end{equation}

so that:

\begin{equation}
\begin{pmatrix}
        \mathcal{A}_{sn,x_E} \\[3mm]
        \mathcal{A}_{sg,x_E} \\
        \end{pmatrix} =
        \vv{\frac{\delta s}{\delta x_E}} =
    \big( \ten{\Id} - \ten{S} \ten{\frac{\partial w_{inter}}{\partial s}} \big)^{-1}  \ten{S} \vv{\frac{\partial w_{inter}}{\partial x_E}}      
\end{equation}

\bigskip

Now turning to $x_I$, we can proceed the same way, expressing at fixed point the model with pertubation::

\begin{equation}
  \left\{
      \begin{aligned}
        &\vv{F}(\vv{s},\vv{x_{inter}})= \vv{0}\\
        &\vv{x_{inter}}= \vv{w_{inter}}(\vv{s}, x_E, x_I + \delta x_I)
      \end{aligned}
    \right.
\end{equation}

and we obtain:

\begin{equation}
    \begin{pmatrix}
        \mathcal{A}_{sn,x_I} \\[3mm]
        \mathcal{A}_{sg,x_I} \\
    \end{pmatrix} = 
    \vv{\frac{\delta s}{\delta x_I}} = 
    \big( \ten{\Id} - \ten{S} \ten{\frac{\partial w_{inter}}{\partial s}} \big)^{-1}  \ten{S} \vv{\frac{\partial w_{inter}}{\partial x_I}}   
\end{equation}

Using that:
\begin{equation}
    \vv{\frac{\partial w_{inter}}{\partial x_E}} = 
    \begin{bmatrix}
        1 \\
        0
    \end{bmatrix} \ \ \ \text{and} \quad
    \vv{\frac{\partial w_{inter}}{\partial x_I}} = 
    \begin{bmatrix}
        0 \\
        1
    \end{bmatrix}
\end{equation}

we finally get a compact expression for the matrix of sensitivities to excitatory perturbations upon external forcings in either excitatory or inhibitory pool:

\begin{equation}
    \begin{pmatrix}
         \mathcal{A}_{sn,x_E} &  \mathcal{A}_{sn,x_I}\\
         \mathcal{A}_{sg,x_E} & \mathcal{A}_{sg,x_I}
    \end{pmatrix} 
=
    \big( \ten{\Id} - \ten{S} \ten{\frac{\partial w_{inter}}{\partial s}} \big)^{-1}  \ten{S}
\end{equation}

\subsubsection{Expressing Open Loop Sensitivities}

In the closed loop situation, from Eq \ref{eq:xinter}, we have that:

\begin{equation}
    \ten{\frac{\partial w_{inter}}{\partial s}} = 
    \begin{pmatrix}
        0 & -J_{gaba} \\
        J_{nmda} & 0
    \end{pmatrix}
\end{equation}

and in the open loop situation, we set: $\ten{\frac{\partial w_{inter}}{\partial s}} = \ten{0}$.

\bigskip

Under perturbation upon $x_E$, Eq~\ref{Eq:deltaXe} is

\begin{equation}
 \big( \ten{\Id} - \underbrace{\ten{S} \ten{\frac{\partial w_{inter}}{\partial s}}}_{\ten{G}} \big) \vv{\delta s}=  \ten{S} \vv{\frac{\partial w_{inter}}{\partial x_E}}   {\delta x_E}    
\end{equation}

where $\ten{G}$ is the matrix of feedback gain due to the closed loop between the two pools.

In the open loop situation, $\ten{G}$ is then nullified while the r.h.s. remains untouched.

In this case, 

\begin{equation}
 \vv{\delta s}^{O}=  \ten{S} \vv{\frac{\partial w_{inter}}{\partial x_E} }  {\delta x_E}    
\end{equation}

represents the effect of perturbation upon fixed point values when the feedback loop has been opened.

The canonical form can then be written as:

\begin{equation}
 \big( \ten{\Id} - \ten{G} \big) \vv{\delta s}= \vv{\delta s}^{O}
\end{equation}

\bigskip

We have:
\begin{equation} 
 \frac{ \vv{\delta s}^{O}}{\delta x_E} = \ten{S} \underbrace{    \vv{\frac{\partial w_{inter}}{\partial x_E}}   }_{\begin{bmatrix}
    1\\
    0 
    \end{bmatrix}
    }
    = \begin{pmatrix}
        \mathcal{A}_{sn,x_E}^{O} \\
        0 \\
    \end{pmatrix}
\end{equation}

Obviously, in the open loop condition, the inhibitory pool is not affected by a perturbation upon $x_E$.

The same way, considering perturbation upon $x_I$, we get:

\begin{equation}
 \frac{ \vv{\delta s}^{O}}{\delta x_I} = \ten{S} \underbrace{    \vv{\frac{\partial w_{inter}}{\partial x_I}}   }_{\begin{bmatrix}
   0\\
    1 
    \end{bmatrix}
    }
    = \begin{pmatrix}
        0\\
        \mathcal{A}_{sg,x_I}^{O}  \\
    \end{pmatrix}
\end{equation}

In the open loop condition gain, the excitatory pool is not affected by a perturbation upon $x_I$.

Hence $\ten{S}$ reads:
\begin{equation}
 \ten{S} = \begin{pmatrix}
        \mathcal{A}_{sn,x_E}^{O} & 0 \\
        0 &  \mathcal{A}_{sg,x_I}^{O}\\
    \end{pmatrix}
\label{SEnLambda}
\end{equation}

and represent the matrix of open loop sensitivities.

\subsubsection{Using Isolated Pool Sensitivities}
\label{sec:UsingIsolatedPools}

Now, lets consider $z_E$ as: 
\begin{equation}
  z_E = -J_{gaba} sg^* + x_E = xn_{inter} = wn_{inter} (sg^*)
\end{equation}

It represents the total amount of forcing at excitatory pool, due to the basic external forcing and the feedback current from the inhibitory pool.

Considering that perturbing here $x_E$ is the same thing as perturbing $z_E$ in the excitatory pool considered as an isolated pool, and that the same is true for the inhibitory pool, we then have 

\begin{equation}
  \left\{
      \begin{aligned}
	\mathcal{A}_{sn,x_E}^{O} & = \varphi n_{_E} (sn^*, -J_{gaba} sg^* + x_E)\\
	\mathcal{A}_{sg,x_I}^{O} &= \varphi g_{_I} (sg^*, J_{nmda} sn^* + x_I)
	\end{aligned}
	\right.
\label{eq:2regPhis}
\end{equation}

where $\varphi n_{_E}$ and $\varphi g_{_I}$ are given by Eq \ref{eq:funSensiPools}
and are to be evaluated at the fixed points $sn^*$ and $sg^*$ yielded by the closed loop system, and taking into account the total amount of external forcing.

\subsubsection{Closed Loop Sensitivities as functions of Open Loop Sensitivities }

We have

\begin{equation}
\ten{\frac{\partial w_{inter}}{ \partial s} }=
    \begin{bmatrix}
    \frac{\partial wn_{inter}}{\partial sn} & \frac{\partial wn_{inter}}{\partial sg} \\[3mm]
    \frac{\partial wg_{inter}}{\partial sn} & \frac{\partial wg_{inter}}{\partial sg} 
    \end{bmatrix}=
    \begin{bmatrix}
    0 & -J_{gaba} \\[3mm]
    J_{nmda} & 0
    \end{bmatrix}
\end{equation}

so, using Eq \ref{SEnLambda}, we get:

\begin{align}
	\big(\ten{\Id} - \ten{S}\ten{\frac{\partial w_{inter}}{\partial s}}\big)& 
=  \ten{\Id}-
    \begin{bmatrix}
    	\mathcal{A}_{sn,x_E}^{O} & 0 \\[3mm]
    	0 & \mathcal{A}_{sg,x_I}^{O}
    \end{bmatrix}
    \begin{bmatrix}
    	0 & -J_{gaba} \\[3mm]
    	J_{nmda} & 0
    \end{bmatrix}\\
    &=
    \ten{\Id}-
    \begin{bmatrix}
    	0 & -J_{gaba}\mathcal{A}_{sn,x_E}^{O} \\[3mm]
    	J_{nmda}\mathcal{A}_{sg,x_I}^{O} & 0
    \end{bmatrix}\\
    &=
    \begin{bmatrix}
    	1 & J_{gaba}\mathcal{A}_{sn,x_E}^{O} \\[3mm]
    	-J_{nmda}\mathcal{A}_{sg,x_I}^{O} & 1
    \end{bmatrix}
\end{align}

Inverting:

\begin{align}
\big(\ten{\Id} - \ten{S}\ten{\frac{\partial w_{inter}}{\partial s}}\big)^{-1} 
    &=
    \frac{1}{1 + J_{gaba}\mathcal{A}_{sn,x_E}^{O} J_{nmda}\mathcal{A}_{sg,x_I}^{O} }
    \begin{bmatrix}
        1 & -J_{gaba}\mathcal{A}_{sn,x_E}^{O} \\[3mm]
        J_{nmda}\mathcal{A}_{sg,x_I}^{O} & 1 
    \end{bmatrix}\\
\end{align}

Finally, 
\begin{equation}
\begin{aligned}
     \begin{pmatrix}
         \mathcal{A}_{sn,x_E} &  \mathcal{A}_{sn,x_I}\\
         \mathcal{A}_{sg,x_E} & \mathcal{A}_{sg,x_I}
    \end{pmatrix} 
    &=
    \big( \ten{\Id} - \ten{S} \ten{\frac{\partial w_{inter}}{\partial s}} \big)^{-1}  \ten{S} \\
    &=
    \frac{1}{1 + J_{nmda} J_{gaba}\mathcal{A}_{sn,x_E}^{O} \mathcal{A}_{sg,x_I}^{O} }\\
    &\times 
    \begin{bmatrix}
        \mathcal{A}_{sn,x_E}^{O} & -J_{gaba}\mathcal{A}_{sn,x_E}^{O} \mathcal{A}_{sg,x_I}^{O} \\[3mm]
         J_{nmda} \mathcal{A}_{sn,x_E}^{O} \mathcal{A}_{sg,x_I}^{O} & \mathcal{A}_{sg,x_I}^{O} 
    \end{bmatrix}
\end{aligned}
\end{equation}

which expresses closed loop sensitivities (sensitivities for the pools when they are coupled) as functions of open loop sensitivities (sensitivities for the pools with only their recurrent coupling).

\subsubsection{Functional expressions of closed loop sensitivities} 
\label{sec:sensiuneregion}

We define the functions $\Phi n_{_E}$, $\Phi g_{_E}$, $\Phi n_{_I}$ and $\Phi g_{_I}$ to explicitly express the dependencies of the closed loop sensitivities for an isolated region:

\begin{equation}
  \left\{
  \begin{aligned}
     \mathcal{A}_{sn,x_E}  &= \Phi n_{_E}(\vv{s^{*}},x_E,x_I) = \frac{\mathcal{A}_{sn,x_E}^{O}}{1 + J_{nmda} J_{gaba}\mathcal{A}_{sn,x_E}^{O} \mathcal{A}_{sg,x_I}^{O} }\\[2mm]
     \mathcal{A}_{sg,x_E}  &= \Phi g_{_E}(\vv{s^{*}},x_E,x_I)= \frac{J_{nmda} \mathcal{A}_{sn,x_E}^{O} \mathcal{A}_{sg,x_I}^{O}}{1 + J_{nmda} J_{gaba}\mathcal{A}_{sn,x_E}^{O} \mathcal{A}_{sg,x_I}^{O} }\\[2mm]
     \mathcal{A}_{sn,x_I}  &= \Phi n_{_I}(\vv{s^{*}},x_E,x_I)= \frac{-J_{gaba}\mathcal{A}_{sn,x_E}^{O} \mathcal{A}_{sg,x_I}^{O}}{1 + J_{nmda} J_{gaba}\mathcal{A}_{sn,x_E}^{O} \mathcal{A}_{sg,x_I}^{O} }\\[2mm]
     \mathcal{A}_{sg,x_I}  &= \Phi g_{_I}(\vv{s^{*}},x_E,x_I)= \frac{\mathcal{A}_{sg,x_I}^{O} }{1 + J_{nmda} J_{gaba}\mathcal{A}_{sn,x_E}^{O} \mathcal{A}_{sg,x_I}^{O} }\\  
  \end{aligned}
  \right.
\end{equation}

\bigskip

where the open loop sensitivities are given in Eq \ref{eq:2regPhis}.

\subsection{Two coupled area system}
\label{twocoupledareas}

We now turn to the coupling between two areas.
We seek the expression of their closed loop sensitivities to perturbation as functions of their sensitivities when isolated, as it has been expressed in the previous sections, hence, corresponding to their response to perturbation when the feedback loops are opened at the system scale.

\subsubsection{Splitting $x_{inter}$ from $x_{intra}$}

At fixed point, the system can be expressed as:

\begin{equation}
  \left\{
      \begin{aligned}
        & 0 = -\beta^E sn_i^{*}+\alpha^E T_{glu}(1-sn_i^{*})rn_i^{*} &\equiv fn_i(sn_i^{*},rn_i^{*}) \\
        & 0 =-\beta^I sg_i^{*}+\alpha^I T_{gaba}(1-sg_i^{*})rg_i^{*} &\equiv fg_i(sn_i^{*},rn_i^{*})
      \end{aligned}
    \right.
\end{equation}

with

\begin{equation}
  \left\{
      \begin{aligned}
        &rn_i^{*}= \frac{a_E xn_i^{*}-b_E}{1-e^{-d_E (a_E xn_i^{*}-b_E)}} &\equiv hn_i(xn_i^{*})\\
        &rg_i^{*}= \frac{a_I xg_i^{*}-b_I}{1-e^{-d_I (a_I xg_i^{*}-b_I)}} &\equiv hg_i(xg_i^{*})
      \end{aligned}
    \right.
\label{eq:2regionsRn}
\end{equation}

where $i \in \{1,2\}$.

\bigskip

In Eq \ref{eq:2regionsRn}, $xn_i^*$ and $xg_i^*$ represent the respective total input current to area $i$.
In order to express closed loop sensitivities of the areas as a function of their open loop sensitivities, we split explicitly this total input current between an internal current within an area, due to intra-pool recurrence, and coupling between pools:

\begin{equation}
  \left\{
      \begin{aligned}
        &xn_{intra,i}^*= W_{+}J_{nmda} sn_i^* - J_{gaba_i} sg_i^* &\equiv wn_{intra,i}(sn_i^*,sg_i^*)\\
        &xg_{intra,i}^*= J_{nmda} sn_i^*-J_{-} sg_i^* &\equiv wg_{intra,i}(sn_i^*,sg_i^*)
      \end{aligned}
    \right.
\end{equation}

and external current due to the external inputs and the coupling between the two areas:

\begin{equation}
  \left\{
      \begin{aligned}
        &xn_{inter,i}^*=  k_{E_{ij}} \kappa_{ij} sn_j^* + B_{E_i} &\equiv wn_{inter,i}(sn^*_{j}) \quad &{j \neq i}\\
        &xg_{inter,i}^*=  (1-k_{E_{ij}}) \kappa_{ij} sn_j^* + B_{I_i} &\equiv wg_{inter,i}(sn^*_{j}) \quad &{j \neq i}
      \end{aligned}
    \right.
\label{eq:2regionsxinter}
\end{equation}

so that total input currents reads:

\begin{equation}
  \left\{
      \begin{aligned}
        &xn_i^{*}= xn_{intra,i}^* + xn_{inter,i}^* = wn_i(xn_{intra,i}^*, xn_{inter,i}^*)\\
        &xg_i^{*}= xg_{intra,i}^* + xg_{inter,i}^*  = wg_i(xg_{intra,i}^*, xg_{inter,i}^*)
      \end{aligned}
    \right.
\end{equation}

\subsubsection{Intermediate variables elimination}

We define

\begin{equation}
    \vv{s}=
    \begin{pmatrix}
    sn_1^* \\[2mm]
    sn_2^* \\[2mm]
    sg_1^* \\[2mm]
    sg_2^* \\
    \end{pmatrix}, \ \ \ 
    \vv{r}=
    \begin{pmatrix}
    rn_1^* \\[2mm]
    rn_2^* \\[2mm]
    rg_1^* \\[2mm]
    rg_2^* \\
    \end{pmatrix}, \ \ \ 
    \vv{x}=
    \begin{pmatrix}
    xn_1^* \\[2mm]
    xn_2^* \\[2mm]
    xg_1^* \\[2mm]
    xg_2^* \\
    \end{pmatrix}, \ \ \ 
    \vv{x_{intra}}=
    \begin{pmatrix}
        xn_{intra,1}^* \\[2mm]
        xn_{intra,2}^* \\[2mm]
        xg_{intra,1}^* \\[2mm]
        xg_{intra,2}^* \\
    \end{pmatrix}, \ \ \ 
    \vv{x_{inter}}=
    \begin{pmatrix}
        xn_{inter,1}^* \\[2mm]
        xn_{inter,2}^* \\[2mm]
        xg_{inter,1}^* \\[2mm]
        xg_{inter,2}^* \\
    \end{pmatrix}
\end{equation}

and

\begin{equation}
    \vv{B_E}=
    \begin{pmatrix}
        B_{E_1}\\[2mm]
        B_{E_2} \\
    \end{pmatrix}, \ \ \
    \vv{B_I}=
    \begin{pmatrix}
        B_{I_1}\\[2mm]
        B_{I_2} \\
    \end{pmatrix}
\end{equation}

so that we can summarize the fixed point as:

\begin{equation}
  \left\{
      \begin{aligned}
        &\vv{f}(\vv{s},\vv{r})= \vv{0}\\
        &\vv{r}= \vv{h}(\vv{x})\\
        &\vv{x}= \vv{w}(\vv{x_{intra}},\vv{x_{inter}})\\
        &\vv{x_{intra}}= \vv{w_{intra}}(\vv{s})\\
        &\vv{x_{inter}}= \vv{w_{inter}}(\vv{s},\vv{B_E},\vv{B_I})
      \end{aligned}
    \right.
    \label{sys2region_1eForm}
\end{equation}

Plugging intermediate variables $\vv{r}$, $\vv{x}$, $\vv{x_{intra}}$  into the first equation, Eq \ref{sys2region_1eForm} can be rewritten as:

\begin{equation}
  \left\{
      \begin{aligned}
        &\vv{F}(\vv{s},\vv{x_{inter}})= \vv{0}\\
        &\vv{x_{inter}}= \vv{w_{inter}}(\vv{s},\vv{B_E} ,\vv{B_I})
      \end{aligned}
    \right.
    \label{2reg_fixedpoint}
\end{equation}
 
where
 
\begin{equation}
  \left\{
      \begin{aligned}
        Fn_i(sn_i^*, sg_i^*,xn_{inter,i}^* ) &= -\beta^E sn_i^* \\ & +\alpha^E T_{glu}(1-sn_i^*) \frac{a_E (W_{+}J_{nmda} sn_i^* - J_{gaba_i} sg_i^* + xn_{inter,i}^*)-b_E}{1-e^{-d_E [a_E( W_{+}J_{nmda} sn_i^* - J_{gaba_i} sg_i^* + xn_{inter,i}^*)-b_E]}}\\
        Fg_i(sn_i^*, sg_i^*,xg_{inter,i}^* ) &=-\beta^I sg_i^* \\ & +\alpha^I T_{gaba}(1-sg_i^*)\frac{a_I (J_{nmda} sn_i^*-J_{-} sg_i^*+ xg_{inter,i}^* )-b_I}{1-e^{-d_I [a_I (J_{nmda} sn_i^*-J_{-} sg_i^*+ xg_{inter,i}^* )-b_I]}}
      \end{aligned}
    \right.
    \label{2reg_nouveau_modele}
\end{equation}
 
and where $\vv{x_{inter}}$, given by Eq \ref{eq:2regionsxinter}, will be the transfer variables.

\subsubsection{Expressing sensitivities w.r.t. forcings}

\paragraph{Perturbation of one forcing $B$}
~\\
Here, we build the general expression of the sensitivity, for any given forcing $B$ among $B_{E_1}$, $B_{E_2}$, $B_{I_1}$ or $B_{I_2}$.

The perturbed form of system \ref{2reg_fixedpoint} reads:

\begin{equation}
  \left\{
      \begin{aligned}
        &\vv{F}(\vv{s},\vv{x_{inter}})= \vv{0}\\
        &\vv{x_{inter}}= \vv{w_{inter}}(\vv{s},B + \delta B)\\
      \end{aligned}
    \right.
\end{equation}

By linearization at fixed point, we get:

\begin{equation}
  \left\{
      \begin{aligned}
        &\ten{\frac{\partial F}{\partial s}} \vv{ \delta s}_{B} + \ten{\frac{\partial F}{\partial x_{inter}}} \vv{ \delta x_{inter}}_{B} = \vv{0}\\
        &\vv{\delta x_{inter}}_{B}= \ten{\frac{\partial w_{inter}}{\partial s} } \vv{\delta s}_{B}+\vv{\frac{\partial w_{inter}}{\partial B}} \delta B\\
      \end{aligned}
    \right.
\end{equation}

Plugging the second equation into the first, we have:

\begin{equation}
    \big( \ten{\frac{\partial F}{\partial s}} + \ten{\frac{\partial F}{\partial x_{inter}}} \ten{\frac{\partial w_{inter}}{\partial s} }\big) \vv{\delta s}_{B}
    =
    -
    \ten{\frac{\partial F}{\partial x_{inter}}} \vv{\frac{\partial w_{inter}}{\partial B}} \delta B
\label{pre-canonique}
\end{equation}

Denoting

\begin{equation}
\ten{S} =  -
    \ten{\frac{\partial F}{\partial s}} ^{-1} \ten{\frac{\partial F}{\partial x_{inter}}}  
    \label{Def_Mat_A}
\end{equation}

expression \ref{pre-canonique} can be written under the canonical form as:

\begin{equation}
    \big(  \ten{\Id} - \ten{S} \ten{\frac{\partial w_{inter}}{\partial s}} \big) \vv{\delta s}_{B}
    = \ten{S} \vv{\frac{\partial w_{inter}}{\partial B}} \delta B
\end{equation}

In this expression $ \ten{\frac{\partial w_{inter}}{\partial s}}$ represents the coupling between the two areas, and is:

\begin{align}
\ten{ \frac{\partial w_{inter}}{\partial s} } &= 
     \begin{pmatrix}
        0 & \frac{\partial wn_{inter,1}}{ \partial sn_{2}} & 0 & 0  \\[2mm]
        \frac{\partial wn_{inter,2}}{ \partial sn_{1}} & 0 &  0 & 0  \\[2mm]
        0 & \frac{\partial wg_{inter,1}}{ \partial sn_{2}} &  0 & 0  \\[2mm]
        \frac{\partial wg_{inter,2}}{ \partial sn_{1}} & 0 &  0 & 0  \\[2mm]        
    \end{pmatrix} 
    &= 
    \begin{pmatrix}
    0 &  k_{E_{12}}\kappa_{12} & 0 & 0\\[2mm]
     k_{E_{21}}\kappa_{21} & 0 & 0 & 0\\[2mm]
    0 & (1-k_{E_{12}})\kappa_{12} & 0 & 0\\[2mm]
    (1- k_{E_{21}})\kappa_{21} & 0 & 0 & 0\\[2mm]
    \end{pmatrix} 
\end{align}

If we set  $\ten{\frac{\partial w_{inter}}{\partial s}} = \ten{0}$ while the r.h.s. remains untouched, we then obtain the effect of perturbation in the open loop case:

\begin{equation}
\vv{\delta s}_{B}^{O} =  \ten{S}  \vv{\frac{\partial w_{inter}}{\partial B}} \delta B
\label{eq:openloop2regions}
\end{equation}

The canonical expression then reads:

\begin{equation}
\boxed{
    \big(  \ten{\Id} - \ten{G}\big) \vv{\delta s}_{B}
    = \vv{\delta s}_{B}^{O}
    }
\label{eq:canoniqueBE1}
\end{equation}

where 
\begin{equation}
\ten{G} = - \ten{\frac{\partial F}{\partial s}} ^{-1} \ten{\frac{\partial F}{\partial x_{inter}}} \ten{\frac{\partial w_{inter}}{\partial s}} = \ten{S} \ten{\frac{\partial w_{inter}}{ \partial s}}
\label{GMat}
\end{equation}

is the feedback gain matrix.

\paragraph{Expressing Open Loop Sensitivities}
~\\
Perturbations are operated upon transfer variables, given by $\vv{w_{inter}}$. 

Let $\vv{x_{inter}}^{O}$ denote the transfer variables in open loop condition when $\ten{\frac{\partial w_{inter}}{ \partial s}} = \ten{0}$ (i.e. $\kappa_{12} = \kappa_{21} = 0$):

\begin{equation}
\vv{x_{inter}}^{O} = 
    \begin{pmatrix}
    B_{E_1} \\
    B_{E_2} \\
    B_{I_1} \\
    B_{I_2}
    \end{pmatrix}
\end{equation}

Since forcing parameters are independent, we have: $\ten{\frac{\partial w_{inter}}{\partial x_{inter}^{O}}} = \ten{\Id}$.

Hence considering each perturbation one by one,
the operator $\vv{\frac{\partial w_{inter}}{\partial B}}$ acts as a selector of a column of $\ten{S}$, so that we can express the sensitivities in closed loop condition as function of sensitivities in open loop condition.

For instance, perturbing $B_{E_1}$, we get:

\begin{equation}
 \vv{\delta s}_{B_{E_1}}^{O}= \ten{S} \vv{\frac{\partial w_{inter}}{\partial B_{E_1}}}  { \delta B_{E_1} }  \Longleftrightarrow
 {\frac{ \vv{\delta s}_{B_{E_1}}^{O}}{\delta B_{E_1}}} = \ten{S} \underbrace{    \vv{\frac{\partial w_{inter}}{\partial B_{E_1}}}   }_{\begin{bmatrix}
    1\\
    0\\
    0 \\
    0
    \end{bmatrix}
    }
\end{equation}

hence the open loop sensitivities of both areas to a perturbation upon the excitatory pool of the first area yields the first column of $\ten{S}$.

Obviously, in the open loop condition, the second area is not perturbed at all.

\bigskip

Furthermore, considering that perturbing here $B_{E_1}$ is the same thing as perturbing $x_{E_1}$ in the first area considered as an isolated area, we have, by definition given in Eq \ref{eq:sensiUneRegion}, that:

\begin{equation}
     \vv{\delta s}_{B_{E_1}}^{O} =
     \begin{pmatrix}
        \delta sn_{1} \\[2mm]
        \delta sn_{2} \\[2mm]
        \delta sg_{1} \\[2mm]
        \delta sg_{2} \\[2mm]
    \end{pmatrix}_{B_{E_1}}^{O}
    = \begin{pmatrix}
        \mathcal{A}_{sn_1,x_{E_1}} \delta B_{E_1} \\[2mm]
        0\\[2mm]
        \mathcal{A}_{sg_1,x_{E_1}}  \delta B_{E_1} \\[2mm]
        0\\[2mm] 
    \end{pmatrix}
\end{equation}

hence the first column of $\ten{S}$ is given by:

\begin{equation}
    \frac{\vv{\delta s}_{B_{E_1}}^{O} }{\delta B_{E_1}} = 
    \begin{pmatrix}
        \mathcal{A}_{sn_1,x_{E_1}}\\
        0\\
        \mathcal{A}_{sg_1,x_{E_1}}\\
        0\\ 
    \end{pmatrix}
\end{equation}
     
Following the same lines of reasoning for the three other perturbations, we finally obtain:
  
\begin{align}
    \ten{S} &= 
    \begin{pmatrix}
        \ten{\mathcal{A}_{sn,x_{E}}} & \ten{\mathcal{A}_{sn,x_{I}}} \\
        \ten{\mathcal{A}_{sg,x_{E}}} & \ten{\mathcal{A}_{sg,x_{I}}}
     \end{pmatrix}
 \end{align}
 
with
\begin{align}
    \ten{\mathcal{A}_{sn,x_{E}}} 
    &= 
    \begin{pmatrix}
        \mathcal{A}_{sn_1,x_{E_1}} & 0 \\
        0 & \mathcal{A}_{sn_2,x_{E_2}}
    \end{pmatrix}\text{,}\quad
    &\ten{\mathcal{A}_{sn,x_{I}}}
    =
    \begin{pmatrix}
        \mathcal{A}_{sn_1,x_{I_1}} & 0 \\
        0 & \mathcal{A}_{sn_2,x_{I_2}}\\
    \end{pmatrix}\\
    \ten{\mathcal{A}_{sg,x_{E}}} 
    &=
    \begin{pmatrix}
        \mathcal{A}_{sg_1,x_{E_1}} & 0 \\
        0 & \mathcal{A}_{sg_2,x_{E_2}} 
    \end{pmatrix}\text{,}\quad
    &\ten{\mathcal{A}_{sg,x_{I}}} 
    =
    \begin{pmatrix}
         \mathcal{A}_{sg_1,x_{I_1}} & 0\\
         0 & \mathcal{A}_{sg_2,x_{I_2}}
    \end{pmatrix}
\end{align}

\subsubsection{Using Isolated Area Sensitivities}

In the same spirit as in Sec \ref{sec:UsingIsolatedPools}, these open loop sensitivities of areas can be expressed by the analytical expression of their sensitivities when considered isolated, yet to be evaluated respectively at fixed points $\vv{s_1}^*$ and $\vv{s_2}^*$ yielded by the closed loop system and taking into account the total amount of external forcing, so we write:  

\begin{equation}
  \left\{
    \begin{aligned}
       \mathcal{A}_{sn_i,x_{E_i}} &= \Phi n_{_E}(\vv{s^{*}} = \vv{s_i^{*}},x_E = xn_{inter,i}^*, x_I = xg_{inter,i}^*)   \\[2mm]
       \mathcal{A}_{sg_i,x_{E_i}} &= \Phi g_{_E}(\vv{s^{*}} = \vv{s_i^{*}},x_E = xn_{inter,i}^*, x_I = xg_{inter,i}^*)  \\[2mm]
       \mathcal{A}_{sn_i,x_{I_i}} &= \Phi n_{_I}(\vv{s^{*}} = \vv{s_i^{*}},x_E = xn_{inter,i}^*, x_I = xg_{inter,i}^*)  \\[2mm]
       \mathcal{A}_{sg_i,x_{I_i}} &= \Phi g_{_I}(\vv{s^{*}} = \vv{s_i^{*}},x_E = xn_{inter,i}^*, x_I = xg_{inter,i}^*)  
  \end{aligned}
  \right.
\end{equation}

where $xn_{inter,i}^*$ and  $x_I = xg_{inter,i}^*$ are defined in Eq \ref{eq:2regionsxinter}.
Analytical expression for the functions $\Phi n_{_E}$, $\Phi g_{_E}$, $\Phi n_{_I}$ and $\Phi g_{_I}$ are explicitly given in section \ref{sec:sensiuneregion}.\\

\paragraph{Closed loop sensitivities  as functions of Open Loop Sensitivities}
~\\
From definition \ref{GMat} for $\ten{G}$, we have:

\begin{equation}
\ten {G} =
    \begin{pmatrix}
        \mathcal{A}_{sn_1,x_{E_1}} & 0 & \mathcal{A}_{sn_1,x_{I_1}} & 0 \\
        0 & \mathcal{A}_{sn_2,x_{E_2}} & 0 & \mathcal{A}_{sn_2,x_{I_2}}\\
        \mathcal{A}_{sg_1,x_{E_1}} & 0 & \mathcal{A}_{sg_1,x_{I_1}} & 0\\
        0 & \mathcal{A}_{sg_2,x_{E_2}} & 0 & \mathcal{A}_{sg_2,x_{I_2}}\\ 
    \end{pmatrix} 
    \begin{bmatrix}
    0 &  k_{E_{12}}\kappa_{12} & 0 & 0\\
     k_{E_{21}}\kappa_{21} & 0 & 0 & 0\\
    0 & (1-k_{E_{12}})\kappa_{12} & 0 & 0\\
    (1- k_{E_{21}})\kappa_{21} & 0 & 0 & 0\\
    \end{bmatrix} 
\end{equation}
 
that we will write as: 
\begin{align}
\ten {G} =
    \begin{bmatrix}
        0 & G_{12} & 0 & 0 \\
        G_{21} & 0 & 0 & 0 \\
        0 & G_{32} & 0 & 0 \\
        G_{41} & 0 & 0 & 0 
    \end{bmatrix}
\end{align}

with

\begin{equation}
    \left\{
    \begin{aligned}
        G_{12} &= \mathcal{A}_{sn_1,x_{E_1}}  k_{E_{12}}\kappa_{12} + \mathcal{A}_{sn_1,x_{I_1}} (1-k_{E_{12}})\kappa_{12} \\
        G_{21} &= \mathcal{A}_{sn_2,x_{E_2}}  k_{E_{21}}\kappa_{21} +  \mathcal{A}_{sn_2,x_{I_2}}  (1- k_{E_{21}})\kappa_{21} \\
        G_{32} &=  \mathcal{A}_{sg_1,x_{E_1}} k_{E_{12}}\kappa_{12} + \mathcal{A}_{sg_1,x_{I_1}}(1-k_{E_{12}})\kappa_{12} \\
        G_{41} &= \mathcal{A}_{sg_2,x_{E_2}} k_{E_{21}}\kappa_{21} + \mathcal{A}_{sg_2,x_{I_2}} (1- k_{E_{21}})\kappa_{21}
    \end{aligned}
    \right.
\end{equation}

\bigskip

For the l.h.s. term in expression \ref{eq:canoniqueBE1}, we then have:

\begin{equation}
    \ten{\Id} - \ten{G} =  
    \left(
    \begin{tabular}{c c|c c}
        1 & $-G_{12}$ & 0 & 0 \\
        $-G_{21}$ & 1 & 0 & 0 \\
        \hline
        0 & $-G_{32}$ & 1 & 0 \\
        $-G_{41}$ & 0 & 0 & 1 
    \end{tabular}
     \right) 
\end{equation}

that we will write as:

\begin{equation}
\ten{\Id} - \ten{G} =
     \left(
    \begin{tabular}{c |c}
        $\ten{B_1}$ & $\ten{B_2}$ \\
        \hline
        $\ten{B_3}$ & $\ten{B_4}$ \\
    \end{tabular}
     \right)
\label{defGs}
\end{equation}

Considering the property that

\begin{align*}
    &\text{If }  
    \ten{M} = 
    \left(
    \begin{array}{cc}
        \ten{A} & \ten{B} \\
        \ten{C} & \ten{D}
    \end{array}
    \right)
    \text{ with $\ten{D}$ invertible}\\
    & \text{ then }
    \ten{M}^{-1} =
     \left(
    \begin{array}{cc}
        \ten{R} & \ten{S} \\
        \ten{T} & \ten{U}
    \end{array}
    \right)
    \text{ with }
    \left\{
    \begin{aligned}
        \ten{R} &= \big( \ten{A}-\ten{B}\ten{D}^{-1}\ten{C}\big)^{-1}\\
        \ten{S} &= -\ten{R}\ten{B}\ten{D}^{-1}\\
        \ten{T} &= -\ten{D}^{-1}\ten{C}\ten{R}\\
        \ten{U} &= \ten{D}^{-1}\big( \ten{\Id}-\ten{C}\ten{S}\big)\\
    \end{aligned}
    \right. 
\end{align*}

we obtain

\begin{align}
    \big(\ten{\Id} - \ten{G}\big)^{-1} &=
    \left(
    \begin{array}{c | c}
         \ten{C_1} & \ten{C_2}\\[2mm] 
         \hline
         \ten{C_3} & \ten{C_4} 
    \end{array} \right) 
    \text{ where }
    \left\{
    \begin{aligned}
        \ten{C_1} &= \big( \ten{B_1}-\ten{B_2}(\ten{B_4}^{-1})\ten{B_3}\big)^{-1} = \ten{B_1}^{-1} \\
        \ten{C_2} &= -\ten{C_1}\ten{B_2}(\ten{B_4}^{-1}) = \ten{0}\\
        \ten{C_3} &= -\ten{B_4}^{-1}\ten{B_3}\ten{C_1} = -\ten{B_3}\ten{C_1} = -\ten{B_3}\ten{B_1}^{-1} \\
        \ten{C_4} &= \ten{B_4}^{-1}\big( \ten{\Id}-\ten{B_3}\ten{C_2}\big) = \ten{\Id} \\
    \end{aligned}
    \right.\\
    &=\left(
    \begin{array}{c | c}
         \ten{B_1}^{-1} & \ten{0}\\[2mm] 
         \hline
         -\ten{B_3}\ten{B_1}^{-1} & \ten{\Id} 
    \end{array} \right) 
\end{align}

\bigskip

To recover sensitivities for the closed loop condition, we then consider:

\begin{align}
    \big(\ten{\Id} - \ten{G})^{-1} \ten{S} &= 
    \left(
    \begin{array}{c | c}
         \ten{B_1}^{-1} & \ten{0}\\[2mm] 
         \hline
         -\ten{B_3}\ten{B_1}^{-1} & \ten{\Id} \\[2mm]
    \end{array} \right) 
    \left(
    \begin{array}{c | c}
        \ten{\mathcal{A}_{sn,x_{E}}} & \ten{\mathcal{A}_{sn,x_{I}}} \\
        \hline
        \ten{\mathcal{A}_{sg,x_{E}}} & \ten{\mathcal{A}_{sg,x_{I}}}
    \end{array} \right)\\
    &=
    \left(
    \begin{array}{c | c}
        \ten{B_1}^{-1} \ten{\mathcal{A}_{sn,x_E}} & \ten{B_1}^{-1} \ten{\mathcal{A}_{sn,x_I}}\\[2mm]
        \hline
        -\ten{B_3}\ten{B_1}^{-1}  \ten{\mathcal{A}_{sn,x_E}} + \ten{\mathcal{A}_{sg,x_E}} & - \ten{B_3}\ten{B_1}^{-1} \ten{\mathcal{A}_{sn,x_I}} +  \ten{\mathcal{A}_{sg,x_I}} \\[2mm]
    \end{array} \right)
\end{align}

so that 

\begin{equation}
\boxed{
     \frac{\vv{\delta s}_{B}}{\delta B}
    = \big(  \ten{\Id} - \ten{G})^{-1} \ten{S} \vv{\frac{\partial w_{inter}}{\partial B}}
    }
\end{equation}

expresses, in full generality for the two-area system, the matrix of sensitivities to a perturbation upon either forcing $B$, and they are expressed as functions of the sensitivities in single-area system, which are in turn expressed as functions of the sensitivities in the single-pool system.

\subsubsection{Sensitivities in a Target-Control system}

We now focus on the question of how sensitivities would drive the response of the excitatory pool of one area to the activation of the excitatory pool of the other one, depending on the connectivity between the two areas.
Hence, we attribute role to each area: the area 1 which excitatory pool is positively perturbed will be called "Control" area (denoted by C), and the area 2 will be called "Target area" (denoted by T).

From now on, the observable will then be denoted as:

\begin{equation}
    \vv{s}=
    \begin{pmatrix}
    sn_C \\[2mm]
    sn_T \\[2mm]
    sg_C \\[2mm]
    sg_T \\
    \end{pmatrix}
\end{equation}

and the transfer variables as:

\begin{align}
    \vv{x}=
    \begin{pmatrix}
    xn_{intra,C}&=&B_{E_C} &+& k_{E_{CT}} \kappa_{CT} sn_T \\[2mm]
    xn_{intra,T}&=&B_{E_T} &+& k_{E_{TC}} \kappa_{TC} sn_C \\[2mm]
    xg_{intra,C}&=&B_{I_C} &+& (1- k_{E_{CT}}) \kappa_{CT} sn_T \\[2mm]
    xg_{intra,T}&=&B_{I_T} &+& (1- k_{E_{TC}}) \kappa_{TC} sn_C\\   
    \end{pmatrix}
\end{align}

and we focus on:

\begin{equation}
    \begin{pmatrix}
        \delta sn_T \\
        \delta sn_C\\
    \end{pmatrix}
\end{equation}
\bigskip

in response to $\delta B_{E_C}$.

To extract the situation of interest, from the general result above, we then pick the case:

\begin{align}
   \vv{\delta s}_{B_{E_C}}
    & =  \big(  \ten{\Id} - \ten{G} \big) ^{-1}\ten{S} \vv{\frac{\partial w_{inter}}{\partial B_{E_C}}} \delta B_{E_C}\\
    & = \left(
    \begin{array}{c | c}
        \ten{B_1}^{-1} \ten{\mathcal{A}_{sn,x_E}} & \ten{B_1}^{-1} \ten{\mathcal{A}_{sn,x_I}}\\[2mm]
        \hline
        -\ten{B_3}\ten{B_1}^{-1}  \ten{\mathcal{A}_{sn,x_E}} + \ten{\mathcal{A}_{sg,x_E}} & - \ten{B_3}\ten{B_1}^{-1} \ten{\mathcal{A}_{sn,x_I}} +  \ten{\mathcal{A}_{sg,x_I}} \\[2mm]
    \end{array} \right)
    \begin{pmatrix}
        1\\0\\
        \hline 0\\0
    \end{pmatrix}
    \delta B_{E_C}\\
    &=
    \left(
    \begin{array}{c}
        \ten{B_1}^{-1} \ten{\mathcal{A}_{sn,x_E}} 
        \begin{pmatrix}
            1\\0
        \end{pmatrix}\\[2mm]
        \hline
        -\ten{B_3}\ten{B_1}^{-1}  \ten{\mathcal{A}_{sn,x_E}} + \ten{\mathcal{A}_{sg,x_E}}
        \begin{pmatrix}
            1\\0
        \end{pmatrix} \\[2mm]
    \end{array} \right)
    \delta B_{E_C} \\
    & \equiv
    \begin{pmatrix}
        \delta sn_C \\
        \delta sn_T \\
        \hline
        \delta sg_C \\
        \delta sg_T \\
    \end{pmatrix}
\end{align}

where the perturbations of interest are in the upper part, and we have:

\begin{align}
 \begin{pmatrix}
        \delta sn_C \\
        \delta sn_T \\
    \end{pmatrix}
    &=        
    \ten{B_1}^{-1} \ten{\mathcal{A}_{sn,x_E}}
        \begin{pmatrix}
            1\\0
        \end{pmatrix} 
        \delta B_{E_C} 
\end{align}

From definition \ref{defGs}, we get

\begin{equation}
    \ten{B_1}^{-1} = 
    \begin{pmatrix}
        1 & -G_{12}\\
        -G_{21} & 1 
    \end{pmatrix} ^{-1} = 
    \frac{1}{1 -G_{12} G_{21}}
    \begin{pmatrix}
        1 & G_{12}\\
        G_{21} & 1 
    \end{pmatrix}
\end{equation}

hence

\begin{align}
 \begin{pmatrix}
        \delta sn_C \\
        \delta sn_T \\
    \end{pmatrix}
    &=
    \frac{1}{1 -G_{12} G_{21} }
    \begin{pmatrix}
        1 & G_{12}\\
        G_{21} & 1 
    \end{pmatrix}
    \begin{pmatrix}
        \mathcal{A}_{sn_C,x_{E_C}} & 0 \\
        0 & \mathcal{A}_{sn_T,x_{E_T}}
    \end{pmatrix} 
    \begin{pmatrix}
            1\\0
    \end{pmatrix}
    \delta B_{E_C}\\
    &=
    \frac{1}{1 -G_{12} G_{21} }
    \begin{pmatrix}
        \mathcal{A}_{sn_C,x_{E_C}} & G_{12} \mathcal{A}_{sn_T,x_{E_T}} \\
         G_{21} \mathcal{A}_{sn_C,x_{E_C}} & \mathcal{A}_{sn_T,x_{E_T}}
    \end{pmatrix} 
    \begin{pmatrix}
            1\\0
    \end{pmatrix} 
    \delta B_{E_C}\\
    &=\frac{1}{1 -G_{12} G_{21} }
    \begin{pmatrix}
        \mathcal{A}_{sn_C,x_{E_C}} \\
         G_{21} \mathcal{A}_{sn_C,x_{E_C}} 
    \end{pmatrix} 
    \delta B_{E_C}
\end{align}

with

\begin{equation}
    \left\{
    \begin{aligned}
        G_{12} &= \mathcal{A}_{sn_C,x_{E_C}}  k_{E_{CT}}\kappa_{CT} + \mathcal{A}_{sn_C,x_{I_C}} (1-k_{E_{CT}})\kappa_{CT} \\
        G_{21} &= \mathcal{A}_{sn_T,x_{E_T}}  k_{E_{TC}}\kappa_{TC} +  \mathcal{A}_{sn_T,x_{I_T}}  (1- k_{E_{TC}})\kappa_{TC} 
    \end{aligned}
    \right.
\end{equation}

For the sake of clarity, $G_{12}$ and $G_{21}$ are denoted $G_{CT}$ and $G_{TC}$ in the main text.

\clearpage

\section{Appendix 2~: Parameters values of reference as found in Naskar et al. \cite{Naskar2021}.}
\label{S2_Appendix}

\begin{table}[h!]
	\caption{Set of parameters.}
	
	\begin{tabular}{llll}
		Parameter	&	Value	&  Unit\\	
		\toprule
		$\beta^E$	&  6.6 	&  s$^{-1}$\\
		\midrule
		$\alpha^E$	&  72   & $s^{-1} mM^{-1}$\\
		\midrule
		$T_{glu}$	&  0.008 & $mM \ s $\\
		\midrule
		$\beta^I$	&  180  & $s^{-1}$\\
		\midrule
		$\alpha^I$	&  530  & $s^{-1} mM^{-1} $\\
		\midrule
		$T_{gaba}$ 	& 0.003 & $mM \ s$\\

		\midrule
		$a_E$	&	310	& $nC^{-1}$ \\
		\midrule
		$b_E$	& 125	& $Hz$  \\
		\midrule
		$d_E$	& 0.16	& $s$  \\
		\midrule
		$a_I$	& 615	& $nC^{-1} $ \\
		\midrule
		$b_I$	& 177	&$ Hz $ \\
		\midrule
		$d_I$	& 0.087	& $s $ \\

		\midrule
		$W_E$	& 1	& ---  \\
		\midrule
		$W_I$	& 0.7	& ---  \\
		\midrule
		$I_0$	& 0.382	& $nA$  \\
		\midrule
		$W_{+}$	& 1.4	& ---  \\
		\midrule
		$J_{nmda}$	& 0.15	& $nA$   \\
		\midrule
		$J_{gaba_i}$ & [0.25;3.0]	& $nA$\\
		\midrule
		$G$	& 0.69	&  --- \\
		\midrule
		$J_{-}$	& 1	& $nA$  \\
		\bottomrule
	\end{tabular}
	\label{tab:params}
\end{table}

\clearpage

\section{Numerical illustrations}

Numerical illustrations of the formal developments are given below using the parameters values of reference listed in Sec \ref{S2_Appendix}.

\clearpage

\newgeometry{textwidth=7.0in,textheight=8.75in}

\newpage

\newpage

\newpage

\newpage

\newpage

\newpage

\newpage

\newpage

\clearpage


\begin{thebibliography}{100}

\bibitem{Naskar2021}
Naskar A, Vattikonda A, Deco G, Roy D, Banerjee A.
\newblock {Multiscale dynamic mean field (MDMF) model relates resting-state
  brain dynamics with local cortical excitatory–inhibitory neurotransmitter
  homeostasis}.
\newblock Network Neuroscience. 2021;5(3):757--782.
\newblock doi:{10.1162/netn\_a\_00197}.

\bibitem{Sanzleon2013}
Sanzleon P, Knock SA, Woodman MM, Domide L, Mersmann J, Mcintosh AR, et~al.
\newblock {The virtual brain: A simulator of primate brain network dynamics}.
\newblock Frontiers in Neuroinformatics. 2013;7(MAY).
\newblock doi:{10.3389/fninf.2013.00010}.

\bibitem{Sanz-Leon2015}
Sanz-Leon P, Knock SA, Spiegler A, Jirsa VK.
\newblock {Mathematical framework for large-scale brain network modeling in The
  Virtual Brain}.
\newblock NeuroImage. 2015;111:385--430.
\newblock doi:{10.1016/j.neuroimage.2015.01.002}.

\bibitem{Amunts2022}
Amunts K, Defelipe J, Pennartz C, Destexhe A, Migliore M, Ryvlin P, et~al.
\newblock {Linking Brain Structure, Activity, and Cognitive Function through
  Computation}.
\newblock eNeuro. 2022;9(2).
\newblock doi:{10.1523/ENEURO.0316-21.2022}.

\bibitem{Schirner2022}
Schirner M, Domide L, Perdikis D, Triebkorn P, Stefanovski L, Pai R, et~al.
\newblock {Brain simulation as a cloud service: The Virtual Brain on EBRAINS}.
\newblock NeuroImage. 2022;251(November 2021):118973.
\newblock doi:{10.1016/j.neuroimage.2022.118973}.

\bibitem{Honey2009}
Honey CJ, Sporns O, Cammoun L, Gigandet X, Thiran JP, Meuli R, et~al.
\newblock {Predicting human resting-state functional connectivity from
  structural connectivity}.
\newblock Proceedings of the National Academy of Sciences of the United States
  of America. 2009;106(6):2035--2040.
\newblock doi:{10.1073/pnas.0811168106}.

\bibitem{Deco2012}
Deco G, Jirsa VK.
\newblock {Ongoing cortical activity at rest: Criticality, multistability, and
  ghost attractors}.
\newblock Journal of Neuroscience. 2012;32(10):3366--3375.
\newblock doi:{10.1523/JNEUROSCI.2523-11.2012}.

\bibitem{Deco2014a}
Deco G, Ponce-Alvarez A, Hagmann P, Romani GL, Mantini D, Corbetta M.
\newblock {How local excitation-inhibition ratio impacts the whole brain
  dynamics}.
\newblock Journal of Neuroscience. 2014;34(23):7886--7898.
\newblock doi:{10.1523/JNEUROSCI.5068-13.2014}.

\bibitem{Hansen2015}
Hansen ECA, Battaglia D, Spiegler A, Deco G, Jirsa VK.
\newblock {Functional connectivity dynamics: Modeling the switching behavior of
  the resting state}.
\newblock NeuroImage. 2015;105:525--535.
\newblock doi:{10.1016/j.neuroimage.2014.11.001}.

\bibitem{Castro2020}
Castro S, El-Deredy W, Battaglia D, Orio P.
\newblock {Cortical ignition dynamics is tightly linked to the core
  organisation of the human connectome}.
\newblock PLoS Computational Biology. 2020;16(7):1--23.
\newblock doi:{10.1371/journal.pcbi.1007686}.

\bibitem{Kobeleva2022}
Kobeleva X, Varoquaux G, Dagher A, Adhikari M, Grefkes C, Gilson M.
\newblock {Advancing brain network models to reconcile functional neuroimaging
  and clinical research}.
\newblock NeuroImage: Clinical. 2022;36(August 2021).
\newblock doi:{10.1016/j.nicl.2022.103262}.

\bibitem{Sporns2005}
Sporns O, Tononi G, K{\"{o}}tter R.
\newblock {The human connectome: A structural description of the human brain}.
\newblock PLoS Computational Biology. 2005;1(4):0245--0251.
\newblock doi:{10.1371/journal.pcbi.0010042}.

\bibitem{Bullmore2009}
Bullmore E, Sporns O.
\newblock {Complex brain networks: Graph theoretical analysis of structural and
  functional systems}.
\newblock Nature Reviews Neuroscience. 2009;10(3):186--198.
\newblock doi:{10.1038/nrn2575}.

\bibitem{Rubinov2010}
Rubinov M, Sporns O.
\newblock {Complex network measures of brain connectivity: Uses and
  interpretations}.
\newblock NeuroImage. 2010;52(3):1059--1069.
\newblock doi:{10.1016/j.neuroimage.2009.10.003}.

\bibitem{Betzel2016}
Betzel RF, Avena-Koenigsberger A, Go{\~{n}}i J, He Y, de~Reus MA, Griffa A,
  et~al.
\newblock {Generative models of the human connectome}.
\newblock NeuroImage. 2016;124:1054--1064.
\newblock doi:{10.1016/j.neuroimage.2015.09.041}.

\bibitem{Elam2021}
Elam JS, Glasser MF, Harms MP, Sotiropoulos SN, Andersson JLR, Burgess GC,
  et~al.
\newblock {The Human Connectome Project: A retrospective}.
\newblock NeuroImage. 2021;244(August):118543.
\newblock doi:{10.1016/j.neuroimage.2021.118543}.

\bibitem{Munakata2011}
Munakata Y, Herd SA, Chatham CH, Depue BE, Banich MT, O'Reilly RC.
\newblock {A unified framework for inhibitory control}.
\newblock Trends in Cognitive Sciences. 2011;15(10):453--459.
\newblock doi:{10.1016/j.tics.2011.07.011}.

\bibitem{Anderson2021}
Anderson MC, Hulbert JC.
\newblock {Active Forgetting: Adaptation of Memory by Prefrontal Control}.
\newblock Annual Review of Psychology. 2021;72:1--36.
\newblock doi:{10.1146/annurev-psych-072720-094140}.

\bibitem{Apsvalka2022}
Ap{\v{s}}valka D, Ferreira CS, Schmitz TW, Rowe JB, Anderson MC.
\newblock {Dynamic targeting enables domain-general inhibitory control over
  action and thought by the prefrontal cortex}.
\newblock Nature Communications. 2022;13(1):1--21.
\newblock doi:{10.1038/s41467-021-27926-w}.

\bibitem{Wessel2024}
Wessel JR, Anderson MC.
\newblock {Neural mechanisms of domain-general inhibitory control}.
\newblock Trends in Cognitive Sciences. 2024;28(2):124--143.
\newblock doi:{10.1016/j.tics.2023.09.008}.

\bibitem{Gagnepain2017}
Gagnepain P, Hulbert J, Anderson MC.
\newblock {Parallel Regulation of Memory and Emotion Supports the Suppression
  of Intrusive Memories}.
\newblock The Journal of Neuroscience. 2017;37(27):6423--6441.
\newblock doi:{10.1523/JNEUROSCI.2732-16.2017}.

\bibitem{Deco2018}
Deco G, Cruzat J, Cabral J, Knudsen GM, Carhart-Harris RL, Whybrow PC, et~al.
\newblock Whole-Brain Multimodal Neuroimaging Model Using Serotonin Receptor
  Maps Explains Non-linear Functional Effects of LSD.
\newblock Current Biology. 2018;28:3065--3074.e6.
\newblock doi:{10.1016/j.cub.2018.07.083}.

\bibitem{Wong2006}
Wong KF, Wang XJ.
\newblock {A recurrent network mechanism of time integration in perceptual
  decisions}.
\newblock Journal of Neuroscience. 2006;26(4):1314--1328.
\newblock doi:{10.1523/JNEUROSCI.3733-05.2006}.

\bibitem{Deco2013a}
Deco G, Ponce-Alvarez A, Mantini D, Romani GL, Hagmann P, Corbetta M.
\newblock {Resting-state functional connectivity emerges from structurally and
  dynamically shaped slow linear fluctuations}.
\newblock Journal of Neuroscience. 2013;33(27):11239--11252.
\newblock doi:{10.1523/JNEUROSCI.1091-13.2013}.

\bibitem{Deco2013}
Deco G, Jirsa VK, McIntosh AR.
\newblock {Resting brains never rest: Computational insights into potential
  cognitive architectures}.
\newblock Trends in Neurosciences. 2013;36(5):268--274.
\newblock doi:{10.1016/j.tins.2013.03.001}.

\bibitem{Deco2014}
Deco G, McIntosh AR, Shen K, {Matthew Hutchison} R, Menon RS, Everling S,
  et~al.
\newblock {Identification of optimal structural connectivity using functional
  connectivity and neural modeling}.
\newblock Journal of Neuroscience. 2014;34(23):7910--7916.
\newblock doi:{10.1523/JNEUROSCI.4423-13.2014}.

\bibitem{Glomb2017}
Glomb K, Ponce-Alvarez A, Gilson M, Ritter P, Deco G.
\newblock {Resting state networks in empirical and simulated dynamic functional
  connectivity}.
\newblock NeuroImage. 2017;159(July):388--402.
\newblock doi:{10.1016/j.neuroimage.2017.07.065}.

\bibitem{Abbott2005}
Abbott LF, Chance FS.
\newblock {Drivers and modulators from push-pull and balanced synaptic input}.
\newblock Progress in Brain Research. 2005;149:147--155.
\newblock doi:{10.1016/S0079-6123(05)49011-1}.

\bibitem{Mary2020}
Mary A, Dayan J, Leone G, Postel C, Fraisse F, Malle C, et~al.
\newblock {Resilience after trauma: The role of memory suppression}.
\newblock Science. 2020;367(6479).
\newblock doi:{10.1126/science.aay8477}.

\bibitem{Wang2019}
Wang P, Kong R, Kong X, Liégeois R, Orban C, Deco G, et~al.
\newblock Inversion of a large-scale circuit model reveals a cortical hierarchy
  in the dynamic resting human brain.
\newblock Science Advances. 2019;5(1):eaat7854.
\newblock doi:{10.1126/sciadv.aat7854}.

\bibitem{Anderson2016}
Anderson MC, Bunce JG, Barbas H. Prefrontal–hippocampal pathways underlying
  inhibitory control over memory; 2016.

\bibitem{Harush2017}
Harush U, Barzel B.
\newblock {Dynamic patterns of information flow in complex networks}.
\newblock Nature Communications. 2017;8(1):1--11.
\newblock doi:{10.1038/s41467-017-01916-3}.

\bibitem{Mohan2022}
Mohan VM, Banerjee A.
\newblock {A perturbative approach to study information communication in brain
  networks}.
\newblock Network Neuroscience. 2022;6(4):1275--1295.
\newblock doi:{10.1162/netn\_a\_00260}.

\end{thebibliography}

\end{document}